\documentclass{aa}
\usepackage{graphicx}
\usepackage[varg]{txfonts}
\usepackage{lipsum}
\usepackage{subcaption}         %
\usepackage{lscape}             %
\usepackage{placeins}           %
\usepackage{amsmath}            %
\usepackage[colorlinks]{hyperref}
\usepackage[normalem]{ulem}
\newcommand{\rsun}{\ensuremath{R_{\sun}}}
\newcommand{\kms}{\mbox{km~s$^{-1}$}}
\newcommand{\Lya}{\mbox{Ly-$\alpha$}}
\newcommand{\bnorm}{\ensuremath{I}}
\newcommand{\bdiff}{\ensuremath{b}}
\newcommand{\rdiff}[1]{\ensuremath{\delta b_{#1}}}

\begin{document}
\title{First detection of acoustic-like flux in the middle solar corona}

\author{%
  V. Andretta\inst{\ref{INAF-OAC}}
  \and
  L. Abbo\inst{\ref{INAF-OATo}}
  \and
  G. Jerse\inst{\ref{INAF-OATs}}
  \and
  R. Lionello\inst{\ref{PSI}}
  \and
  G. Naletto\inst{\ref{UniPd},\ref{CNR-IFN}}
  \and
  G. Russano\inst{\ref{INAF-OAC}}
  \and
  D. Spadaro\inst{\ref{INAF-OACt}}
  \and
  M. Stangalini\inst{\ref{ASI}}
  \and
  R. Susino\inst{\ref{INAF-OATo}}
  \and
  M. Uslenghi\inst{\ref{INAF-IASF}}
  \and
  R. Ventura\inst{\ref{INAF-OACt}}
  \and
  A. Bemporad\inst{\ref{INAF-OATo}}
  \and
  Y. De Leo\inst{\ref{INAF-OACt},\ref{UniGraz}}
  \and
  S. Farina\inst{\ref{INAF-IASF}}
  \and
  G. Nistic\`o\inst{\ref{UniCal}}
  \and
  M. Romoli\inst{\ref{UniFi},\ref{INAF-OAA}}
  \and
  Th. Straus\inst{\ref{INAF-OAC}}
  \and
  D. Telloni\inst{\ref{INAF-OATo}}
  \and
  L. Teriaca\inst{\ref{MPS}}
  \and
  A. Burtovoi\inst{\ref{UniFi},\ref{INAF-OATo}}
  \and
  V. Da Deppo\inst{\ref{CNR-IFN}}
  \and
  S. Fineschi\inst{\ref{INAF-OATo}}
  \and
  F. Frassati\inst{\ref{INAF-OATo}}
  \and
  M. Giarrusso\inst{\ref{INAF-OACt}}
  \and
  C. Grimani\inst{\ref{UniUrbino},\ref{INFN-Fi}}
  \and
  P. Heinzel\inst{\ref{CAS}}
  \and
  F. Landini\inst{\ref{INAF-OATo}}
  \and
  D. Moses\inst{\ref{NASA-HQ}}
  \and
  G. Nicolini\inst{\ref{INAF-OATo}}
  \and
  M. Pancrazzi\inst{\ref{INAF-OATo}}
  \and
  C. Sasso\inst{\ref{INAF-OAC}}
}

\institute{%
  INAF -- Osservatorio Astronomico di Capodimonte, 
  Salita Moiariello 16, I-80131 Naples, Italy \\
  \email{vincenzo.andretta@inaf.it}
  \label{INAF-OAC}
  \and
  INAF -- Osservatorio Astrofisico di Torino, Turin, Italy
  \label{INAF-OATo}
  \and
  INAF -- Osservatorio Astronomico di Trieste, Trieste, Italy
  \label{INAF-OATs}
  \and
  Predictive Science Inc., San Diego, CA 92121, USA
  \label{PSI}
  \and
  Universit\`a di Padova -- Dip. Fisica e Astronomia ``Galileo Galilei'', Padua, Italy
  \label{UniPd}
  \and
  CNR -- Istituto di Fotonica e Nanotecnologie, Padua, Italy
  \label{CNR-IFN}
  \and
  INAF -- Osservatorio Astrofisico di Catania, Catania, Italy
  \label{INAF-OACt}
  \and
  Agenzia Spaziale Italiana, Rome, Italy
  \label{ASI}
  \and
  INAF -- Istituto di Astrofisica Spaziale e Fisica Cosmica, Milan, Italy
  \label{INAF-IASF}
  \and
  Institute of Physics, University of Graz, Graz, Austria
  \label{UniGraz}
  \and
  Universit\`a della Calabria -- Dip. Fisica, Italy
  \label{UniCal}
  \and
  Universit\`a di Firenze -- Dip. Fisica e Astronomia, Florence, Italy
  \label{UniFi}
  \and
  INAF -- Osservatorio Astrofisico di Arcetri, Florence, Italy
  \label{INAF-OAA}
  \and
  Max-Planck-Institut f\"ur Sonnensystemforschung, G\"ottingen, Germany
  \label{MPS}
  \and
  Universit\`a di Urbino ``Carlo Bo'' -- DiSPeA, Urbino, Italy
  \label{UniUrbino}
  \and
  INFN -- Sez. Firenze, Florence, Italy
  \label{INFN-Fi}
  \and
  Astronomical Institute of the Czech Academy of Sciences, Ond\v rejov, Czech Republic
  \label{CAS}
  \and
  NASA HQ, Washington DC, USA
  \label{NASA-HQ}
}

\date{Received DAY MONTH YEAR; accepted DAY MONTH YEAR}

\abstract
{%
  Waves are thought to play a significant role in the heating of the solar atmosphere and the acceleration of the wind. Among the many types of waves observed in the Sun, the so-called $p$-modes with a 3~mHz frequency peak dominate the lower atmosphere. In the presence of magnetic fields, these waves can be converted into magnetohydrodynamic modes, which then leak into the corona through magnetic conduits.  High-resolution off-limb observations have indeed revealed signatures of ubiquitous and global 3~mHz oscillations in the corona, although limited to low heights and to incompressible modes.
}
{%
  We present high-cadence, high-resolution observations of the corona in the range 1.7 -- 3.6~\rsun\ taken in broad-band 580--640~nm visible light by the Metis coronagraph aboard Solar Orbiter. These observations were designed to investigate density fluctuations in the middle corona.%
} 
{%
  The data were acquired over several days in March 2022, October 2022, and for two days in April 2023.  We selected representative regions of the corona on three sample dates. %
  Analysis of the data in those regions revealed the presence of periodic density fluctuations.  By examining several time-distance diagrams, we determined the main properties (apparent propagation speed, amplitude) of those fluctuations.  We also show power spectra in selected locations in order to determine the dominant frequencies.
}
{%
  We found wave-like, compressible fluctuations of low amplitude --- of the order of 0.1~\% of the background --- in several large-scale regions in the corona at least up to 2.5~\rsun. We also found that the apparent propagation speeds of these perturbations typically fall in the range 150--450 \kms. A power spectrum analysis of time series revealed an excess power in the range 2--7~mHz, often with peaks at 3 or 5~mHz, i.e.\ in a range consistent with $p$-mode frequencies of the lower solar atmosphere.
}
{%
}

\keywords{%
  Sun: corona --
  Sun: oscillations --
  Waves
}

\titlerunning{%
    First detection of acoustic-like flux in the middle solar corona%
}
\authorrunning{V. Andretta et al.}
  
\maketitle
\nolinenumbers

\section{Introduction}\label{sec:introduction}

  The role of waves in heating the solar atmosphere  \citep[see for instance][and references therein]{Narain-Ulmschneider:1996,Klimchuk:2006,Erdelyi-Ballai:2007,Arregui:2015} and generating the solar wind \citep{Sharma-Morton:2023} is widely recognised.
  The magnetically structured coronal plasma supports the propagation of a large variety of Magnetohydrodynamic (MHD) oscillations and waves, which constitute the natural response of the magnetic structure to the forcing action of the photospheric driver \citep{Edwin-Roberts:1983}.  They include standing MHD oscillations such as fast kink, fast sausage and slow longitudinal acoustic modes, as well as propagating slow and fast waves which propagate with sound speed, magneto-acoustic speed, up to Alfv\'enic speed.  These waves perturb the coronal plasma, producing variations in the plasma density, temperature, flow and (frozen-in) magnetic field, then making it possible their detection through the physical effects they determine on the host structures.

  Waves and oscillatory phenomena are observed in the visible light, EUV, X-ray and radio bands almost ubiquitously in the solar corona \citep{DePontieu-etal:2007,Tomczyk-etal:2007,KrishnaPrasad-etal:2012,Morton-etal:2012,Nistico-etal:2013}. They usually permeate individual magnetic structures, such as loops, polar plume and interplume regions, streamers, filaments, prominences, jets and coronal holes, with periods ranging from a few seconds to several hours (sometimes up to days) and typical wavelengths from a few million to several hundred million meters \citep[see e.g.][for a complete review of the different types of waves and oscillations hosted by a large variety of magnetic structures]{Jess-etal:2023}.

  In addition, the leakage of photospheric and chromospheric oscillations into the low corona can also lead to the generation of coronal waves \citep[e.g.][]{DeMoortel-etal:2002,Sych-etal:2009,Yuan-etal:2011,Shen-Liu:2012}, mainly recognised as fast and slow outward propagating magneto-acoustic waves.

  Slow magneto-acoustic waves are among the most studied and frequently detected wave motions in the solar corona \citep[e.g.][for recent comprehensive reviews]{Nakariakov-Kolotkov:2020,Banerjee-etal:2021,Wang-etal:2021}. They have been commonly detected in plume and interplume regions in polar coronal holes \citep[e.g.][]{DeForest-Gurman:1998,Gupta-etal:2010}, legs of long fan-like loops in active regions \citep{Yuan-Nakariakov:2012}, and also in plume-like structures in equatorial coronal holes and quiet Sun regions \citep[e.g.][]{Tian-etal:2011,KrishnaPrasad-etal:2012}. These waves are usually associated with quasi-periodic EUV and soft X-ray intensity perturbations but quasi-periodic variations in the polarised brightness of the visible light in the polar coronal holes have been also ascribed to these waves \citep[e.g.][]{Ofman-etal:1997,Ofman-etal:2000}.

  Their typical oscillation periods range from a few minutes to a few tens of minutes while sometimes longer oscillation periods are observed too \citep[e.g.][]{KrishnaPrasad-etal:2012,Mandal-etal:2018}. Their amplitudes, typically ranging from tens to a few percent of the background intensity \citep[e.g.][]{DeMoortel:2006,DeMoortel:2009}, were found to decay quickly with height as they travel along the supporting structures, with propagating speeds ranging from about 70 to 235 \kms.

  Damping of slow waves has been studied extensively in polar plumes and active region loops both theoretically and observationally. Typical damping lengths in individual structures are found to be of the order of 10 – 20 Mm \citep[e.g.][]{DeMoortel-etal:2002,Marsh-etal:2011,KrishnaPrasad-etal:2014}. This result raises a puzzling question about the nature of periodic compressive perturbations detected at much larger heights by the Ultraviolet Coronagraph Spectrometer \citep[UVCS --][]{Kohl-etal:1995} on board SOHO \citep[SOlar and Heliospheric Observatory --][]{Domingo-etal:1995}, both in polar \citep{Ofman-etal:1997,Ofman-etal:2000,Morgan-etal:2004,Bemporad-etal:2008,Telloni-etal:2009a} and equatorial regions \citep{Telloni-etal:2009b} and more recently \citep{Telloni-etal:2013,Telloni-etal:2014} by the inner coronagraphs COR1 onboard the STEREO \citep[Solar TErrestrial RElations Observatory --][]{Thompson-etal:2003,Kaiser-etal:2008} A and B spacecraft.

  Overall, the existence of discrepancies between observations and theoretical predictions suggests that the current theory of damping of slow waves may be missing critical elements; data collected so far are insufficient to allow us to drawn any conclusion.
  Internal, resonance modes are often postulated to be a significant source for oscillations in the upper solar atmosphere. Indeed, $p$-modes can penetrate through magnetic regions
where they are converted into magneto-acoustic modes and leak to the upper layers of the solar atmosphere
\citep[see for instance][]{Centeno-etal:2009}.
However, they can also
readily develop into shocks and dissipate energy at chromospheric heights \citep{Vecchio-etal:2009}. Due to the presence of the acoustic cutoff at around $5$ mHz imposed by its vertical stratification, the atmosphere acts as a frequency filter allowing the propagation of acoustic waves with higher frequency only. As a result, there is a dominant frequency at about $5$ mHz in the chromosphere and the upper layers \citep{Felipe-Sangeetha:2020} of the atmosphere. However, the acoustic cutoff can also be significantly reduced due to either the presence of inclined field lines with respect to the propagation direction of the wave \citep{Jefferies-etal:2006,Stangalini-etal:2011} or radiative losses \citep{Khomenko-etal:2008}, thus allowing the upward propagation of acoustic waves even with frequency lower than $\sim 5$ mHz.

Evidence of acoustic power leakage into the corona has been obtained
by exploiting high spatial and temporal resolution data from the \textit{Coronal Multi-Channel Polarimeter} \citep[CoMP,][]{Tomczyk-etal:2008}, \cite{Tomczyk-etal:2007} and later, e.g.\ \citet{Morton-etal:2016} have found the presence of ubiquitous transverse oscillations in the corona up to 1.3~\rsun. Interestingly, the presence of a power enhancement at $3$ mHz ($\sim 5$ min period) was also found, and therefore it was argued that these ubiquitous waves could be the result of the mode conversion of leaking $p$-modes into transverse waves close to the transition region (i.e. kink modes) and that this process is acting at global scales. 
  More specifically, clear signatures of acoustic power leakage in the upper solar atmosphere have been reported to date only
in the lower corona, and as transverse oscillations only, i.e.\ with no signature of compressible waves.
  Although the first observations of periodic and quasi-periodic phenomena in the solar corona date back to the 1980s \citep{Aschwanden:1987}, decisive contributions in the investigation of coronal waves and transients have been possible thanks to the improved spatial and temporal resolution of multi-wavelength observations carried out by the instrumentation on board SOHO, TRACE \citep[Transition Region And Coronal Explorer --][]{Handy-etal:1999}, STEREO and SDO \citep[Solar Dynamics Observatory --][]{Pesnell-etal:2012}.  Nevertheless, many fundamental questions remain still open. New constraints by data-intensive observations for the validation or rejection of the existing theoretical models and a spur for new theoretical works are needed.
  
  We expect, therefore, that the unprecedented high spatial and temporal resolution of the instrumentation on board the currently operating fleet of new generation solar missions could provide a unique opportunity for a more comprehensive and detailed understanding of the origin, excitation and dissipation mechanisms, mode coupling, energy transfer and balance for a large variety of MHD waves hosted in a large sample of individual structures and in extended regions of the solar corona.
  In this context, visible light observations of the Metis coronagraph \citep{Antonucci-etal:2020,Fineschi-etal:2020} onboard Solar Orbiter
  \citep{Mueller-etal:2020}, %
  has the potential of providing valuable and original contributions, particularly to our understanding of propagation and dissipation of density fluctuations and waves in the middle corona, a region of the solar atmosphere where imaging observations and studies on density fluctuations and waves are so far very limited.

We report here the first, unexpected, detection of coronal density oscillations with frequency below the cutoff, in the five-minute range (3~mHz) between 1.7 and 3.6 solar radii, right in the critical region where the solar wind is believed to be accelerated \citep[e.g.][]{Abbo-etal:2010,Romoli-etal:2021}.
  These
periodic variations are best seen in the brightest structures of the solar corona, but in some cases it was possible to detect them in regions with relatively fainter signal.  In this work, we describe and discuss representative examples of these periodic brightness variation.

\section{Observations}\label{sec:observations}

In the course of its elliptical orbit around the Sun, the Solar Orbiter spacecraft approaches the Sun at distances smaller than 0.3~au \citep{Mueller-etal:2020}.  For several days around the first three perihelia of the mission nominal phase, which occurred respectively on 25 March 2022, 8 October 2022, and 10 April 2023, the Metis coronagraph %
was carrying out a sequence of high-cadence imaging observations of the corona with its visible-light channel.  The goal of these observations was to explore the variability of the solar corona in the regions where the solar wind is accelerated, at temporal and spatial scales never observed before.

\begin{figure}[th]
  \centering
  \includegraphics[width=\linewidth,trim=25 25 250 50]{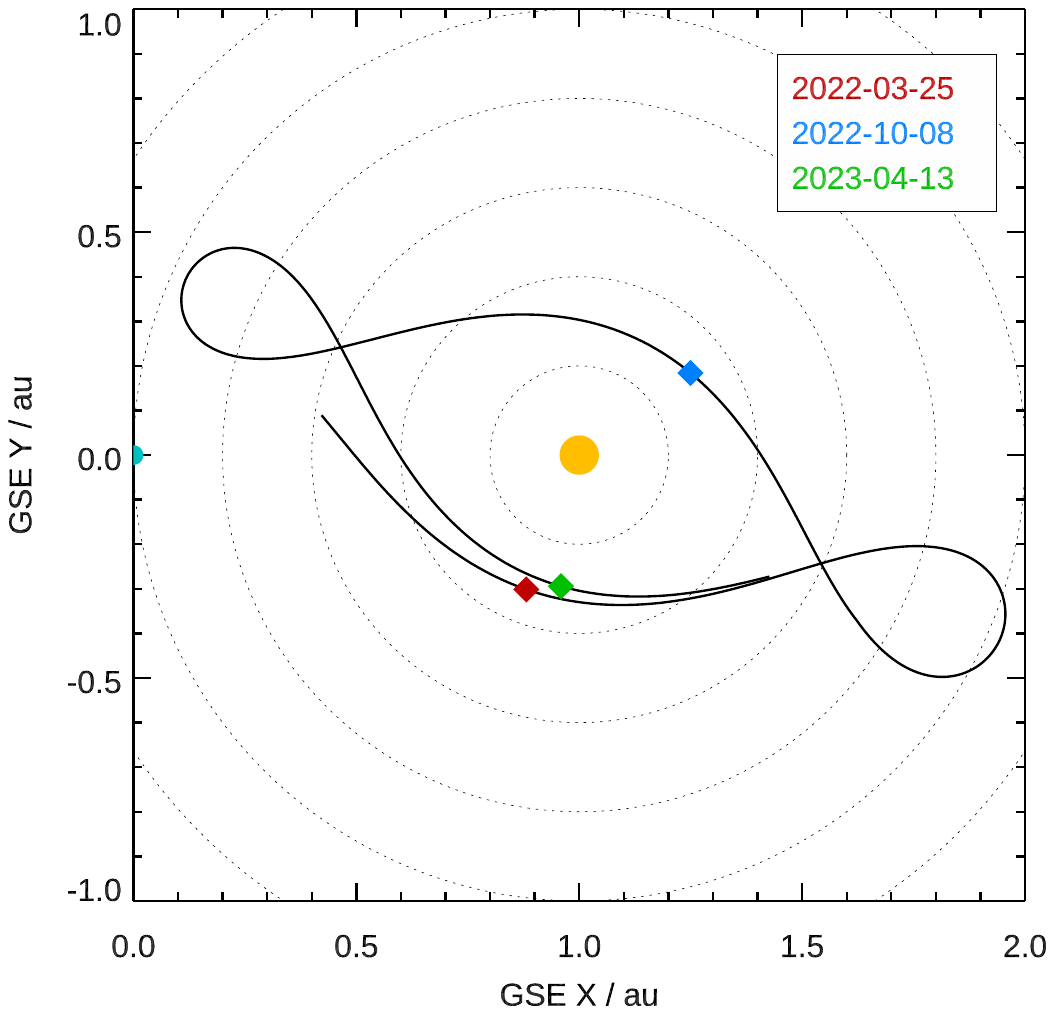}
  \caption{Solar Orbiter trajectory between 1 March 2022 and 30 April 2023 projected on the ecliptic plane in Geocentric Solar Ecliptic (GSE) coordinate system, and its positions during the high-cadence observations analysed in this work. The dotted circles are separated by 0.2~au.}
  \label{Fig:orbit}
\end{figure}

The three sets of Metis observations discussed here were obtained over the course of several days, around the dates of the first three perihelia of the Solar Orbiter Nominal Mission Phase.  The instrumental acquisition settings were identical in all cases, except for the duration of the observations.  The position of the Solar Orbiter spacecraft in the ecliptic plane during the observations is shown in Fig.~\ref{Fig:orbit} in the Geocentric Solar Ecliptic (GSE) coordinate system (defined such that $X$ is the Earth-Sun line, and $Z$ is aligned with the ecliptic north of date).

The Visible-Light (VL) channel of the Metis instrument acquires brightness images in the 580 -- 640 nm wavelength range, and can operate by means of four different acquisition schemes as described in \cite{Fineschi-etal:2020} and \cite{Antonucci-etal:2020}. In Appendix~\ref{sec:metis_processing} we briefly summarise the differences between the main Metis acquisition schemes.

In this work, we focus our analysis on data acquired with the VL-tB (``total brightness'') scheme, which provides high-cadence sequences of total brightness images of the corona.  During the observations taken around the first three perihelia, the VL detector was configured to acquire data at maximum cadence (20~s) with a 2$\times$2 pixels binning, corresponding to a spatial scale on the plane of the sky of $\ga 4400$ km~pixel$^{-1}$ (at 0.3~au from the Sun).  The duration of these VL-tB acquisitions was of 41~minutes, except for a run in April 2023 for which the duration was double (82~minutes).  The estimated mean signal-to-noise ratio in each image of these data sets exceeds the value of 200 over most of the field of view.

These high-cadence observations were taken around the first perihelion once a day, from 22 to 27 March 2022; from 8 to 13 October and then once on 27 October around the second perihelion; on 12 and 13 April 2023 shortly after the third perihelion.  Since the phenomena described in this work are visible in all these observations, we selected only one date for each orbit; the data sets analysed in this work are listed in Table~\ref{Tab:data}. It is important to note that the time gap between the two data sets taken on 13 April 2023 is only $\sim 140$~s, ensuring that the combined observations span more than two hours.

When observing during Solar Orbiter closest approach, at about $0.28$~au, the Metis field-of-view (FOV) ranges from $\sim 1.7$~\rsun\ (internal occulter edge) up to $\sim 3.6$~\rsun\ (detector corners); along the ecliptic plane, the outer edge of the FOV is at $\sim 3.1$~\rsun\ \citep{Antonucci-etal:2020}. These values scale linearly with the heliocentric distance at the epoch of observations (given in Table~\ref{Tab:data}).

The acquired data were processed and calibrated according to \cite{Romoli-etal:2021} and \cite{DeLeo-etal:2023}. Furthermore, we noticed in these high cadence acquisitions an overall variation of the detector response at a frequency of about ~11 mHz.  We investigated this phenomenon, likely due to detector electronics, and then devised an experimental procedure, described in Appendix~\ref{sec:metis_processing}, to mitigate this effect. We however verified that the application of this procedure to the data analysis did not significantly alter the results of this work.

\section{Data analysis and results}\label{sec:results}

The oscillations we discuss in this work are best seen in the brightest structures of the solar corona but also, in some cases, in regions with lower mean brightness.  For a more detailed discussion, we made a selection of data sets and regions of interest (ROIs) we believe to be representative of the typical locations where these periodic brightness variations are detectable.
\begin{table*}
  \centering
  \caption{Data sets analysed}
  \label{Tab:data}
  \begin{tabular}{c c c c c c c c c}
    \hline\hline
    Index & Session n. & N.\ of images & Date & Start time & End time & Sun dist. & \multicolumn{2}{c}{Carrington coord.} \\
          &            &               &      &            &          &           &  Longitude  &  Latitude \\
          &            &               &      &            &          & au        &    degrees  &  degrees  \\
    \hline
        1 & 208407 & 120 & 2022-03-25 & 20:10:31 & 20:51:43 & 0.323 & 186.9 &  -0.7 \\
        2 & 228103 & 120 & 2022-10-08 & 03:40:01 & 04:21:13 & 0.311 & 259.9 &  -6.6 \\
        3 & 310303 & 120 & 2023-04-13 & 00:34:31 & 01:15:43 & 0.298 & 182.1 &  -0.4 \\
        4 & 310304 & 239 & 2023-04-13 & 01:18:01 & 02:40:24 & 0.298 & 181.8 &  -0.3 \\
    \hline
  \end{tabular}
  \tablefoot{Each data set is identified by a session number, which uniquely identifies in the Metis data base the set of images acquired within the same acquisition session.  The  table also lists the time intervals covered and the Solar Orbiter coordinates during the observations. Times are UTC on board.}
\end{table*}

\subsection{The solar context}\label{sec:results:context}

\begin{figure*}[th]
  \centering
  \includegraphics[width=0.32\textwidth,trim=160 40 50 40,clip]{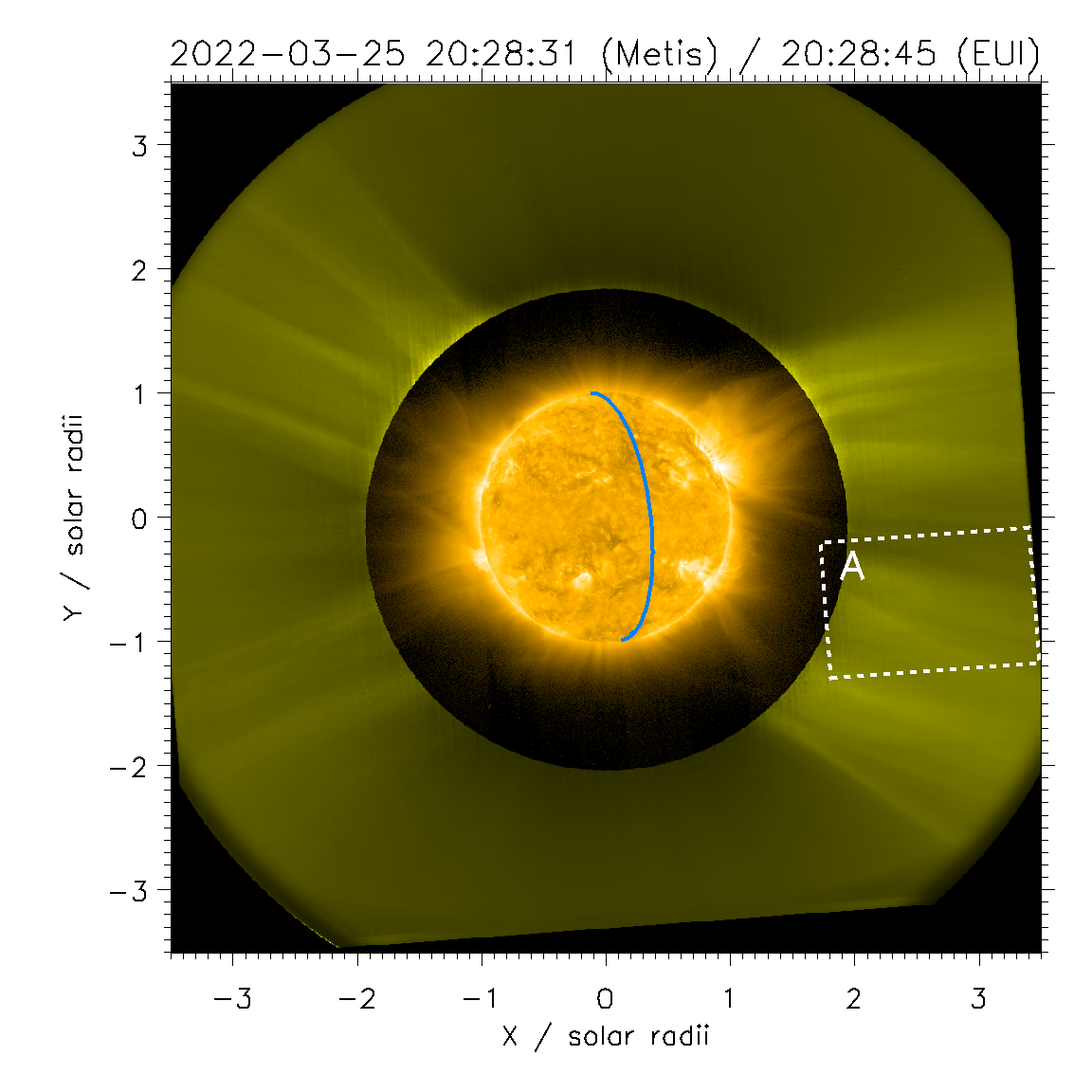}
  \includegraphics[width=0.32\textwidth,trim=160 40 50 40,clip]{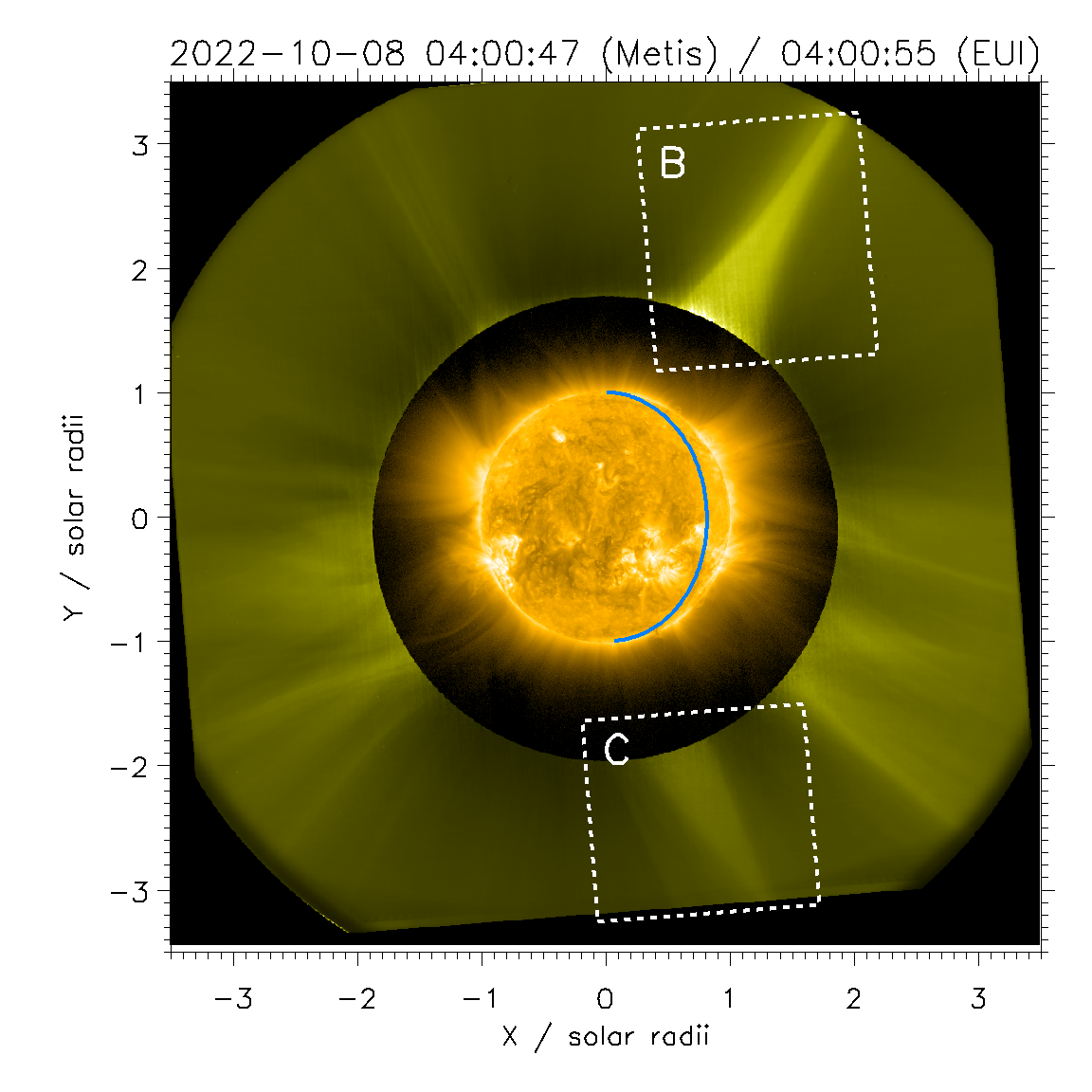}
  \includegraphics[width=0.32\textwidth,trim=160 40 50 40,clip]{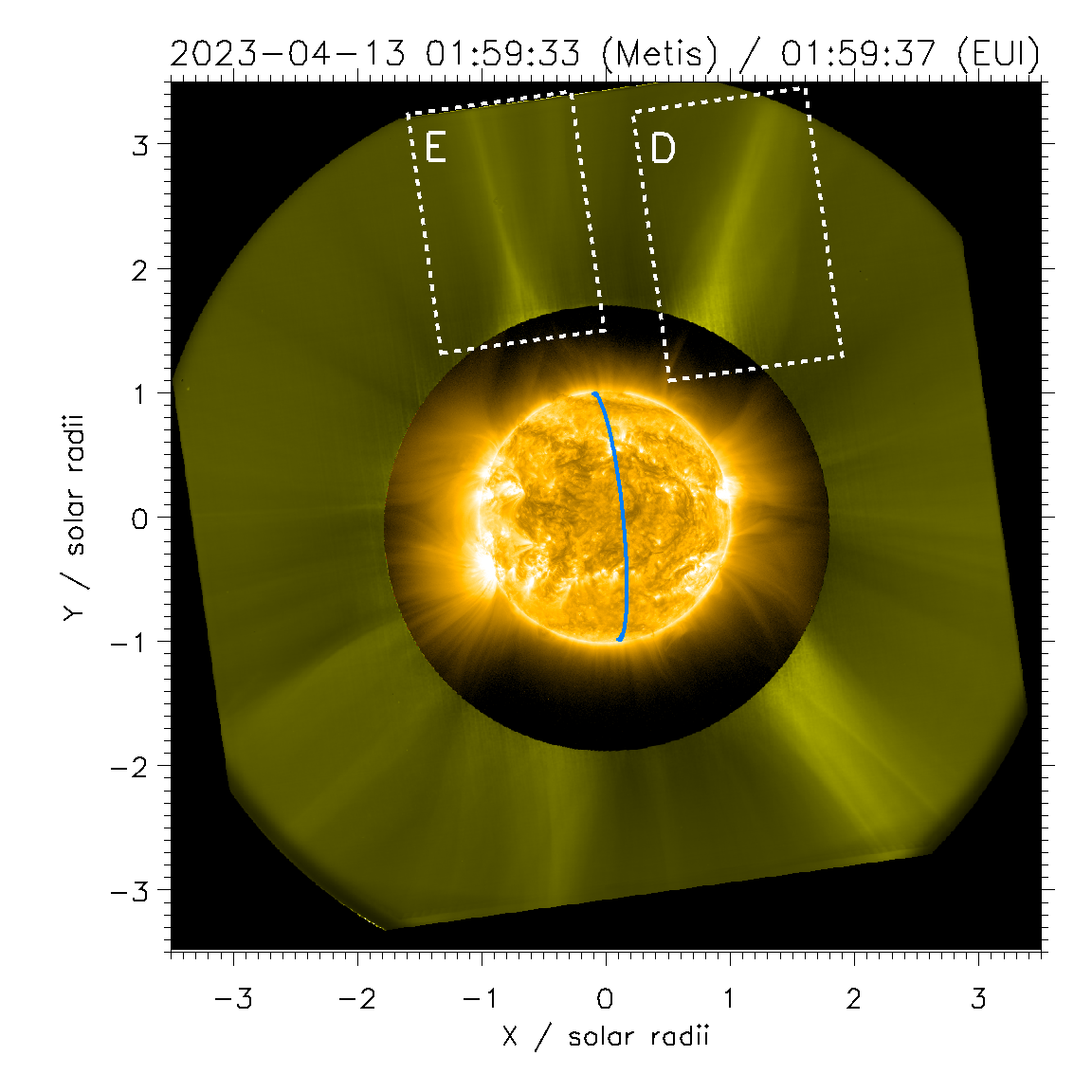}
  \caption{Metis and Extreme Ultraviolet Imager (EUI) \citep{Rochus-etal:2020} composite images of the corona during the first three perihelia of the Solar Orbiter mission, from left to right.  Labelled boxes mark the ROIs discussed in this work.  The EUI image shown for context within the occulted Metis area was obtained in the FSI174 band.  The blue line shows the position of solar limb seen from Earth. Metis images have been normalised by the average coronal intensity profile of that date (obtained as described in Sec.~\ref{sec:results:processing}), while EUI/FSI174 images are displayed on a logarithmic scale.}
  \label{Fig:overview}
\end{figure*}

Figure~\ref{Fig:overview} shows the configuration of the solar corona during the observation and the ROIs we selected for further analysis, labelled with capital letters, \textit{A} to \textit{E}.
In Appendix~\ref{sec:roi_desc} we briefly discuss the properties and the magnetic topology of these ROIs. A summary of those properties is given in Table~\ref{Tab:ROIs}. In brief: Two of these regions (\textit{B} and \textit{D}) can be classified as classical helmet streamers, two (\textit{C} and \textit{E}) can be identified as pseudo-streamers, while region \textit{A} is at the edge of a region characterised by closed magnetic field configuration.  In Sec.~\ref{sec:discussion} we comment on the relevance of the magnetic topology in the formation of these perturbations.
\begin{table*}
  \centering
  \caption{Summary of properties of the ROIs discussed in this work.}
  \label{Tab:ROIs}
  \begin{tabular}{c c r r r l}
    \hline\hline
    Label & Date       & Position Angle & \multicolumn{2}{c}{Carrington coord. (centre)} & Notes\\
          &            &                & Longitude  &  Latitude              & \\
          &            &   degrees      &   degrees  &  degrees               & \\
    \hline
    \textit{A}     & 2022-03-25 &   255          &   276      &  -15                   & Open-field lines at the \\ 
                   &            &                &            &                        & boundary of a streamer \\
    \textit{B}     & 2022-10-08 &   331          &   350      &   63                   & Bright streamer  \\
    \textit{C}     & 2022-10-08 &   198          &   343      &  -74                   & Pseudo-streamer \\
    \textit{D}     & 2023-04-13 &   335          &   270      &   64                   & Streamer \\
    \textit{E}     & 2023-04-13 &    19          &    94      &   69                   & Pseudo-streamer \\
    \hline
  \end{tabular}
\end{table*}

\subsection{Image processing.}\label{sec:results:processing}

Movies obtained from the sequence of Metis images show remarkably coherent oscillations in several coronal structures resulting from density fluctuations apparently propagating outwards. These periodic density perturbations, which will call ``waves'' from now on, can be made readily visible in movies by using relatively simple image processing, such as normalised base or running differences, described in more detail below.

For each data set, we produced normalised base-difference images by employing a variant of the SiRGraF algorithm \citep{Patel-etal:2022}: starting from the series of brightness images, $B(k)$, where $k$ is the time index in the sequence, we computed a base image, $B_\mathrm{m}$, by taking the pixel-by-pixel 1$^\mathrm{st}$ percentile.  From this base image, we then computed a mean radial profile and then the corresponding normalisation image, $B_\mathrm{norm}(r)$, where $r$ is the distance from solar disk centre, by averaging the $B_\mathrm{m}$ values in the azimuthal direction.  The base image is then $\bdiff(k) = \left[B(k)-B_\mathrm{m}\right]/B_\mathrm{norm}$.  We also used the same radial profile to obtain a normalised brightness image, $\bnorm$, with reduced radial contrast: $\bnorm(k) = B(k)/B_\mathrm{norm}$.

As an alternative image enhancing approach, we computed the pixel-by-pixel time average of $2s+1$ images: $\bar{B}_s(k) = \sum_{k-s}^{k+s} B(k)/(2s+1)$, then computing the normalised running difference $\rdiff{s}(k) = \left[\bar{B}_s(k) - \bar{B}_s(k-2s-1)\right]/B_\mathrm{norm}$.

Some images are affected by debris passing in front of the telescope aperture; also, some stars are seen transiting through the field of view.  No attempt to correct for both features has been made.

\begin{figure}[h]
  \centering
  \includegraphics[width=0.95\linewidth,trim= 5 160 85 145,clip]{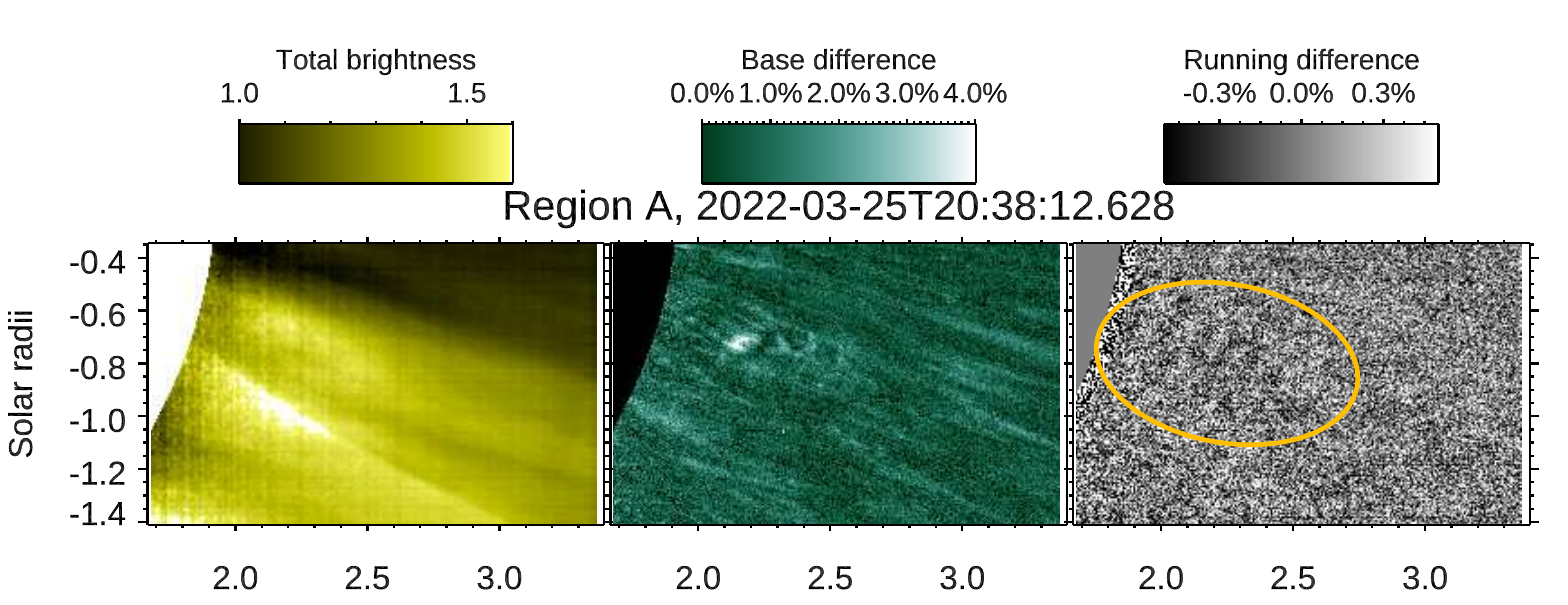}
  \includegraphics[width=0.95\linewidth,trim= 5 110 85 175,clip]{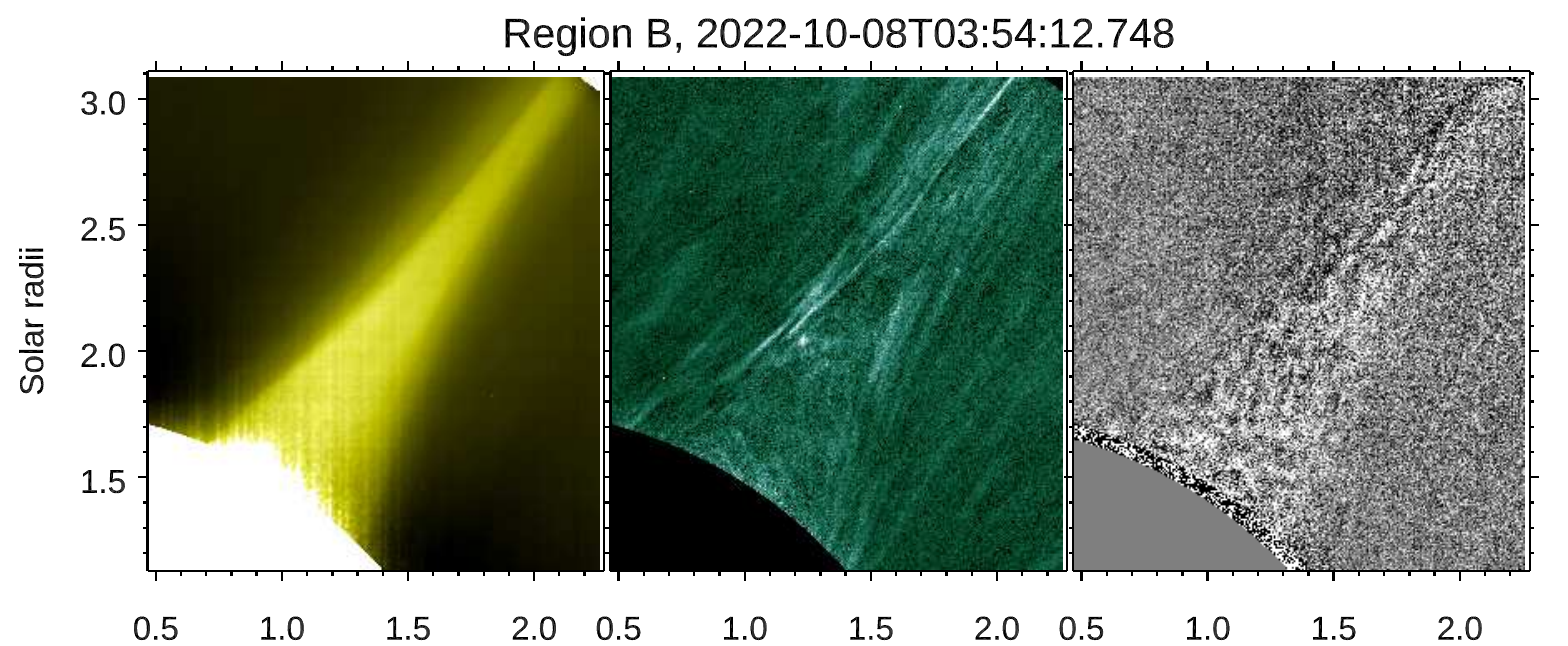}
  \includegraphics[width=0.95\linewidth,trim= 5 135 85 185,clip]{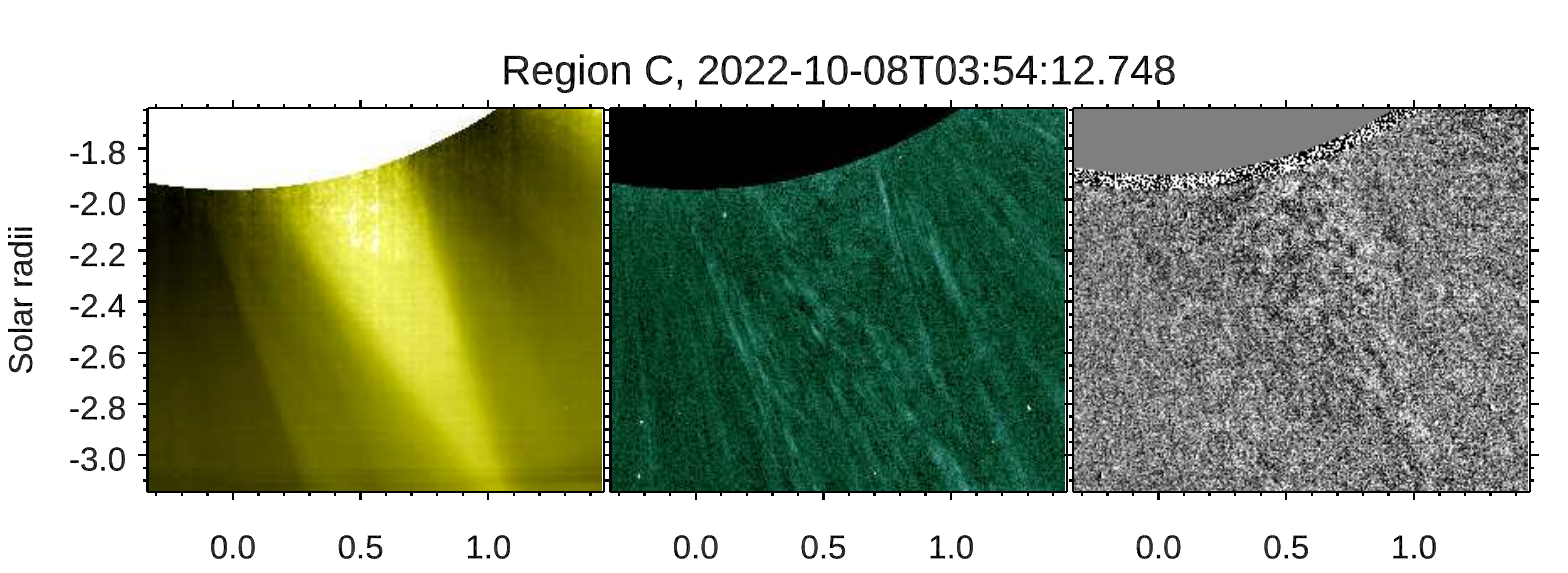}
  \includegraphics[width=0.95\linewidth,trim= 5  60 85 115,clip]{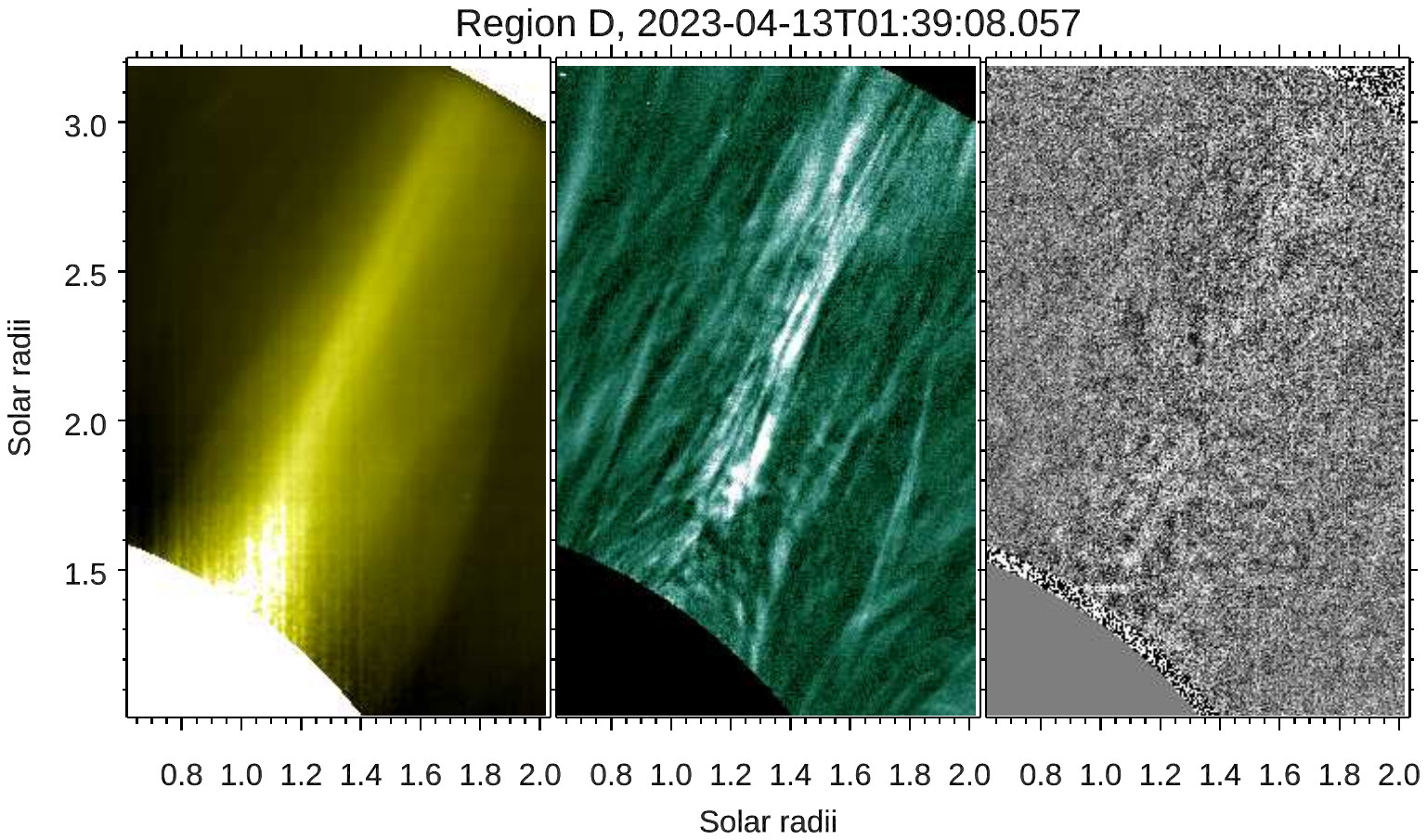}
  \includegraphics[width=0.95\linewidth,trim= 5  20 85 115,clip]{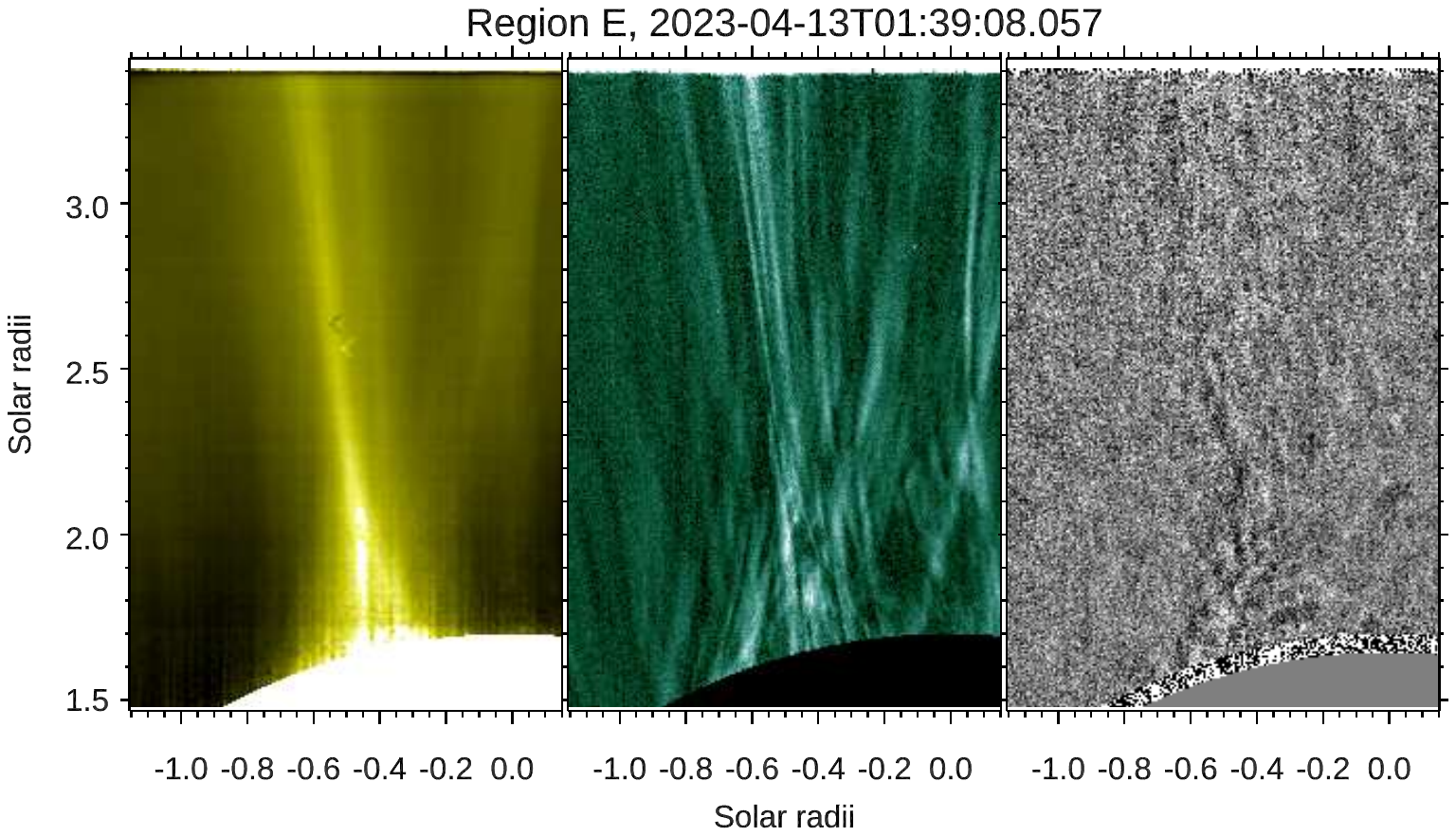}
  \caption{%
    Snapshots from the movies for the five ROIs discussed in this work. For each ROI, $\bnorm$ (left panel), $\bdiff$ (middle panel), and $\rdiff{1}$ (right panel) --- defined in Sec.~\ref{sec:results:processing} --- are displayed with colour bars. }
  \label{Fig:snapshots}
\end{figure}

We provide here a movie for each of the ROIs listed in Table~\ref{Tab:ROIs}, displaying the three quantities defined above: $\bnorm(k)$ (left panel), $\bdiff(k)$ (middle panel), and $\rdiff{s}(k)$ (right panel).  For regions \textit{D} and \textit{E}, the movies encompass both data set \#3 and \#4, for a total duration of more than two hours.  In general, we found that the oscillations we are reporting are best seen with $s\ge 1$; in the movies provided, we adopted $s=1$, corresponding to running differences of a sequence of time averages of three images and therefore with an effective integration time for $\bar{B}(k)$ of 60~s.  All the movies produced are available online; Figure~\ref{Fig:snapshots} shows representative snapshots from these movies. In the case of region \textit{A} (top-most panel of Fig.~\ref{Fig:snapshots}), we encircle the area where it is easier to see in the movies the propagating waves we intend to discuss.

For display purposes, prior to creating the movies with the above quantities, we applied the denoising procedure described in Appendix~\ref{sec:metis_processing}.  We also experimented with other denoising techniques, such as Image Noise Gate \citep{DeForest:2017} or Wavelet-Optimised Whitening (WOW) \citep{Auchere-etal:2023}. However, we found it difficult to strike a good balance between effective noise reduction and preservation of the faint periodic features we are investigating.  We therefore decided not to implement any further noise reduction in this work.

\subsection{Main properties of the waves.}\label{sec:results:properties}

The analysis of time-distance diagrams along selected linear paths within each ROIs provides more quantitative information on these waves.  In the following, we label the selected paths with a Greek letter and, when necessary for clarity, by the label of the ROI, e.g. \textit{D}-$\beta$ refers to path $\beta$ within region \textit{D}.

\begin{figure}[h]
  \centering
  \includegraphics[width=\linewidth,trim=10 225 85 160,clip]{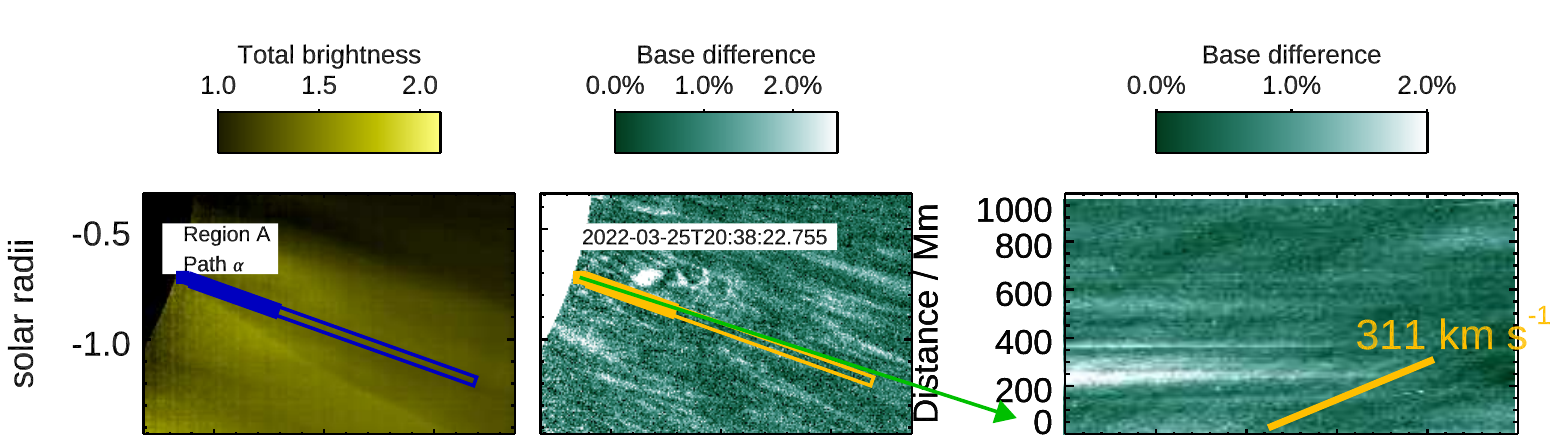}
  \includegraphics[width=\linewidth,trim=10 165 85 160,clip]{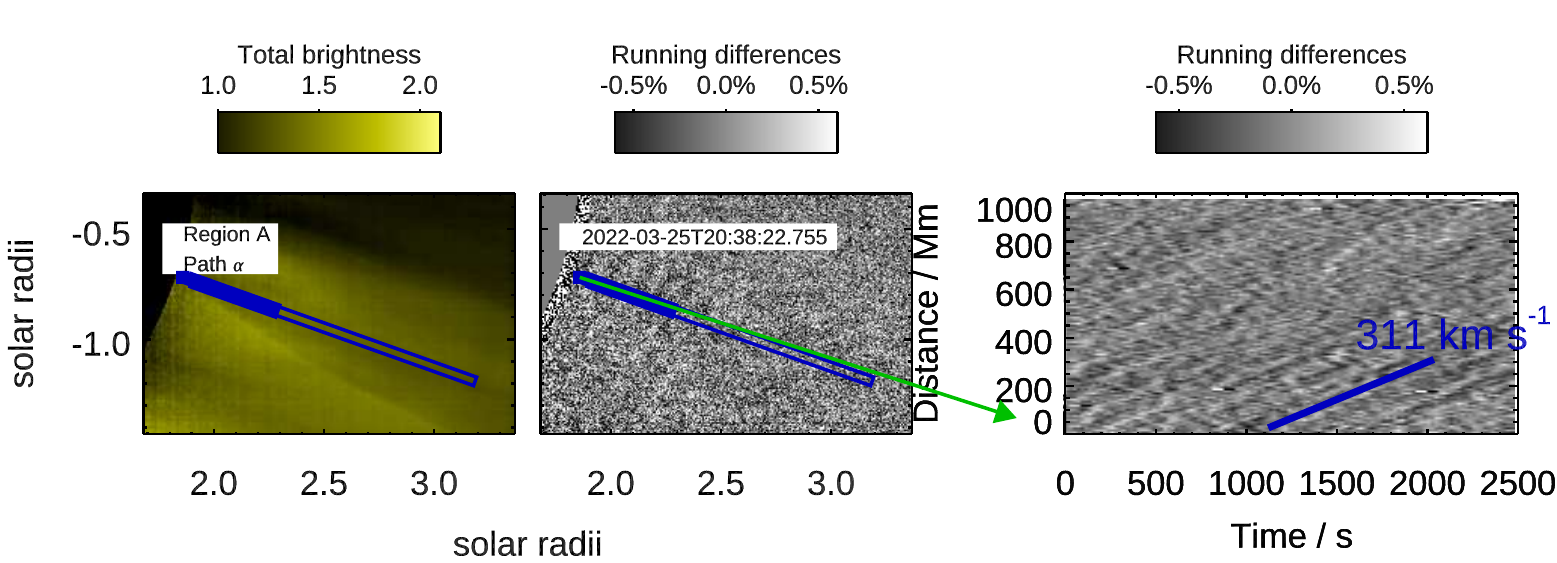}
  \caption{%
  Time-distance diagrams obtained from $\bdiff(k)$ (top panels) and $\rdiff{1}(k)$ (bottom panels) along a selected path, labelled as $\alpha$, in region \textit{A}. The left and middle panels displays a sample image from the series (\bnorm\ is shown in the left-hand panels). In both panels, the strip around the paths utilised to compute the time-distance diagrams is marked as a rectangle. The filled portion of the rectangle corresponds to the spatial extent of the feature marked in the time-distance diagram of the right panels. Right panel: Time-distance diagram along the selected path.  The arrow shows the correspondence between origin of the time-distance diagram and the first point of the path.  A solid line, labelled with the corresponding apparent speed, is drawn over one of the periodic ridges visible in the time-distance diagrams. %
  }
  \label{Fig:time-distance_A}
\end{figure}

Figure~\ref{Fig:time-distance_A} shows the time-distance diagram along a path within region \textit{A} observed on 25 March 2022.  The path, labelled as $\alpha$ and indicated by a solid line in the left-hand panels, has been chosen to be approximately parallel to the apparent propagation direction of the waves.  For each image, the mean value in a strip $\sim 30$~Mm wide along the path was computed, thus producing the time-distance diagrams shown on the right-hand panels of the figure.

We computed the time-distance diagrams for both \bdiff\ and \rdiff{1} images.  In both cases, periodic, inclined ``ridges'' are seen, although they are most easily seen in the running difference time-distance diagram.  The slope of these ridges corresponds to an apparent speed along the path of $\sim 300$~\kms.  The spacing in the time direction of these ridges is in the range 200 -- 300~s. 

These features seem to fade at $\sim 500$~Mm along the path, corresponding to a height in the corona of about 2.5~\rsun, although some traces of these features can be identified at larger distances. 
The amplitude of these perturbations is of the order of $0.1$\% of the background. We recall that the base- and running-difference images shown here have been normalised with respect to a mean radial profile. Since the normalisation profile, $B_\mathrm{norm}$, falls off rapidly with height in the corona,%
a constant-amplitude perturbation of the order of $0.1$\% when reaching the corona at the edge of the instrument FOV would corresponds to only a few DNs (Digital Numbers).  %
The apparent fading of these features with height can at least in part be explained by this observational effect.

\begin{figure}[h]
  \centering
  \includegraphics[width=\linewidth,trim=20 155 85 120,clip]{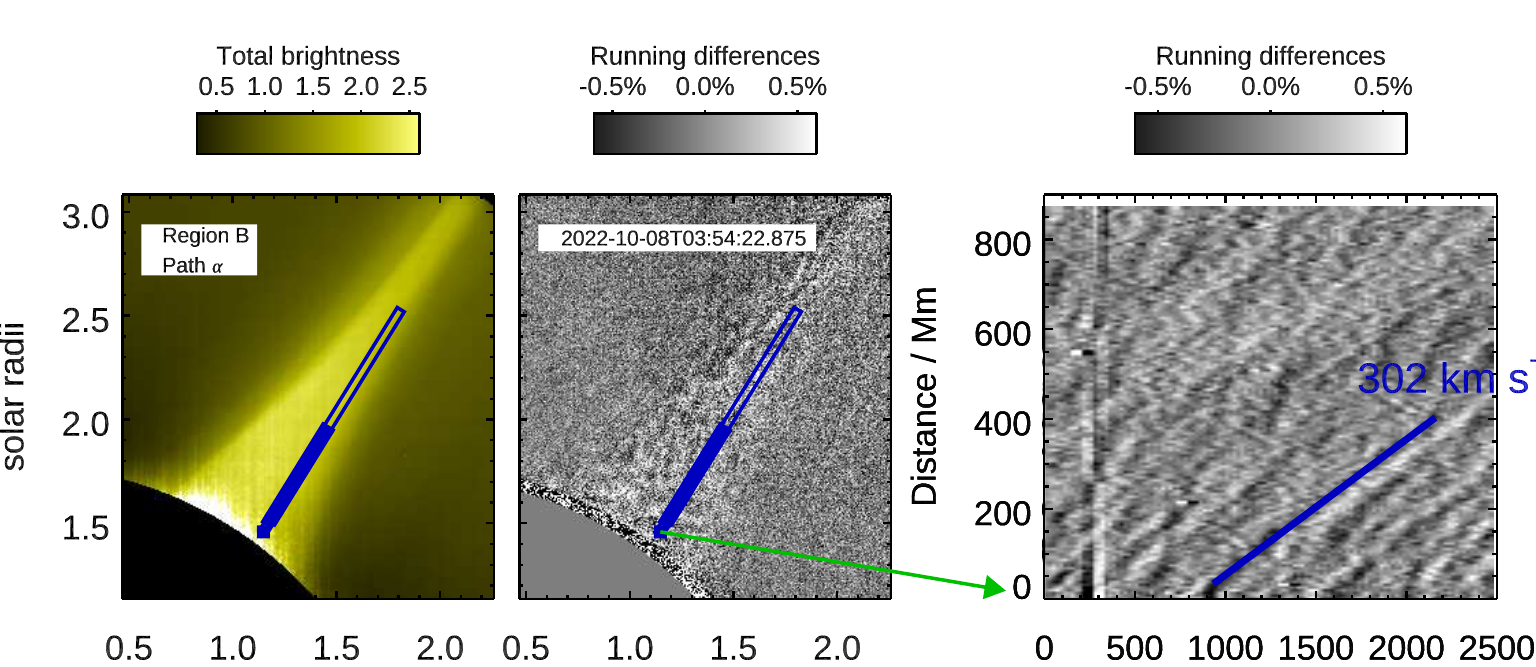}
  \includegraphics[width=\linewidth,trim=20 155 85 200,clip]{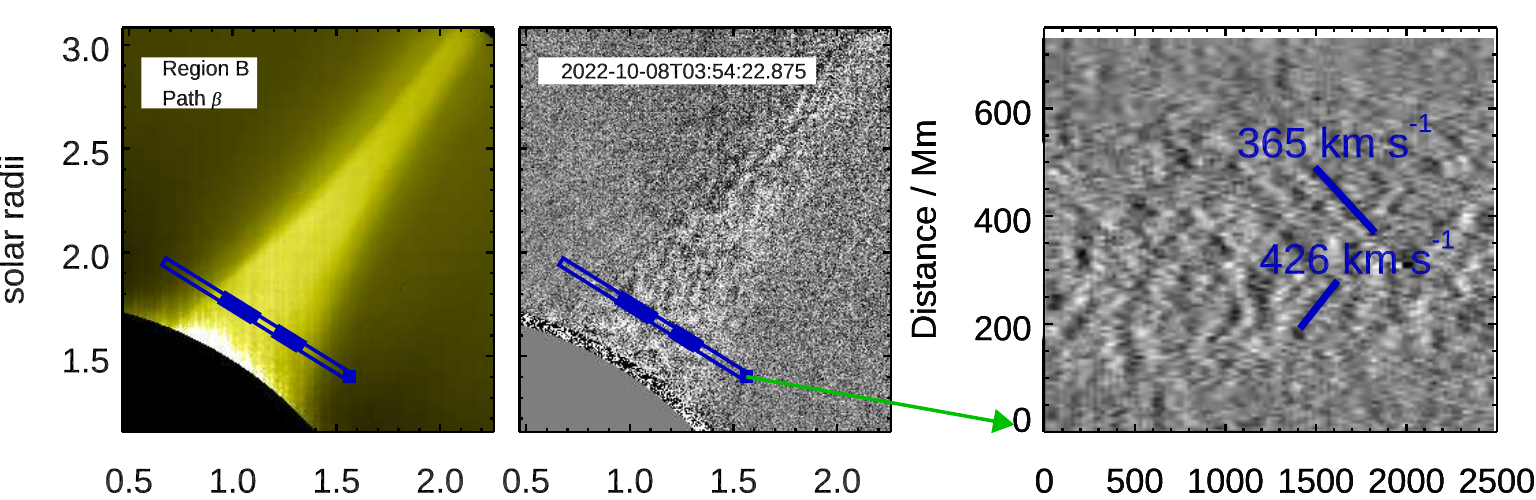}
  \includegraphics[width=\linewidth,trim=20 145 85 215,clip]{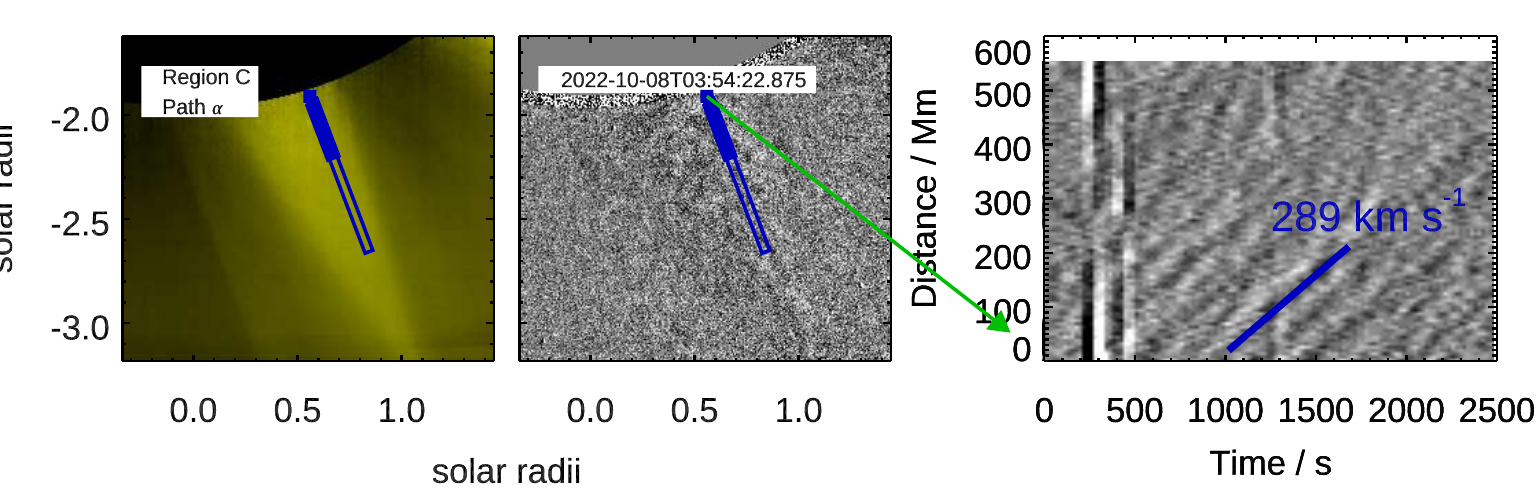}
  \caption{Time-distance diagrams (right panels) from paths $\alpha$ and $\beta$ in region \textit{B} (top and centre panels respectively), and from path $\alpha$ in region \textit{C} (bottom panels), with the same format and annotations as in Fig.~\ref{Fig:time-distance_A}. %
  }
  \label{Fig:time-distance_B-C}
\end{figure}

Figure~\ref{Fig:time-distance_B-C} shows the time-distance diagrams along sample paths within regions \textit{B} and \textit{C} observed on 8 October 2022. Region \textit{B} includes a bright streamer, while region \textit{C} includes a pseudo-streamer (see Sec.~\ref{sec:roi_desc}). In the case of region \textit{B}, two paths were chosen: one slightly off the axis of the streamer (top panels, path $\alpha$) and the other perpendicular to the streamer axis at about 2\,\rsun\ (centre panels, path $\beta$). As in Fig.~\ref{Fig:time-distance_A}, examples of diagonal ridges in the time-distance diagrams are annotated together with the corresponding speed. The vertical stripes in the time-distance diagrams between 300 and 500~s correspond to spurious signal due to debris transiting in front of the telescope aperture.

Paths \textit{B}-$\alpha$ and \textit{B}-$\beta$ together highlight a peculiar pattern of waves apparently propagating not only in the radial direction but also across the streamer axis, producing what seem to be interference patterns.  The time-distance diagram of path \textit{B}-$\beta$, in particular, exhibits a characteristic criss-cross pattern. Examples are marked with blue lines, with the corresponding apparent propagation speeds (between $\sim 350$ and $\sim 450$~\kms) annotated in the figure.  

\begin{figure}[h]
  \centering
  \includegraphics[width=\linewidth,trim=20 135 85 105,clip]{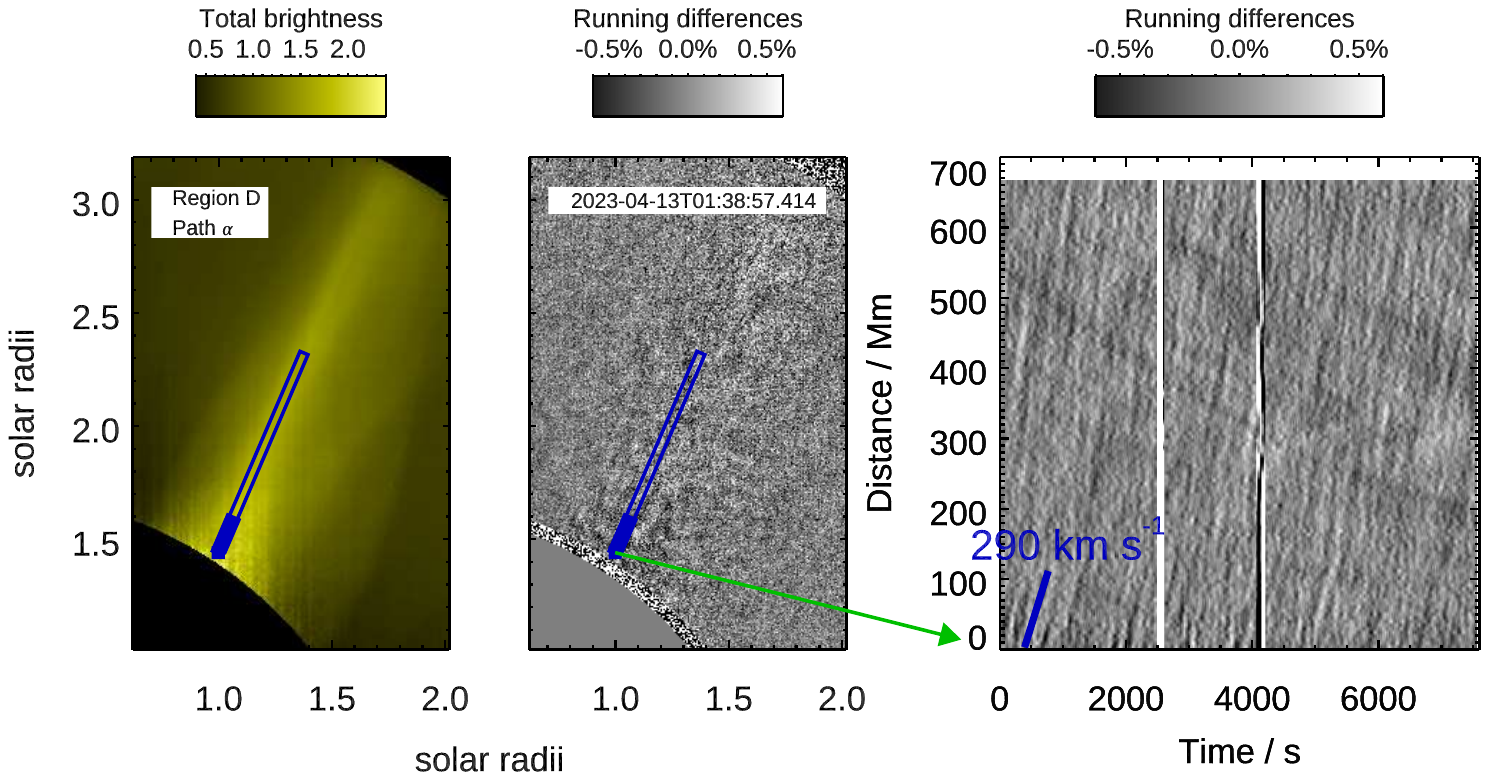}
  \includegraphics[width=\linewidth,trim=20 135 85 180,clip]{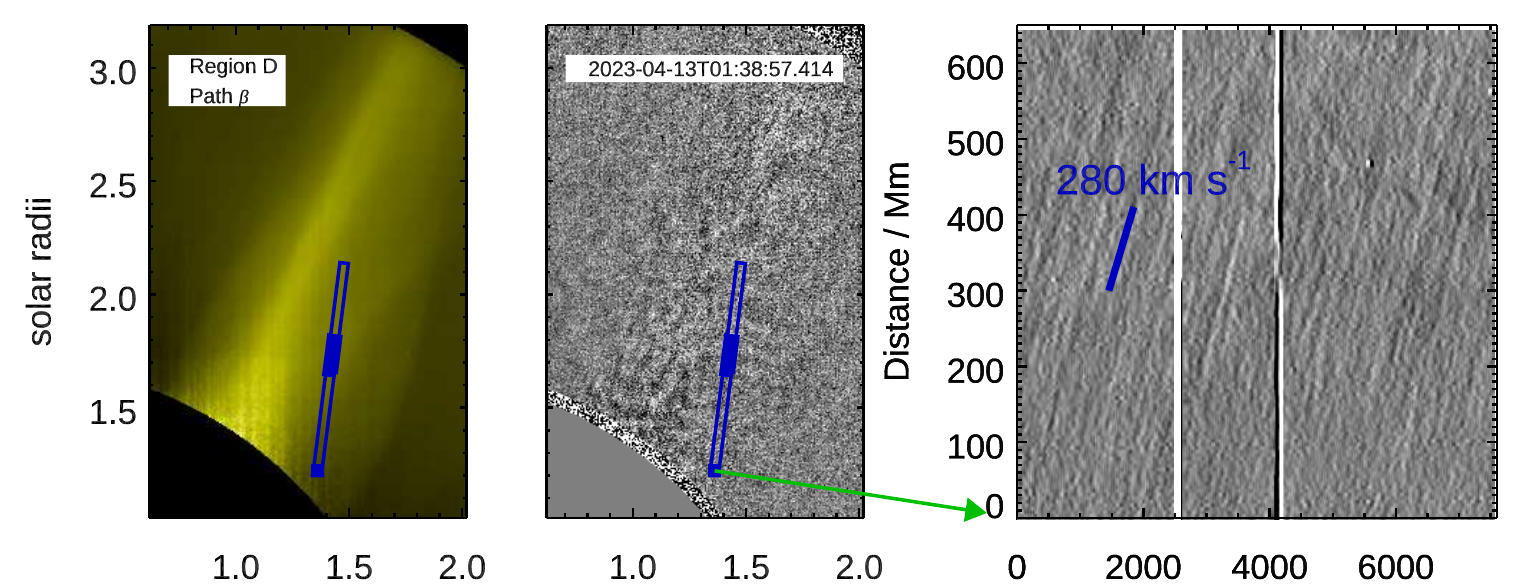}
  \includegraphics[width=\linewidth,trim=20 135 85 175,clip]{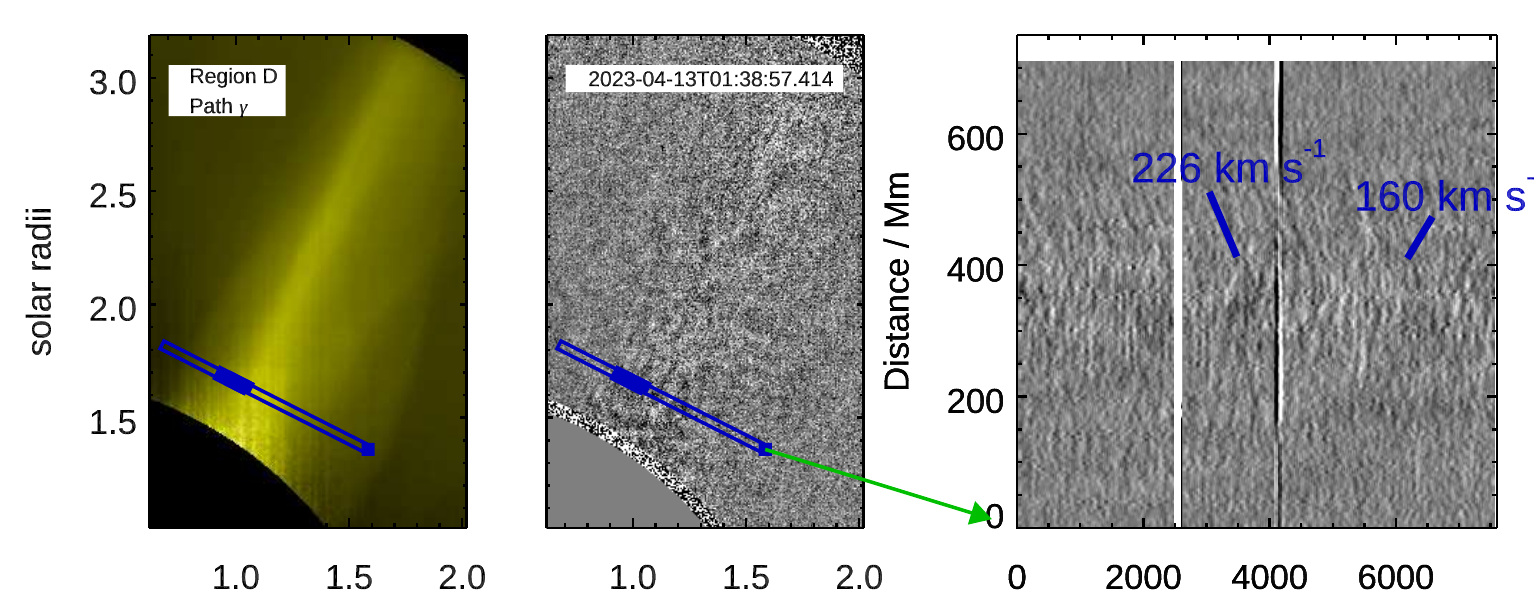}
  \includegraphics[width=\linewidth,trim=20 105 85 180,clip]{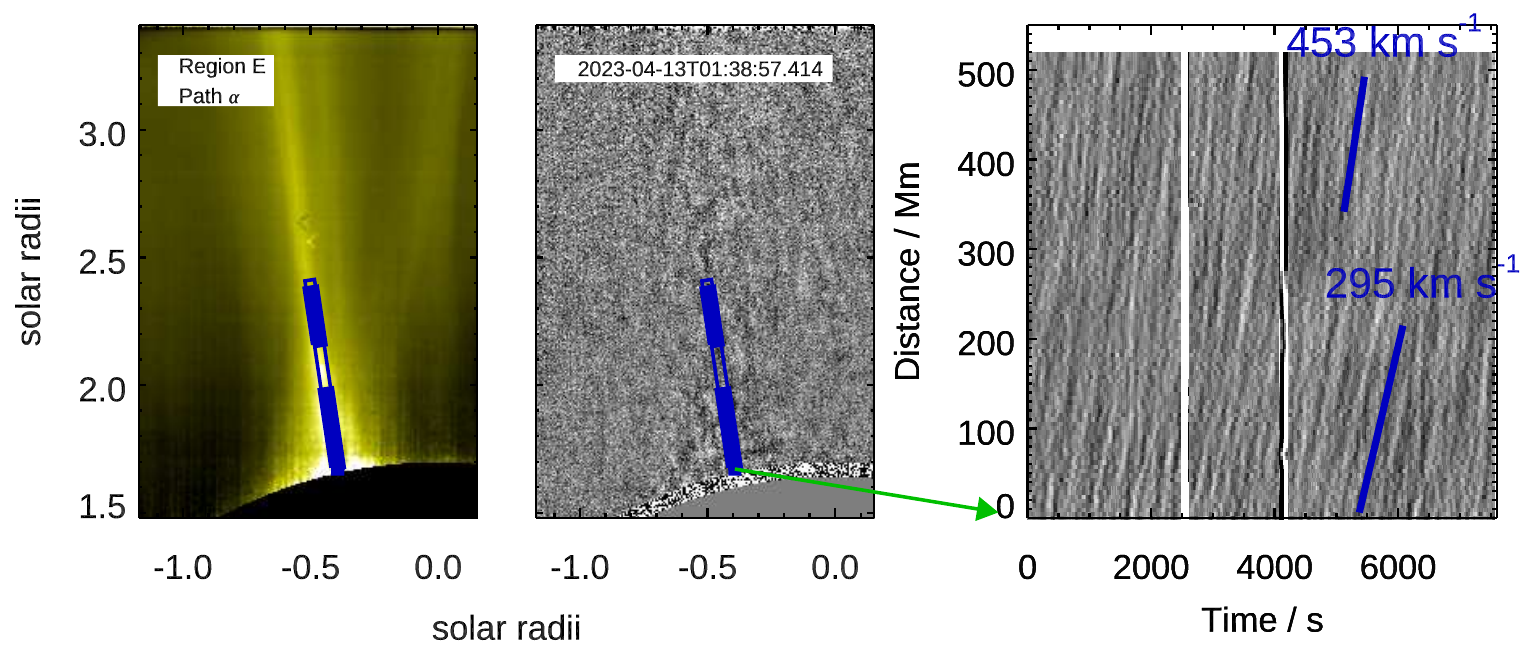}
  \caption{Time-distance diagrams in region \textit{D} from paths $\alpha$, $\beta$, and $\gamma$ (first three rows), and from path $\alpha$ in region \textit{E} (last row), with the same format and annotations as in Figures~\ref{Fig:time-distance_A}.} 
  \label{Fig:time-distance_D-E}
\end{figure}

Figure~\ref{Fig:time-distance_D-E} shows the time-distance diagram along sample paths within region \textit{C} and \textit{E} observed on 13 April 2023. In this case, we joined data sets \#3 and \#4 (Table~\ref{Tab:data}) to obtain a single time-distance diagram covering about two hours of observations.  The gap between those two data sets is apparent as a white vertical stripe at about 2500~s. The vertical stripe at about 4100~s corresponds instead to images affected by transiting debris.  In the case of the streamer of region \textit{C}, we chose a path along its axis (top panels, path $\alpha$), and another path at a side (middle panels, path $\beta$), where features resembling plane-parallel waves can be seen.  A criss-cross pattern analogous to the one of path \textit{B}-$\beta$ is also visible in path \textit{D}-$\gamma$.

It is worth noting that the time-distance diagram of path \textit{C}-$\alpha$ displays not only upward propagating features at about $\sim 300$~\kms\ or higher, but also numerous features moving downwards at lower speeds, and decelerating.  In the movies, these signatures correspond to loop-like features moving downwards along the streamer axis. We interpret these features as signatures of coronal inflows or the downward moving part of inflow-outflow pairs already detected and studied in the past \citep{Sheeley-etal:1997,Sheeley-Wang:2002,Sheeley-Wang:2007,Lynch:2020}. Choosing a path closer to the streamer axis in region B would also show the same kind of features.  A work is in progress to analyse these features in more detail.  The main point of note is that these downward-moving features are related to reconnection processes occurring in the streamer current sheet.  Indeed, these features are not seen in the pseudo-streamers.  In contrast, the periodic perturbations we are discussing in this work are detected also far from the streamer axis and in pseudo-streamers.

In Table~\ref{Tab:speeds} we summarise the estimated projected speeds (slopes) of representative features (ridges) in the time-distance diagrams for the selected paths.  Along transverse paths \textit{B}-$\beta$ and \textit{D}-$\gamma$ mentioned above, negative values refer to apparent projected speeds towards the origin of the segment. In all these data, the absolute values of the apparent propagation speeds estimated from the time-diagrams fall in the range $\sim 150$ -- $450$~\kms.  

It is important to recall that if the propagation front is not perpendicular to the path, the slopes of the corresponding features in the time-distance diagrams will be higher.  Therefore, the values listed in Table~\ref{Tab:speeds} are to be considered as upper limits.  We believe this effect could explain the two distinct slopes measured in the time-distance diagram of path \textit{E}-$\alpha$ (lowest panels of Fig.~\ref{Fig:time-distance_D-E}): the higher value measured above $\sim 400$~Mm along the path is likely due to an angle between the propagation direction at those heights and the path \textit{E}-$\alpha$.  We experimented with different angles of the path at those heights, obtaining indeed lower speeds in the resulting time-distance diagrams.  The implication is that the propagation direction of these perturbation change with height in the pseudo-streamer of region \textit{E}. A sudden change in propagation speed between 1.9 and 2.4 \rsun\ cannot be excluded, however.  A more thorough investigation of a number of similar instances will be needed, but it is outside the scope of this work and will therefore be the subject of a follow-up analysis. 
\begin{table}
  \centering
  \caption{%
  Slopes of representative features in the time-distance diagrams. %
  }
  \label{Tab:speeds}
  \begin{tabular}{l r @{$\pm$} l l}
    \hline\hline
    Path Label           & \multicolumn{2}{l}{Speed} & Mean Height \\
                         & \multicolumn{2}{l}{\kms}  & \rsun \\
    \hline
    \textit{A}-$\alpha$      &  311 &  5 & 2.2 \\ %
    \textit{B}-$\alpha$      &  302 &  3 & 2.2 \\ %
    \textit{B}-$\beta$ (1)   &  425 & 40 & 2.0 \\ %
    \textit{B}-$\beta$ (2)   & -365 & 15 & 2.0 \\ %
    \textit{C}-$\alpha$      &  289 &  6 & 2.2 \\ %
    \textit{D}-$\alpha$      &  290 & 10 & 1.8 \\ %
    \textit{D}-$\beta$       &  280 & 10 & 2.2 \\ %
    \textit{D}-$\gamma$ (1)  &  160 &  7 & 1.9 \\ %
    \textit{D}-$\gamma$ (2)  & -226 &  7 & 1.9 \\ %
    \textit{E}-$\alpha$ (1)  &  295 &  5 & 1.9 \\ %
    \textit{E}-$\alpha$ (2)  &  453 & 20 & 2.3 \\ %
    \hline
  \end{tabular}
  \tablefoot{%
    The slopes of features in the time-distance diagrams were obtained from
    the selected paths within the ROIs.  The mean distance from disc centre of
    the segment in the time-distance diagram utilised to determine the slope
    is also listed. %
  }
\end{table}

\subsection{From time to spectral domain: time series and power spectra.}\label{sec:results:power}

The properties of these waves can be further investigated by analysing the time variability of the corresponding perturbations.  Since the amplitudes of these perturbations in these images is small, we chose to analyse in detail suitable average time series in representative locations within the ROIs.

More specifically, for each of the paths used to compute the time-distance diagrams shown in Figs~\ref{Fig:time-distance_A},~\ref{Fig:time-distance_B-C}, and~\ref{Fig:time-distance_D-E}, we selected a 100~Mm long segment.  The chosen length of the segment corresponds roughly to the distance covered in 300~s by a feature moving at 300~\kms.  We then averaged together the time series of all the pixels within the segment by taking into account the phase shift introduced by the apparent propagation speed, $V$, determined from the time-distance diagrams.  In the time-diagrams, this corresponds to a transformation of the time coordinate, $t$, at each distance, $d$, from $t$ to $t+(d-d_\circ)/V$, where $d_\circ$ is the distance of the central point of the selected segment.  Frames affected by sudden brightness variations due to debris passing in front of the telescope aperture were excluded in the computation of the time-distance diagrams.

\begin{figure*}[h]
  \centering
  \includegraphics[width=0.49\linewidth,trim= 50 40 40 50,clip]{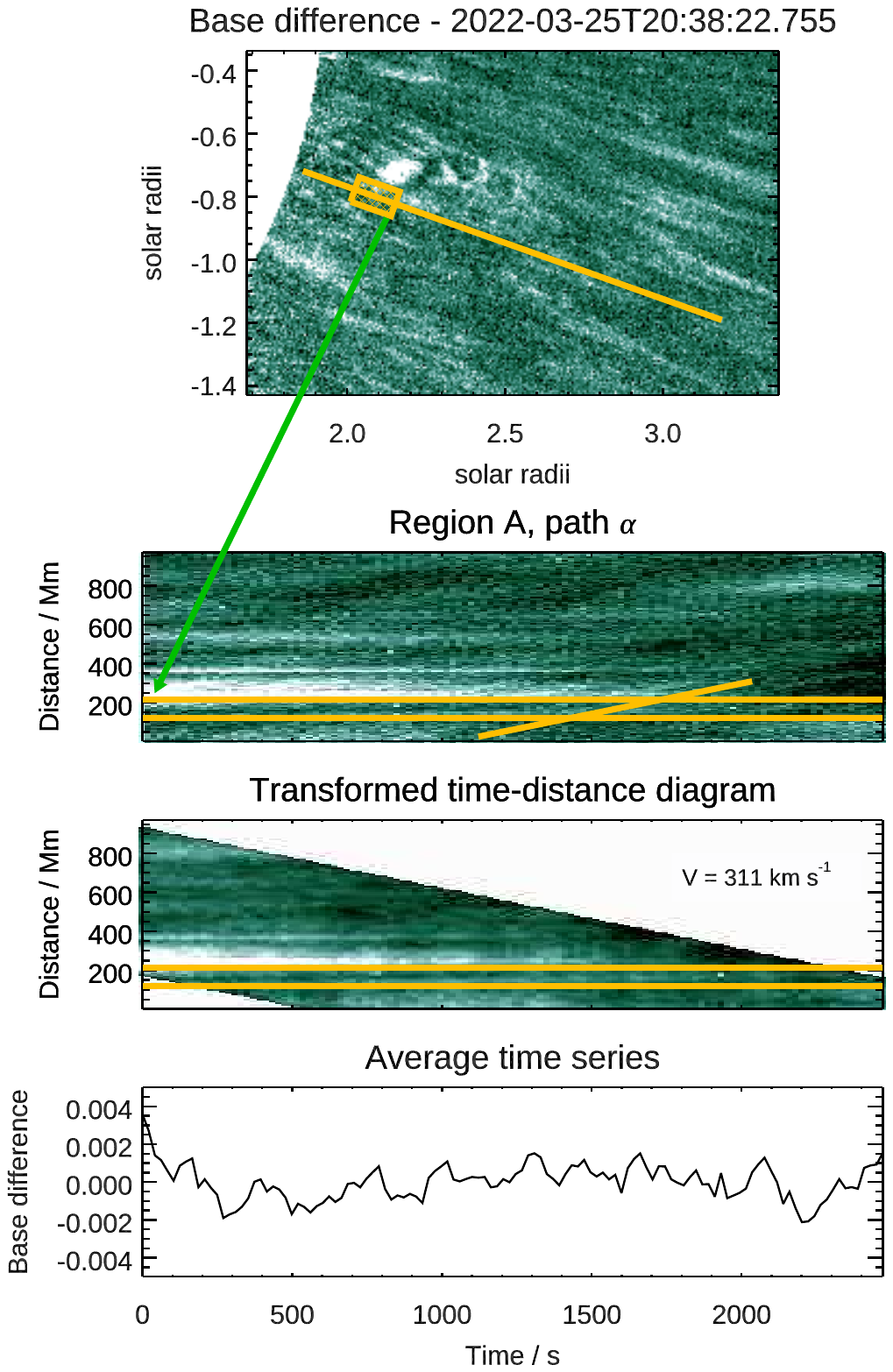}
  \includegraphics[width=0.49\linewidth,trim= 50 40 40 50,clip]{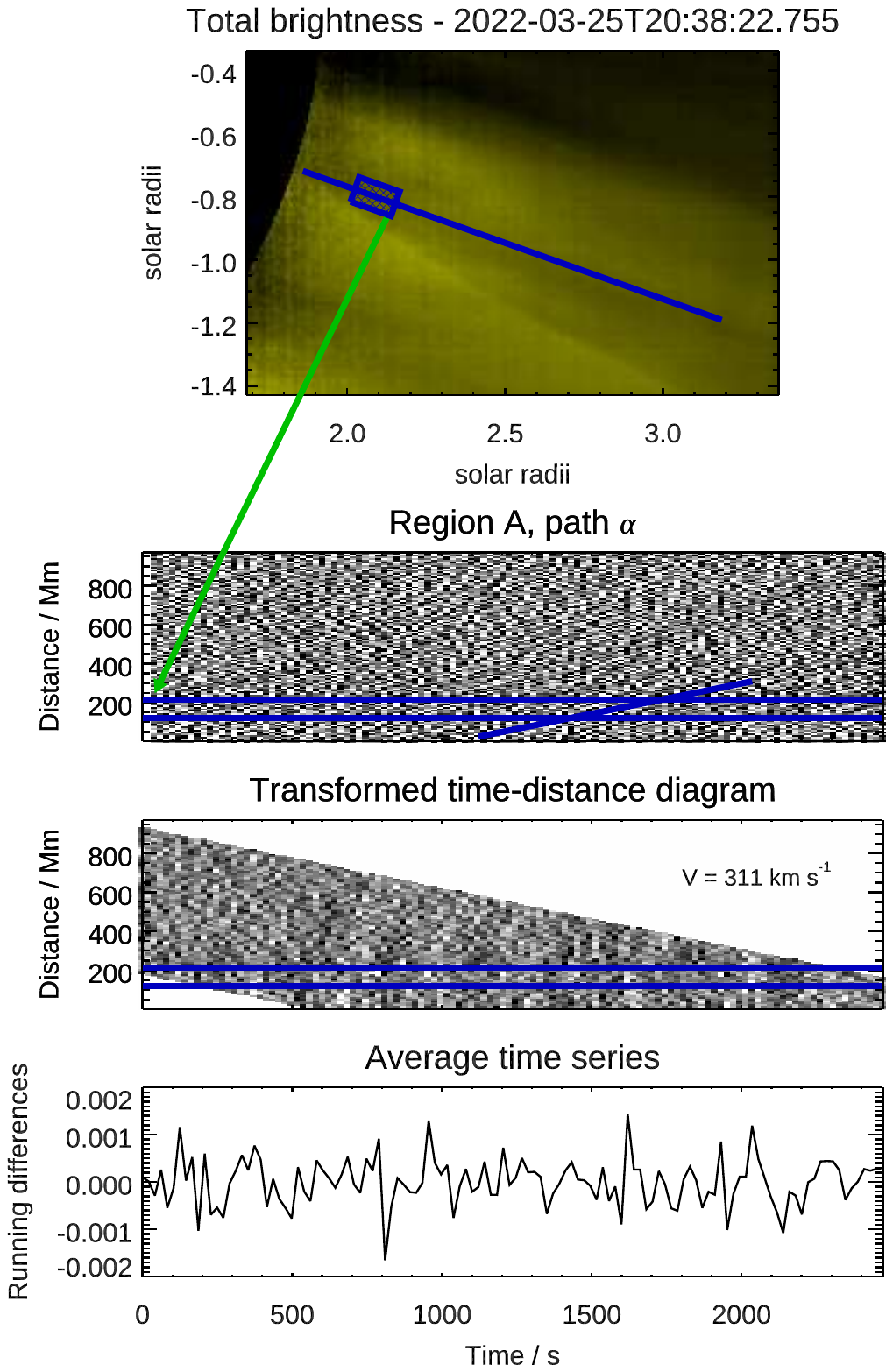}
  \caption{%
    Illustration of the procedure, described in Sec.~\ref{sec:results:power}, adopted to extract mean time series from a time-distance diagram along a path; the procedure in particular refers to a segment chosen along path \textit{A}-$\alpha$ (see also Fig.~\ref{Fig:time-distance_A}). The left column refers to the procedure applied to base difference images \bdiff\ (top left panel), while the right column refers to running difference images \rdiff{0} (the image on the top right panel is however a total brightness image, $\bnorm$, to show the main coronal features in the ROI). The segment where the average time series is obtained is enclosed in a box along the selected path in the top panels. The time-difference diagrams are shown in the second row; the range of distances corresponding to the box shown in the top panels are marked with horizontal lines, while one of the periodic ridges in the diagram is annotated as in Fig.~\ref{Fig:time-distance_A} together with the speed (listed in Table~\ref{Tab:speeds}) adopted for the transform procedure. The third row shows the transformed time-distance diagrams and the last row shows the time series averaged over the chosen segment. %
  }
  \label{Fig:warped_time-distance1}
\end{figure*}

The procedure is illustrated in Fig.~\ref{Fig:warped_time-distance1} for both normalised base and running differences (left and right panels, respectively), in analogy with the images shown in Fig~\ref{Fig:time-distance_A}.  In this case, however, we applied the procedure to time-distance diagrams obtained from $\rdiff{0}$ instead of $\rdiff{1}$, i.e. no temporal averaging was applied to the image series.  

\begin{figure}[h]
  \centering
  \includegraphics[width=0.9\linewidth,trim= 50 40 40 50,clip]{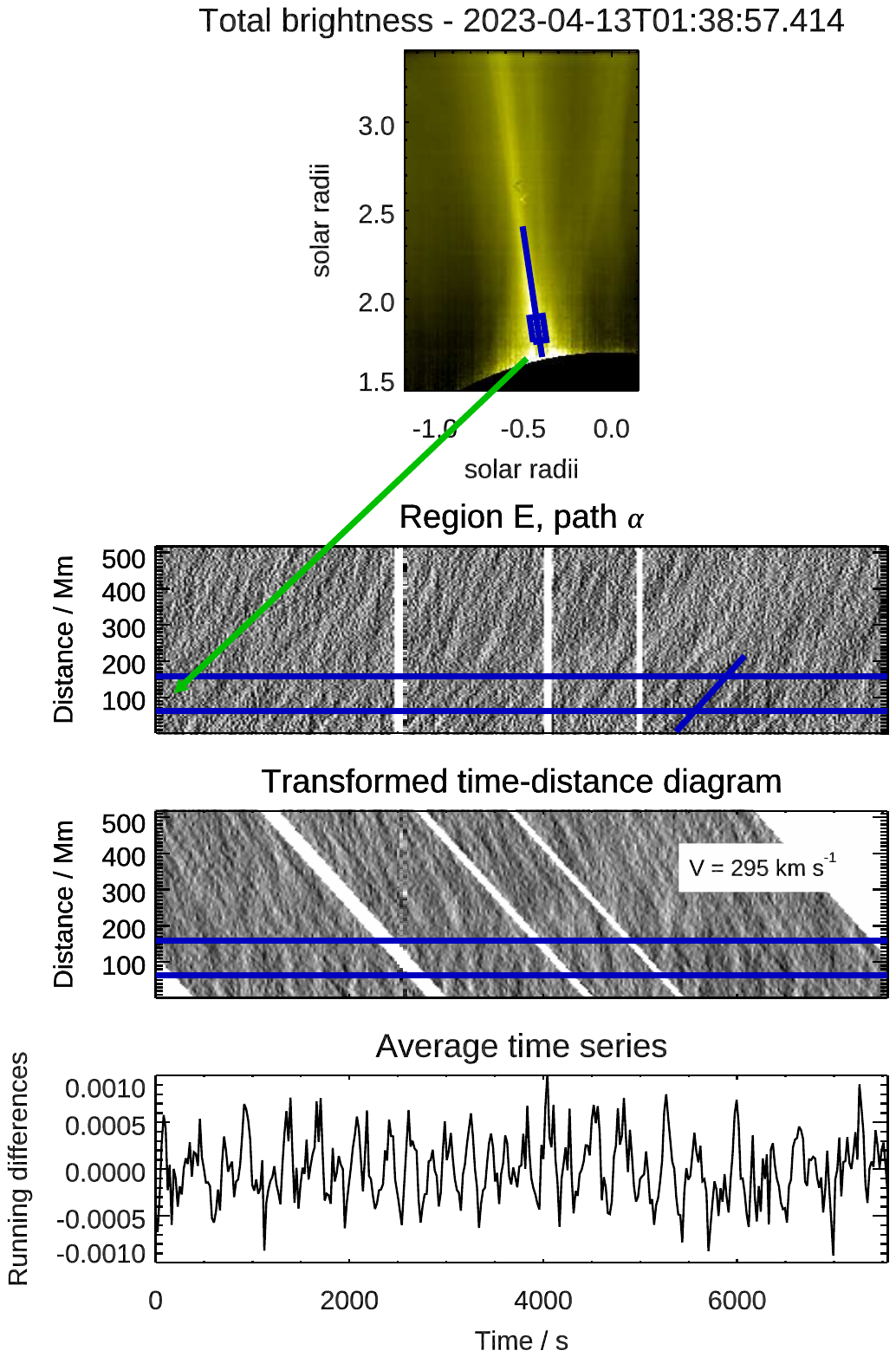}
  \caption{Illustration of the procedure adopted to extract mean time series from time-distance along path \textit{E}-$\alpha$ (see also in the bottom panels of Fig.~\ref{Fig:time-distance_D-E}) with the same format and annotations as Fig.~\ref{Fig:warped_time-distance1}. In this case, only the result for running-difference images is shown.}
  \label{Fig:warped_time-distance2}
\end{figure}

The application of the procedure to \bdiff\ is not always effective because larger amplitude, slower changes in the corona tend to hide these smaller amplitude, periodic variations.  Hence, in the following we consider only transformed time-distance diagrams obtained from normalised running differences, $\rdiff{0}$.  Another example is shown in Fig.~\ref{Fig:warped_time-distance2}, demonstrating the application of the procedure to the two consecutive data sets taken on 13 April 2023 (data sets \#3 and \#4 of Table~\ref{Tab:data}).  

With the adoption of the proper propagation speed, after this transformation the inclined ridges in the original time-diagram become vertical bands, thus making possible extracting the wave periodic signal by a simple spatial averaging over the chosen segment.  The resulting average time series is shown in the lower panel.  %

\begin{figure*}[h]
  \centering
  \includegraphics[width=\linewidth,trim= 0 15 100 50,clip]{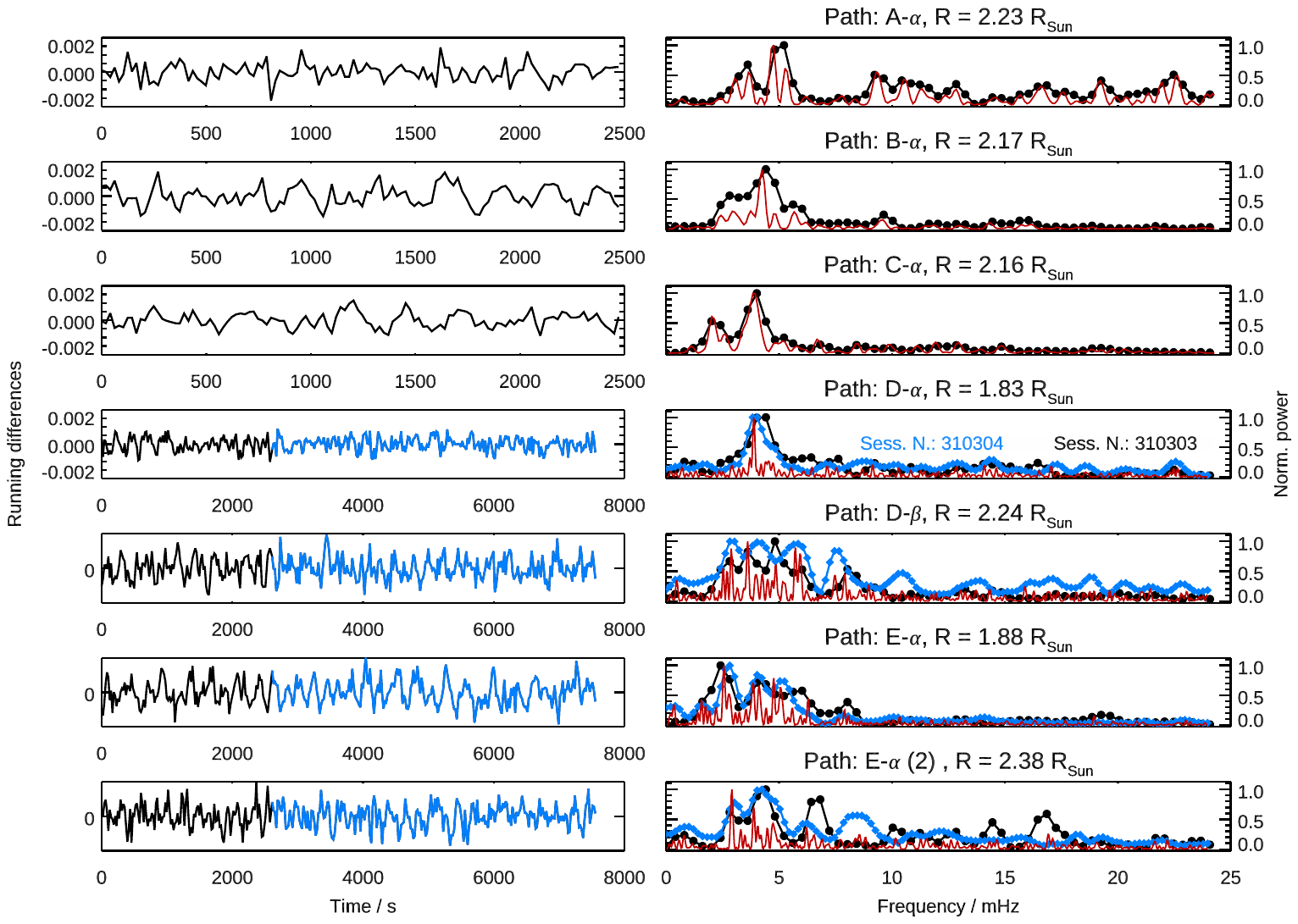}
  \caption{Left panels: Summary of time series (left panels) derived with the procedure described in Sec.~\ref{sec:results:power} in the middle of selected features among those identified in Figs~\ref{Fig:time-distance_A},~\ref{Fig:time-distance_B-C}, and~\ref{Fig:time-distance_D-E} and listed in Table~\ref{Tab:speeds}. In the case of the 13 April 2023 observations, the time series obtained from data set \#4 (session no. 310304) are in shown in blue.  Right panels: The corresponding power spectra for each data set are shown with the same colour coding. The Lomb normalised periodograms for the full time series are drawn with red lines.  The power spectra and periodograms are normalised to their respective maximum values.}
  \label{Fig:power_spectra_paths}
\end{figure*}

We have applied this procedure to the mid points of the features listed in Table~\ref{Tab:speeds} and shown in Figs~\ref{Fig:time-distance_A},~\ref{Fig:time-distance_B-C}, and~\ref{Fig:time-distance_D-E}.  The power spectra of the resulting time series are shown in Fig.~\ref{Fig:power_spectra_paths}.  In the case of the data sets \#3 and \#4 taken on 13 April 2023, since the gap is not a multiple of the cadence, it is not in principle possible to apply a Fast-Fourier Transform (FFT) to the combined time series.  We therefore show the power spectra of the two data sets separately.  It is however possible to compute Lomb periodograms \citep{Lomb:1976,Scargle:1982} for unevenly spaced data over the full time interval (see also \citet{Jess-etal:2023} for more details). These normalised periodograms are also shown in Fig.~\ref{Fig:power_spectra_paths}.

The spectra shown in Fig.\ref{Fig:power_spectra_paths} exhibit a power enhancement in the range 2--7~mHz, with peaks around 3 or 5 mHz, or both, corresponding to periodicities of 5 and 3 minutes respectively.  The power peaks are especially marked in the case of features \textit{A}-$\alpha$ (Fig.~\ref{Fig:time-distance_A}), \textit{C}-$\alpha$ (Fig.~\ref{Fig:time-distance_B-C}, a path along the axis of the pseudo-streamer), and \textit{D}-$\alpha$ (Fig.~\ref{Fig:time-distance_D-E}, along the axis of a streamer).

\section{Discussion}\label{sec:discussion}

Inspection of movies and analysis of time-distance diagrams indicate that wave patterns with periods of several minutes are found in several, extended regions of the corona.  No correlation with magnetic topology is evident: these waves are present with similar characteristics both in helmet streamers and in pseudo-streamers, for instance.  The wave fronts appear to be several Mm wide at least, in many cases filling a significant fraction of the apparent extent of streamers or pseudo-streamers as observed by Metis. The wave trains last as long as the duration of the observations (42~minutes for the first two perihelia in 2022, 125~minutes for the third perihelion, on April 2023).  These wave patterns are detectable in all the Metis high-cadence data we have checked so far, being visible for several days as long as the host structures (streamers or pseudo-streamers, for instance) remain identifiable.  

The amplitudes of these waves are of the order of $0.1$\% of the background signal. Since the background coronal brightness rapidly falls off with distance from the Sun, in the Metis data we have analysed these oscillations normally drop below the noise above 2.5 solar radii.  On the other hand, these waves are detected whenever the signal is sufficiently high, regardless of the magnetic topology of the host structure.  Hence, we speculate they permeate a large fraction or even the entire corona: in fainter structures, their signal is simply hidden in the noise.

  Whether or not the wave-like phenomenon described here could be related to or share some properties with the recurrent density enhancements widely detected in both remote-sensing and in situ observations \citep[e.g.][]{Sheeley-etal:1997,Viall-Vourlidas:2015,Kepko-etal:2016,Kepko-etal:2020,SanchezDiaz-etal:2017a,SanchezDiaz-etal:2017b,DeForest-etal:2018,Ventura-etal:2023} across the solar cycle, is a matter that deserves some attention.

  Both the observed spatial scales and inferred density/brightness amplitudes of periodic density structures reveal a broad spectrum of intermittent compact features characterised by a wide range of observed sizes, density contrast values, and propagation speeds, from the largest ones, the so-called ``Sheeley blobs'' to smaller and fainter substructures detected down to the resolution limits of existing instruments \citep{Viall-etal:2010,Viall-Vourlidas:2015,SanchezDiaz-etal:2017b,DeForest-etal:2018}.

  The recurrence timescales found for density enhancements detected by remote-sensing imaging range from several to few hours \citep[e.g.][]{Viall-Vourlidas:2015,Viall-etal:2010,SanchezDiaz-etal:2017a,Viall-Vourlidas:2015,Ventura-etal:2023} during the minimum and ascending phases of solar activity, and from about 1~h to 20~minutes \citep[e.g.][]{DeForest-etal:2018} for small density features detected during solar maximum, with a clear evolution towards shorter timescales (smaller structures) as the solar activity progresses to its maximum and the complexity of the global magnetic topology increases. Statistical analyses of 25 years of in situ observations \citep[see][]{Viall-etal:2008,Kepko-etal:2020,Kepko-etal:2024} have revealed that the distribution of the occurrence rates for solar wind periodic density structures is more complex than one might have expected on the basis of remote sensing observations only.  These studies have also shown that, in addition to a narrow band of recurrent frequencies between 0.1 and 1~mHz, peaking at about 0.2 mHz ($\sim$ 80 -- 90 min), which is consistent with individual frequencies detected in remote-sensing observations, a separate broad band between 1 and 3~mHz, peaking near 2.1 mHz ($\sim$ 8 min), is present.

  Given the wide range of observed frequencies, amplitudes, time-scales, and propagation speeds reported in literature for recurrent density enhancements in the solar corona, an overlap with the properties of the wave-like perturbations reported in this work can in principle be found in some instances \citep[e.g.][]{Viall-etal:2008,DeForest-etal:2018}.  However, the typical power spectra reported here (see Fig.~\ref{Fig:power_spectra_paths}) appear quite different.

  Moreover, it is widely acknowledged that the recurring density features reported in literature are not propagating waves \citep[e.g.][]{Kepko-etal:2002,Kepko-etal:2003,Viall-etal:2009,Viall-Vourlidas:2015}, but are more likely the result of (interchange) reconnection processes occurring in regions where the connectivity of the magnetic field lines experiences a strong gradient \citep[e.g.][]{Reville-etal:2022}.  This interpretations implies significant variations in occurrence rates and other properties in coronal regions of different magnetic topologies.  It is clear from the analysis shown in this work, that the measured properties of the wave-like features do not show a clear correlation with the topology of the ROIs analysed.

  Furthermore, the appearance of interference-like patterns in all the ROIs examined here except ROI~\textit{A}, would support the hypothesis that we are detecting a genuine wave-like phenomenon.  Also, while the quasi-periodic density features are commonly observed to propagate nearly radially outward, or at least following the magnetic field, the wave-like phenomena reported here are clearly detected even across the axis of a streamers (i.e. along path \textit{D}-$\gamma$).

  Finally, our data show numerous features moving upwards and downwards in streamers, but not in pseudo-streamers (see movies attached ROIs \textit{B} and \textit{D} and Fig~\ref{Fig:time-distance_D-E}). We interpret these features as signatures of coronal inflow-outflow pairs -- already detected and studied in the past \citep{Sheeley-etal:1997,Sheeley-Wang:2002,Sheeley-Wang:2007,Lynch:2020,SanchezDiaz-etal:2017b,Lynch:2020} -- which have been found to be associated with the leading edge of periodic density structures.

  Therefore, we think there is strong evidence that we are observing two different but concomitant phenomena: propagation of waves and periodic density structures flowing in the solar wind. Whether these two classes of phenomena are structurally and/or genetically linked to each others is a question that needs further work -- already in progress -- to be addressed in more detail.

  If it is confirmed that the wave-like phenomena reported in this work are of a different nature than the quasi-periodic density structures commonly observed in the solar wind, this raises the question of whether their source lies deep in the solar atmosphere or they are there produced locally in the solar corona.  In the latter case, a possibility could be in-situ generation of slow waves via parametric decay instability (PDI).  A recent example of an application of this approach in the interpretation of measurements of density fluctuations in the solar corona and wind is given by \cite{Chiba-etal:2025}.

  More generally, an outward, large-amplitude Alfv\'en wave, being unstable to density perturbations, can decay into a back-scattered Alfv\'en wave of lower amplitude and frequency and into a compressive magneto-acoustic wave that propagates in the same direction as the original (mother) wave.  This compressive component, thus associated with non-zero density fluctuations, leads to an enhancement of the spectral power at high frequencies.  We note, however, that while this process may indeed produce intermittent, quasi-coherent packets under certain circumstances, it is generally expected that the solar corona hosts a continuous spectrum of Alfv\'en waves susceptible to parametric decay.  Consequently, the magneto-sonic waves generated via this mechanism would also display a continuous spectral distribution, as hypothesised by \cite{Bruno-etal:2014} to account for the enhanced density fluctuation spectra observed by the Helios~2 spacecraft, rather than well-defined wave packets with a characteristic wavelength-period such as those reported in this work.

  Therefore, while at this stage an in-situ generation of these wave-like density perturbations cannot yet be excluded, the above arguments, together with the near coincidence of the observed frequencies with the typical frequencies of acoustic internal modes, lead us to favour a scenario in which these wave-like perturbations are essentially due to the leakage into the corona of those modes.  Further analysis, including theoretical and numerical calculations, is needed and will be addressed in future studies.
Regarding the observed amplitude of the observed waves, we note that the brightness signal is the result of the integral of the Thomson plasma emissivity along the line of sight over distances that we estimate to be of the order of a significant fraction of the solar radius.  Considering that the spatial scales of these waves is of the order of several tens of Mm (obtained by simply taking the ratio of the observed propagation speed to the frequency) and that they cover large-scale structures in the plane of the sky, the observed low amplitudes might be the result of the cancellation of an oscillatory signal integrated for several wavelengths along the line of sight. 

We also note that the wave signal observed in Metis data corresponds to electron density perturbations detected through Thomson scattering measurements, in contrast with wave-like oscillations with similar periods detected in loops of the lower corona \citep{Tomczyk-etal:2007}, which are seen only in transverse velocity signals from spectral lines of highly charged ions. Those oscillations can more likely be understood as transverse (kink) MHD modes \citep{VanDoorsselaere-etal:2008}.

The apparent propagation speeds of the waves observed by Metis are in the range 150 -- 450~\kms. As discussed in the previous section, these values may depend on the relative angle between the path used to derive the time-distance diagram and the actual propagation direction of the waves.  The frequencies of these waves typically fall in the range 2--7~mHz, with prominent peaks often found at 3 or 5~mHz, or both.  

Considering that the background coronal plasma is being accelerated to reach wind speeds of the order of a few hundred \kms, the propagation speed in the plasma co-moving frame should be $V-w$, where $w$ is the plasma radial velocity, if $V$ and $w$ are parallel. Analogously, the oscillation frequencies in the plasma co-moving frame could be lower than observed by a factor $1-w/V$. Assuming a wind speed of the order of $w = 160$~\kms\ as measured around the minimum of solar activity by \cite{Romoli-etal:2021}, the speeds listed in Table~\ref{Tab:speeds} imply frequencies in the moving plasma lower by factors 30\% to 50\% than observed.  It is worth noting that the measurements of \cite{Romoli-etal:2021}, as well as the more detailed results of \cite{Antonucci-etal:2023}, refer to heliocentric distances greater than 4~\rsun, beyond the range of distances covered by the data analysed here.

  We estimated the outflow wind speeds in the Metis plane of the sky by means of MHD calculations, described in Appendix~\ref{sec:roi_desc}.  We found that, indeed, the outflow speeds in the regions analysed are below 150~\kms\ in the case of region $C$ and even lower ($<$50~\kms) in all other cases.

  In order to obtain an alternative, empirical estimate of the outflow wind speeds, we applied the same technique utilised by the above-mentioned works \cite{Romoli-etal:2021} and \cite{Antonucci-etal:2023}, which relies on simultaneous Metis observations in the VL and UV channel.  The latter instrumental channel provides measurements of the \ion{H}{i} \Lya\ line radiance which, together with electron densities provided by polarised brightness ($pB$) measurements from the VL channel, yield maps of plasma outflow velocities by means of an application of the Doppler Dimming technique.  Further details are given in Appendix~\ref{sec:roi_desc}. Here, it suffices to report that our estimates of the wind speed in the bright structures for which both UV and $pB$ data are available are of the order of ~250~\kms along the axis of the pseudo-streamer of region \textit{C}, less than 200~\kms\ for the streamer of region \textit{D}, and less than $\sim$100~\kms\ for the pseudo-streamer of region \textit{E}.  These values are about 50\% larger than the estimates from MHD modelling and are affected by significant uncertainties, but confirm that the regions analysed are characterised by slow (mostly subsonic) wind outflow.
  
  It is also worth comparing the speed measurements reported in Table~\ref{Tab:speeds} with other relevant speeds in the corona.  We show in Appendix~\ref{sec:roi_desc} that the sound speed in the regions considered range from $\sim$150 to $\sim$200~\kms.  We found instead a larger variability in the Alfv\'en speed, which ranges from $\sim$500~\kms\ in most cases to $\sim$1000~\kms\ or higher in the pseudo-streamer of regions $C$. It is worth noting that the measured propagation speeds fall in between the sound and Alfv\'en speeds. Also, they not seem to correlate with the Alfv\'en speed. For instance, measured propagation speeds in regions $C$ and in two of the three paths selected in region $D$ are almost identical, while the Alfv\'en speed varies by more than a factor two between the two regions. These points merit to be addressed in future work based on a larger measurements dataset.

In any case, these frequencies cover the frequency range typical of internal $p$-modes.  However, we note that absolute values of the measured propagation speeds across the axis of the streamer in region \textit{B} (path \textit{B}-$\beta$) are higher than the speed along the streamer axis (path \textit{B}-$\alpha$).  The opposite occurs in paths \textit{D}-$\alpha$ (along streamer axis) and \textit{D}-$\gamma$ (across streamer axis).  We believe this is an indication that the picture of waves transported by an outward moving plasma is too simplistic.  We therefore think the possibility of a systematic investigation on the relation between the properties of these waves and the local wind speed is an important research avenue to explore.

To conclude, we summarise the main properties of the waves detected in high-cadence Metis observation as follows:
\begin{itemize}
  \item They are electron density perturbations;
  \item They are pervasive of bright coronal structures, perhaps of the entire corona, regardless of the magnetic topology;
  \item They persist in the same regions for several days, at least.
  \item The wave trains are visible with no evident damping for the duration of the observations analysed so far, i.e. for at least two hours %
  and for lengths of several tens of Mm;
  \item They are low-amplitude perturbations (about 0.1\% of the coronal background); considering the large integration volume of the observed signal, it is likely that the actual amplitudes are larger;
  \item The measured propagation speeds fall in the range 150 -- 450~\kms;
  \item The typical frequencies of these waves (even after correcting for the wind flow speed) are consistent with photospheric $p$-mode frequencies.
\end{itemize}

We regard the latter point in particular to be a clear indication that these wave-like density perturbations are the signature of acoustic flux leakage in the middle corona\footnote{We adopt here the definition of ``middle corona'' of \cite{West-etal:2023}}.  These results represent indeed the first detection of acoustic power at distances larger than 1.7~\rsun\ from their sources on the solar surface.

The presence of such acoustic leakage might be associated to a reduced cutoff frequency due for instance to inclined field lines with respect to the propagation direction of the wave; however, this does not appear to be the case according to the results of \cite{Jefferies-etal:2006} and \cite{Stangalini-etal:2011}.  Another, albeit admittedly unlikely, possibility is to postulate a mapping of resonance cavities for acoustic waves, of the kind detected by \cite{Jess-etal:2020} above sunspots, with frequency of the order of 3~mHz.  We believe further in-depth investigations are required to uncover the sources and mechanisms behind the observations described in this work.

\section{Summary and conclusions}\label{sec:conclusions}

We report in this work the first detection of wave-like, density fluctuations in the corona above 1.7~\rsun.  These fluctuations are characterised by enhanced power in the 2--7~mHz range, with peaks corresponding to periods close to 5 minutes.

This discovery was made possible thanks to observations taken with unprecedented temporal cadence and spatial resolution by the Metis coronagraph \citep{Antonucci-etal:2020,Fineschi-etal:2020} aboard Solar Orbiter \citep{Mueller-etal:2020} during its closest approach to the Sun.  These frequencies, typical of the global acoustic $p$-modes observed in the lower solar atmosphere, have been measured in a series of high cadence Metis observations.  These oscillations appear in many structures in the corona observed at three epochs with widely different coronal configurations.  In many cases, the wave-like patterns persist in the same structure for several days at least.  We interpret these waves as the signature of acoustic flux leakage in the corona.

The presence of such compressible waves has broad implications for the energy transfer in the outer layers of the Sun’s atmosphere, the heating of the plasma and the acceleration of the solar wind, and opens new avenues to the use of these waves as a diagnostic tool in the middle corona. It is also worth noting that the presence of waves in the solar corona and their role in supplying energy to the outer solar atmosphere and the solar wind is still widely debated.

Among recent results, \cite{Kepko-etal:2020,Kepko-etal:2024} report in-situ measurements in the solar wind at 1~au of periodic density structures with frequencies between 1 and 3~mHz arranged in trains lasting for hours or even days. Elemental composition arguments indicate that the origin of these solar wind density perturbations is in the solar atmosphere.  %

Incompressible kink oscillations in magnetic tubes with periods of the order of 5 minutes were detected in the lower corona \citep{Tomczyk-etal:2007,Morton-etal:2016}. However, no clear signature of $p$-modes were found at the distances from the solar photosphere probed by Metis. Nevertheless, the mode conversion mechanism identified by \cite{Morton-etal:2019} suggests a possible way for the 5-minute energy flux to overcome the filtering by the temperature and density gradients of the solar atmosphere, and be transmitted higher up in the corona.

Metis can test these hypotheses by measuring the power spectrum and spatial coherence and extension of the coronal oscillations, demonstrating how large volumes of the corona are affected. Metis also shows that these oscillations are compressible (coronal density fluctuations), and this places strong constraints on excitation and propagation of these oscillations.

\begin{acknowledgements}
  Solar Orbiter is a space mission of international collaboration between ESA and NASA, operated by ESA. 
  Metis was built and operated with funding from the Italian Space Agency (ASI), under contracts to the National Institute of Astrophysics (INAF) and industrial partners. Metis was built with hardware contributions from Germany (Bundesministerium f\"ur Wirtschaft und Energie through DLR), from the Czech Republic (PRODEX) and from ESA.
  The EUI instrument was built by CSL, IAS, MPS, MSSL/UCL, PMOD/WRC, ROB, LCF/IO with funding from the Belgian Federal Science Policy Office (BELSPO/PRODEX PEA 4000112292 and 4000134088); the Centre National d'Etudes Spatiales (CNES); the UK Space Agency (UKSA); the Bundesministerium f\"ur Wirtschaft und Energie (BMWi) through the Deutsches Zentrum f\"ur Luft- und Raumfahrt (DLR); and the Swiss Space Office (SSO).
  RL was supported through NASA Grant 80NSSC20K0192.
  This study was carried out with partial support from the \textit{Space It Up} project funded by the Italian Space Agency and the Ministry of University and Research (MUR) under contract n. 2024-5-E.0 - CUP n.I53D24000060005.
  VA wishes to thank E. Antonucci for useful discussions.
\end{acknowledgements}

\bibliographystyle{aa} %

\FloatBarrier
\begin{appendix}
  \section{Instrumental effects in Metis total-brightness acquisitions}
  \label{sec:metis_processing}

  The most frequently employed acquisition scheme in Metis operations is the so called VL-pB scheme described in \cite{Fineschi-etal:2020} and \cite{Antonucci-etal:2020}. In short, the VL-pB scheme consists in a series of 4 interlaced polarised images which are then used to build, through the on-ground processing pipeline, a polarised brightness ($pB$) image.  This acquisition scheme, however, is limited to a minimum exposure time of 15~s per frame, which implies a maximum cadence of one~$pB$ image per minute. To achieve higher cadences, two other different schemes are employed: VL-tB (``total brightness'') and VL-FP (``fixed polarisation'').  In both schemes, the VL detector acquires and delivers frames at constant rate and detector integration time.  In the VL-FP mode, the polarimeter is set to a fixed polarisation angle. This is the fastest mode, allowing cadences up to one frame per second.  We found however that the signal-to-noise ratio from this kind of acquisitions during the first three perihelia is very low, making a scientific analysis very challenging. Moreover, only 60 images can be acquired in a single acquisition session. In the VL-tB acquisition scheme, on the other hand, detector frames are acquired by switching the polarisation angle in the middle of the detector integration time, thus providing a single, total (unpolarised) brightness frame. This mode permits a maximum cadence of 1 frame per 20~s (the cadence of the data analysed in this work).  There are no bounds on the total duration of each acquisition sequence, the main limit being the spacecraft on-board memory and telemetry constraints in the planning of Solar Orbiter operations.
  \begin{figure}[h!]
    \centering
    \includegraphics[width=0.99\linewidth,trim=15 75 75 75]{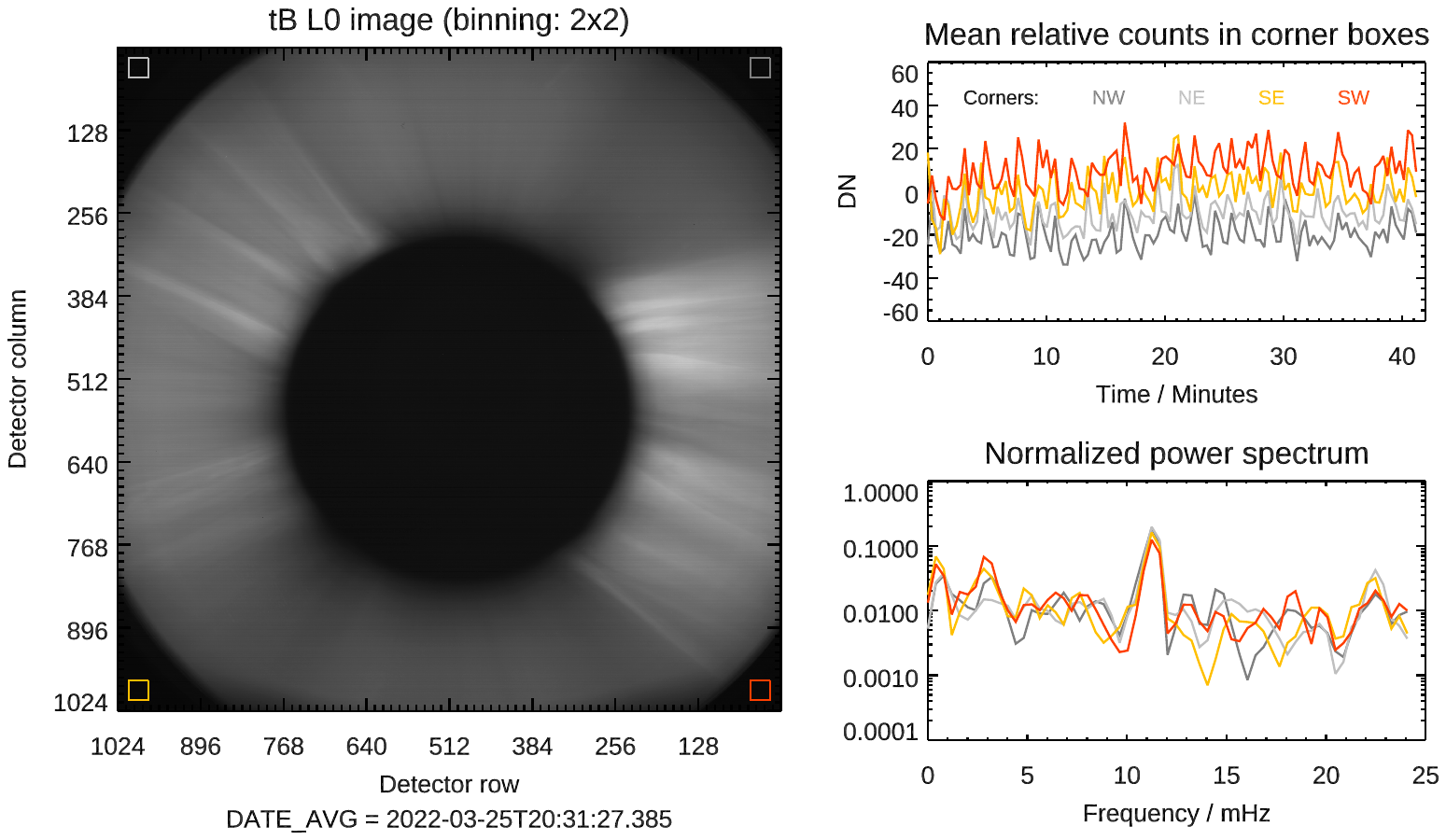}
    \includegraphics[width=0.99\linewidth,trim=15 75 75 75]{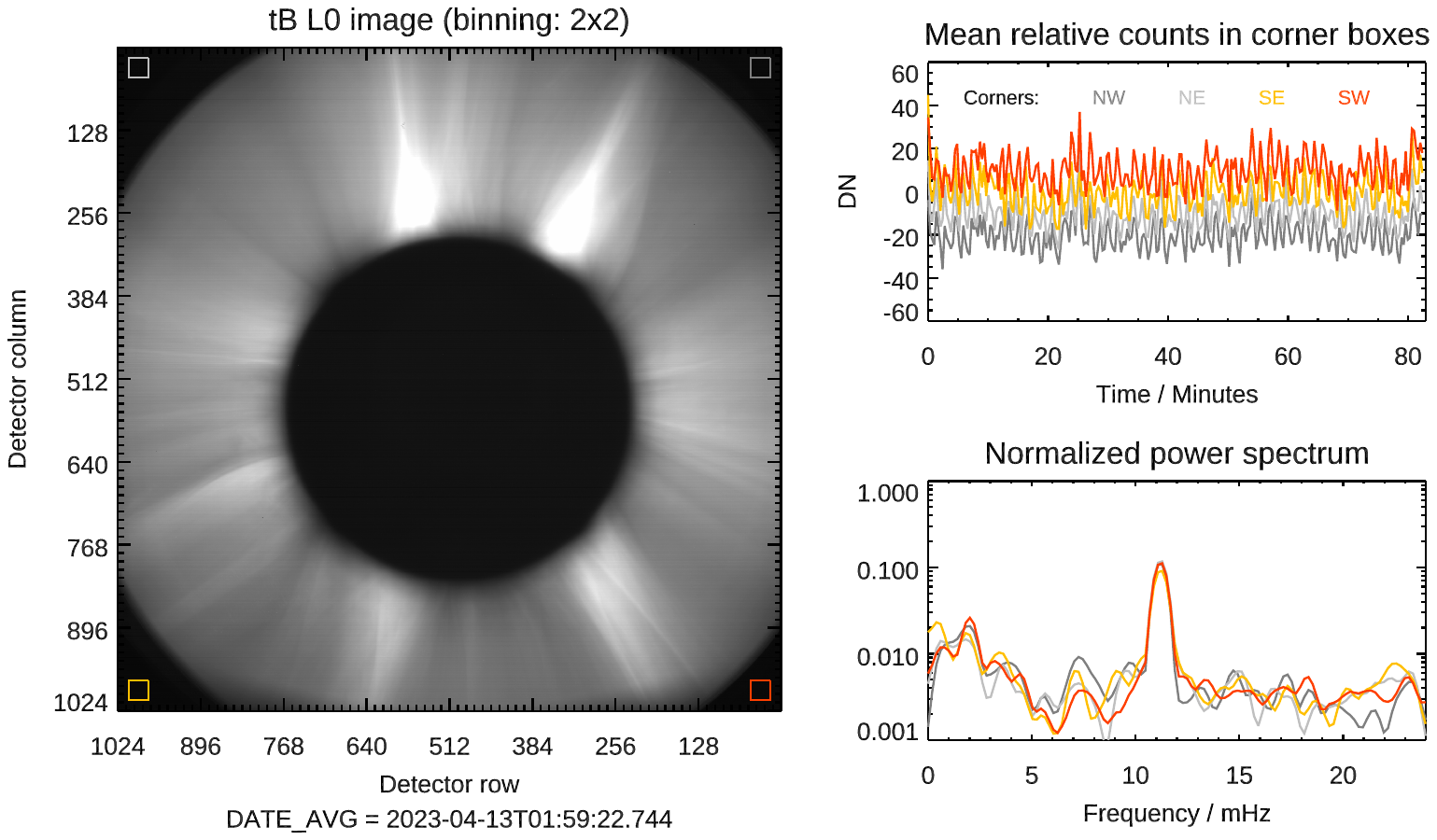}
    \caption{Left panel: position of sample boxes in detector regions not illuminated by the telescope. Top-right panel: Temporal variation of the median counts in those sample boxes (an offset of 15 DNs has been added to each time sequence for display clarity); bottom-panel panel: The corresponding normalised power spectra.}
    \label{Fig:instr_power}
  \end{figure}
  
  In the course of analysis of the data sets described in this work, we noticed periodic variations of the overall signal level.  To investigate the source of these variations, we examined in detail regions of the detectors that are not illuminated by the telescope, and in particular the four extreme corners which are not only shaded by the telescope field stop but are also behind the detector optical baffle. Thus, unlike the central occulted circular region where some residual solar signal might still be present, in those corners straylight is effectively absent and the detector signal is only given by its bias level and dark current. 

  Figure~\ref{Fig:instr_power} shows in the left hand panels the location of sample boxes of $31\times31$ pixels in representative images belonging to two of the datasets analysed in this work (data sets \#1 and \#4 of Table~\ref{Tab:data}).  The right hand panels shows the time series of the median counts in the sample boxes and the corresponding power spectrum.  The time series exhibit variations -- in phase among the four sample box -- with a period of about 90~s; this period corresponds to a frequency peak at 11.2 $\pm$ 0.2~mHz clearly visible in the power spectra.  Actually, such a temporal variation is present with the same phase over the entire detector, and we verified that it is present in all the analysed data sets, as well as in other data sets acquired with the same integration time and cadence. 

  In order to investigate this instrumental effect, we conducted a test on the Metis ground reference model by acquiring sets of dark images with the same DIT (Detector Integration Time), NDIT (number of on-board summed frames), and cadence parameters. Consistently, the same frequency was observed. Changing the acquisition parameters resulted in frequency variations, or even absence of this spurious frequency, suggesting a correlation with the timing control exerted by the Metis Processing and Power Unit (MPPU) on the detector.
  
\begin{figure}[h!]
    \centering
    \includegraphics[width=0.99\linewidth,trim=100 50 100 50]{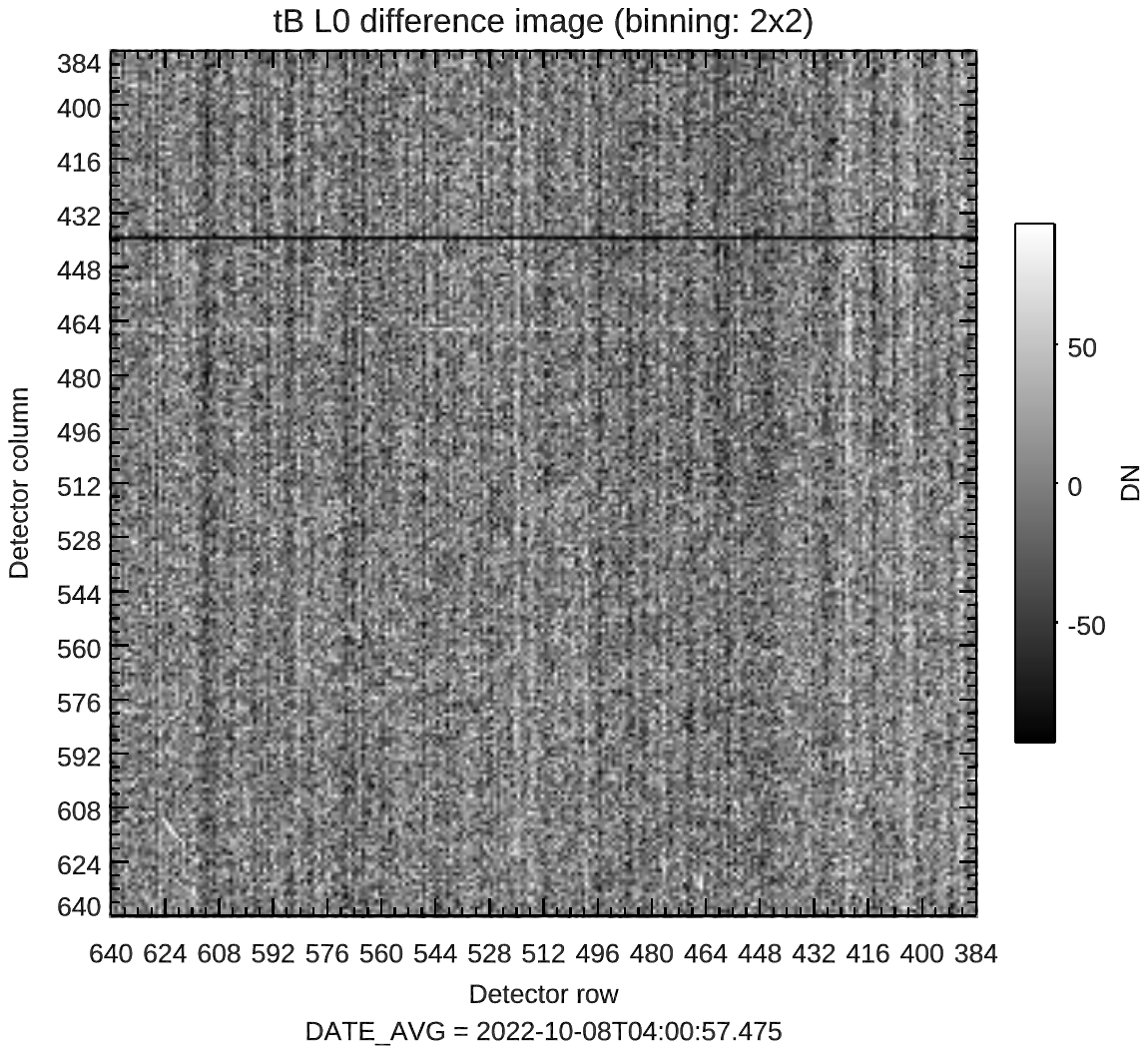}
    \caption{Difference of a sample image with respect to the previous in the central, occulted region of the detector.}
    \label{Fig:instr_hifreq}
  \end{figure}

  In addition, detector rows%
  \footnote{%
    The orientation of the VL detector in the Metis optical system is such that the solar rotation axis is normally approximately parallel to detector rows.  For convenience, the images displayed in this work are oriented so that the solar rotation axis is approximately vertical with north at the top (the same orientation as L2 Metis data), and therefore the noise pattern in detector rows appears as nearly vertical lines in images and movies shown.
  }
  are affected by a weaker but still noticeable row-pattern noise. An effective way to highlight this spatial noise is to simply take the difference of two consecutive images, as shown for example in Fig.~\ref{Fig:instr_hifreq}: the average signal is removed, leaving a residual proportional to the row-pattern noise.

  The row pattern noise is likely attributed to the specific architecture of the sensor electronics. The 2048$\times$2048 pixel array is read two rows at a time by a block of 4096 column amplifiers and sample-and-hold stages. For each pair of rows, the pixel signal is sampled concurrently, and noise in the reference voltages can introduce a uniform contribution across all pixels.

  We therefore devised a procedure to remove these instrumental effects from the data. We assume that for each frame $k$ in the sequence the detected value $V(i,j,k)$ at pixel $(i,j)$, where $i$ and $j$ are the detector column and row indices respectively, can be decomposed as follows:
    \begin{displaymath}
      V(i,j,k) =
      S(i,j,k) +
      N_\mathrm{O}(i,j,k) +
      N_\mathrm{RP}(j,k) + 
      N_\mathrm{D}(k) \; ;
    \end{displaymath}
    where $S$ is the coronal signal, $N_\mathrm{D}$ is the overall variation of the detector signal (at 11.2~mHz in these data sets), $N_\mathrm{RP}$ is the row pattern and $N_\mathrm{O}$ includes all other noise sources. To describe the following procedure we adopt the notation $M_d[A]$ for the median of multidimensional array $A$ over its dimension indexed by $d$; for example, we denote with $M_i[A(i,j,k)]$ the median value of array $A(i,j,k)$ along index $i$. 
    \begin{enumerate}
    \item For each pixel we compute the median value of the sequence, thus obtaining a reference image:
      $
      V_\mathrm{ref}(i,j)
      =
      M_k[V](i,j)
      =
      M_k[S](i,j) + M_k[N_\mathrm{O}](i,j) + M_k[N_\mathrm{RP}](j) + M_k[N_\mathrm{D}]
      \: ;$
      for sequences covering a sufficiently long time, the time average of the noise components $M_k[N_\mathrm{O}](i,j)$, $M_k[N_\mathrm{RP}](j)$, and $M_k[N_\mathrm{D}]$ can be assumed to be negligible, thus:
      $
      V_\mathrm{ref}(i,j)
      =
      M_k[S](i,j)
      \: .$
    \item For each image of the sequence, we compute the median value of the difference from the reference image for each column in a range of columns $[i_1,i_2]$ chosen to cover the selected ROI among those shown in Fig.~\ref{Fig:overview}:
      \begin{multline*}
        \delta V(j,k)
        = 
        M_i[V(i,j,k)-V_\mathrm{ref}(i,j)](j,k)   \\
        = 
        M_i[S(i,j,k)-V_\mathrm{ref}(i,j)](j,k) \\
        + M_i[N_\mathrm{O}(i,j,k)](j,k) + N_\mathrm{RP}(j,k) + N_\mathrm{D}(k) \; .
      \end{multline*}
      We then assume that the noise from sources other than the row-pattern and full-detector variations, when averaged over a sufficiently wide range of columns, becomes negligible. %
      We also assume that residual solar variations averaged over the selected column range vanish.  This is a reasonable assumption in regions of the corona characterised by small scale variations of a relatively constant background, as it is the case of the selected ROIs, but it may be invalid in the case, for example, of eruptive events crossing the ROI.  Under these assumptions, we therefore obtain:
      \begin{displaymath}
        \delta V(j,k)
        =
        N_\mathrm{RP}(j,k) +
        N_\mathrm{D}(k)
        \; .
      \end{displaymath}
    \item We then create an image by replicating this pattern as a function of time and detector row over all the detector columns:
      \begin{displaymath}
        \delta N(i,j,k) =
        \delta V(j,k)
        \; .
      \end{displaymath}
    \item The noise pattern image, $\delta N$, is finally removed from the original image:
      \begin{displaymath}
        V_\mathrm{denoised}(i,j,k) = V(i,j,k) - \delta N(i,j,k)
        \; .
      \end{displaymath}
    \end{enumerate}

  The procedure described above is applied to raw data. The standard processing pipeline is then applied to these corrected data, i.e.: dark, bias and flat-field correction, and removal of the vignetting function. This approach operates under the assumption that the noise is additive.

  The procedure seems to work well when restricted portions of the detectors are corrected, for example in the considered ROI.  The high-noise row-oriented pattern is still visible in regions near the occulter edge, most likely because the removal of the vignetting function amplifies small residuals from the correction.

  The procedure is less effective when there are strong solar variations in the ROI, for instance during the transit of coronal mass ejections, mainly because the reference image may be altered by these high-amplitude solar variations.  This problem could probably be addressed by adopting more sophisticated filters than a simple pixel-by-pixel temporal median filter to obtain the reference image for step \#1 of the procedure.

  We have also experimented with a version of the procedure to be applied to the entire detector, by using only non-illuminated portions of the detector: However, not all detector rows include non-illuminated pixels; furthermore, there are indications that the amplitude of the noise is weakly dependent on the signal level. Consequently, while the 11.2~mHz peak is still effectively removed, the row-oriented pattern noise is only partially suppressed.
  
  Considering that this correction procedure is still in its testing phase, we have verified that the analysis of the solar perturbations discussed here is not altered.  In particular, we have verified that the power spectra shown in Fig.~\ref{Fig:power_spectra_paths} are not substantially altered in the frequency range discussed in this work.

  \section{The solar corona in the regions of interest}
  \label{sec:roi_desc}

  \subsection{Magnetic field configuration}
  \label{sec:roi_desc:topology}

  \begin{figure}[h!]
    \centering
    \includegraphics[width=0.975\linewidth,trim=55 90 55 75]{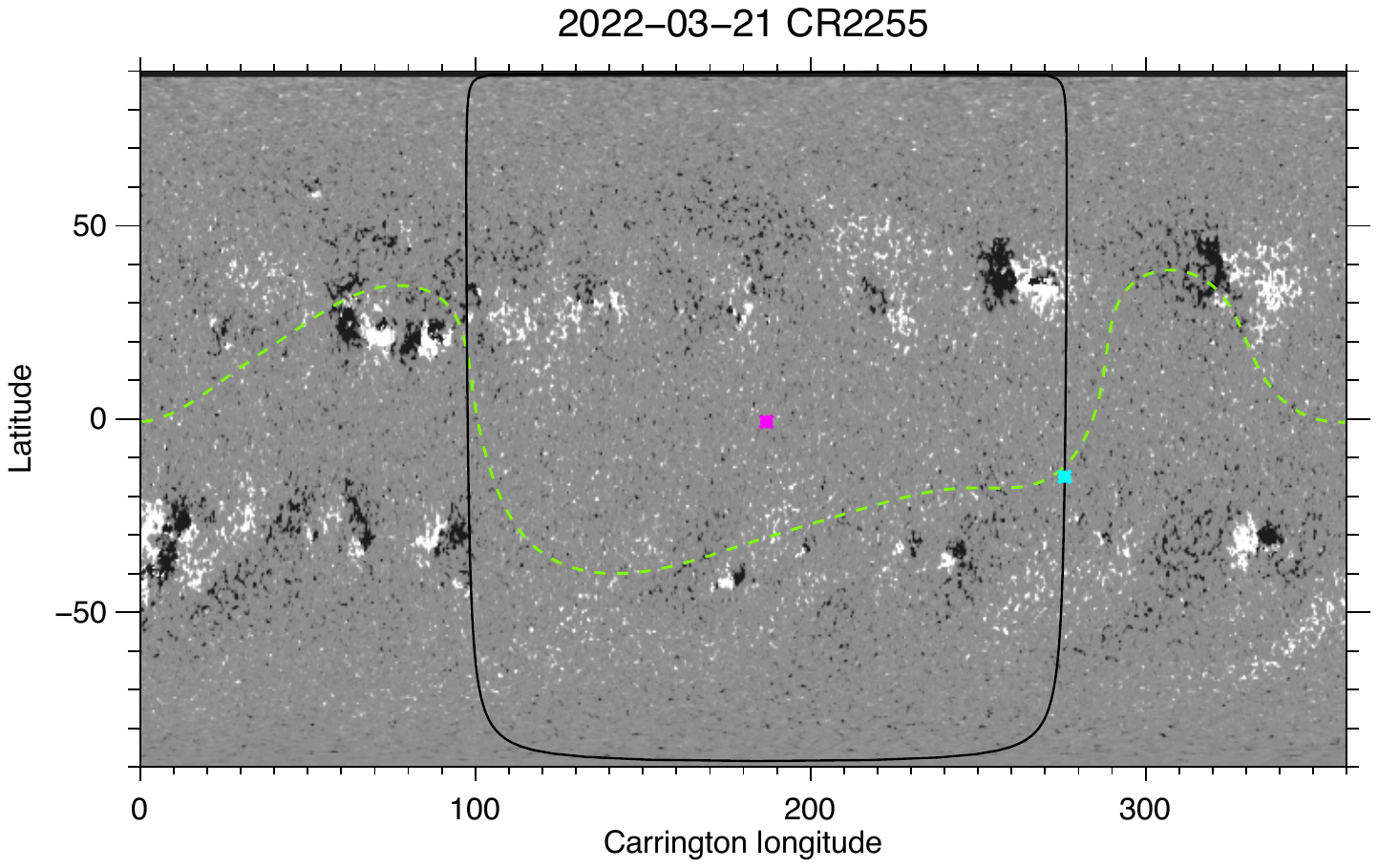}
    \includegraphics[width=0.975\linewidth,trim=55 90 55 75]{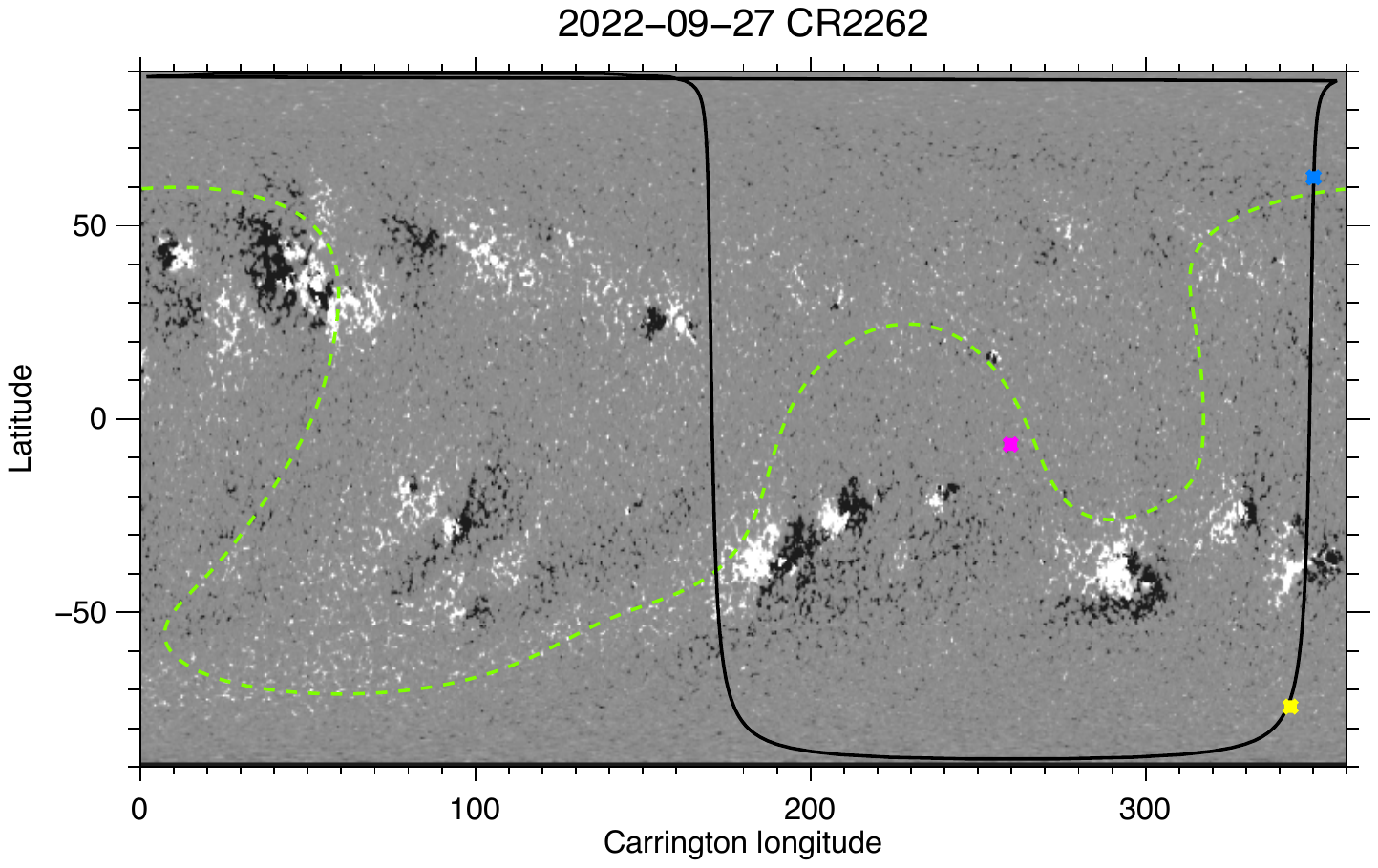}
    \includegraphics[width=0.975\linewidth,trim=55 90 55 75]{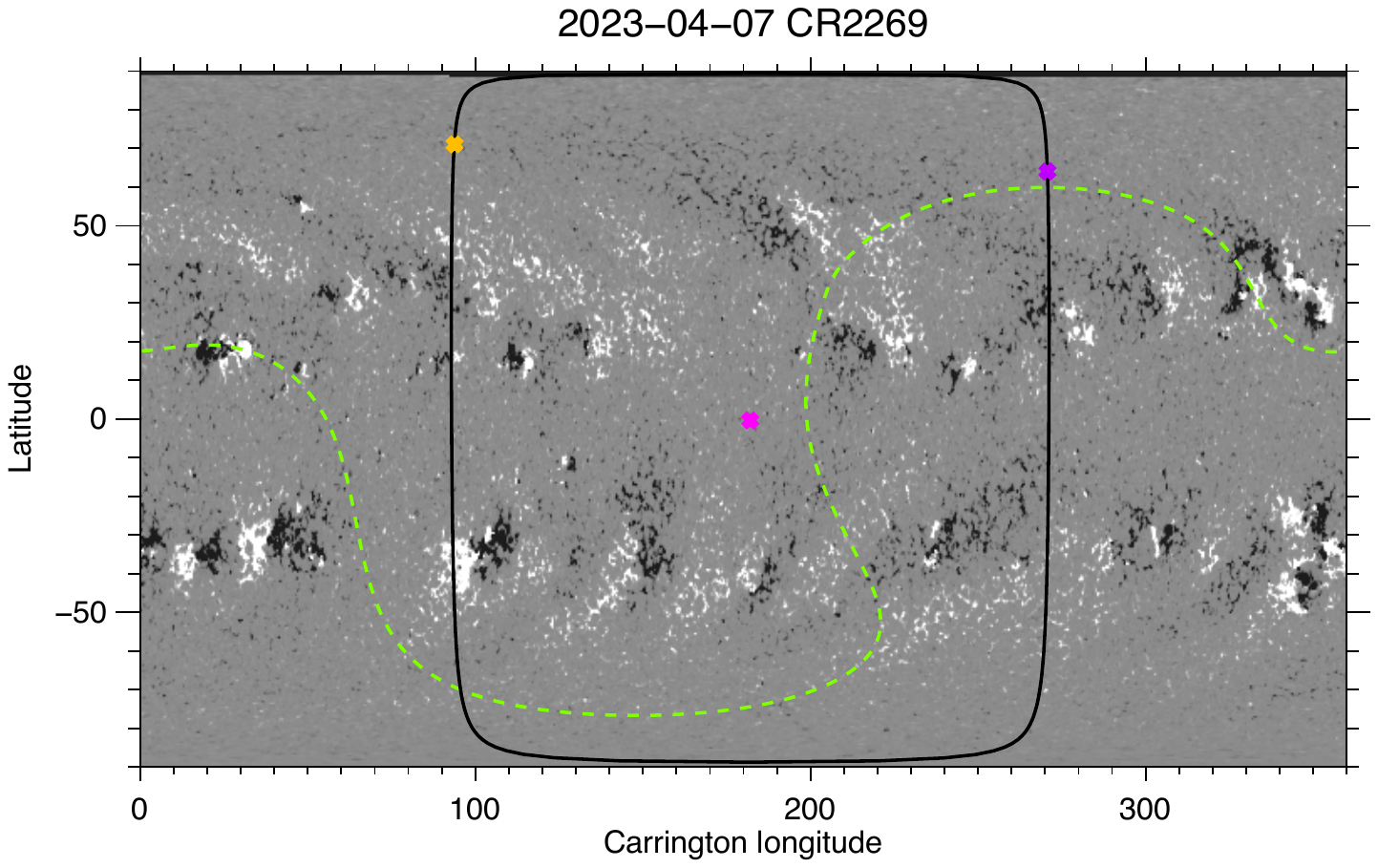}
    \caption{Synoptic charts of the photospheric line-of-sight magnetic field for the Carrington Rotations corresponding to the observation discussed in this work, obtained from HMI full-disk magnetograms.  The limb of the visible disk as seen by Solar Orbiter is shown as a black line. The dashed green lines outline the HCS produced by the magnetic reconstruction as provided by the Magnetic Connectivity Tool.
      Pink dots indicate the Sun centre as seen by Solar Orbiter. Coloured star points are the projection on the solar disk of the central position of each region of interest defined on the Metis plane of the sky, with the following colour codes: cyan for \textit{A}, light blue for \textit{B}, yellow for \textit{C}, purple for \textit{D} and orange for \textit{E}. }
    \label{Fig:synop}
  \end{figure}
    
  We examined the topology of the magnetic field in the ROIs aided by extrapolations computed at the epoch of the observations first by checking the synoptic magnetic maps for the Carrington Rotations covering the three dates considered in this work.  Figure~\ref{Fig:synop} shows synoptic magnetograms from the Helioseismic and Magnetic Imager (HMI) aboard the Solar Dynamics Observatory ~\citep[SDO, ][]{Scherrer-etal:2012} along with the position of the Heliospheric Current Sheet (HCS) at 2.5~\rsun\ computed with the \textit{Magnetic Connectivity Tool}, a web-based tool\footnote{\href{http://connect-tool.irap.omp.eu}{\textbf{http://connect-tool.irap.omp.eu}}} provided by the Solar Orbiter Data Analysis Working Group (MADAWG) to support to Solar Orbiter operations \citep{Rouillard-etal:2020}.  The projected positions on the disk of the centre of the ROIs suggest that regions \textit{C} and \textit{E}, far from the HCS and in between active region and polar coronal holes, might be classified as pseudo-streamers.  On the other hand, the other three ROIs would more likely cover bipolar streamers.

\begin{figure*}[h!]
    \centering
    \includegraphics[width=0.2688\linewidth]{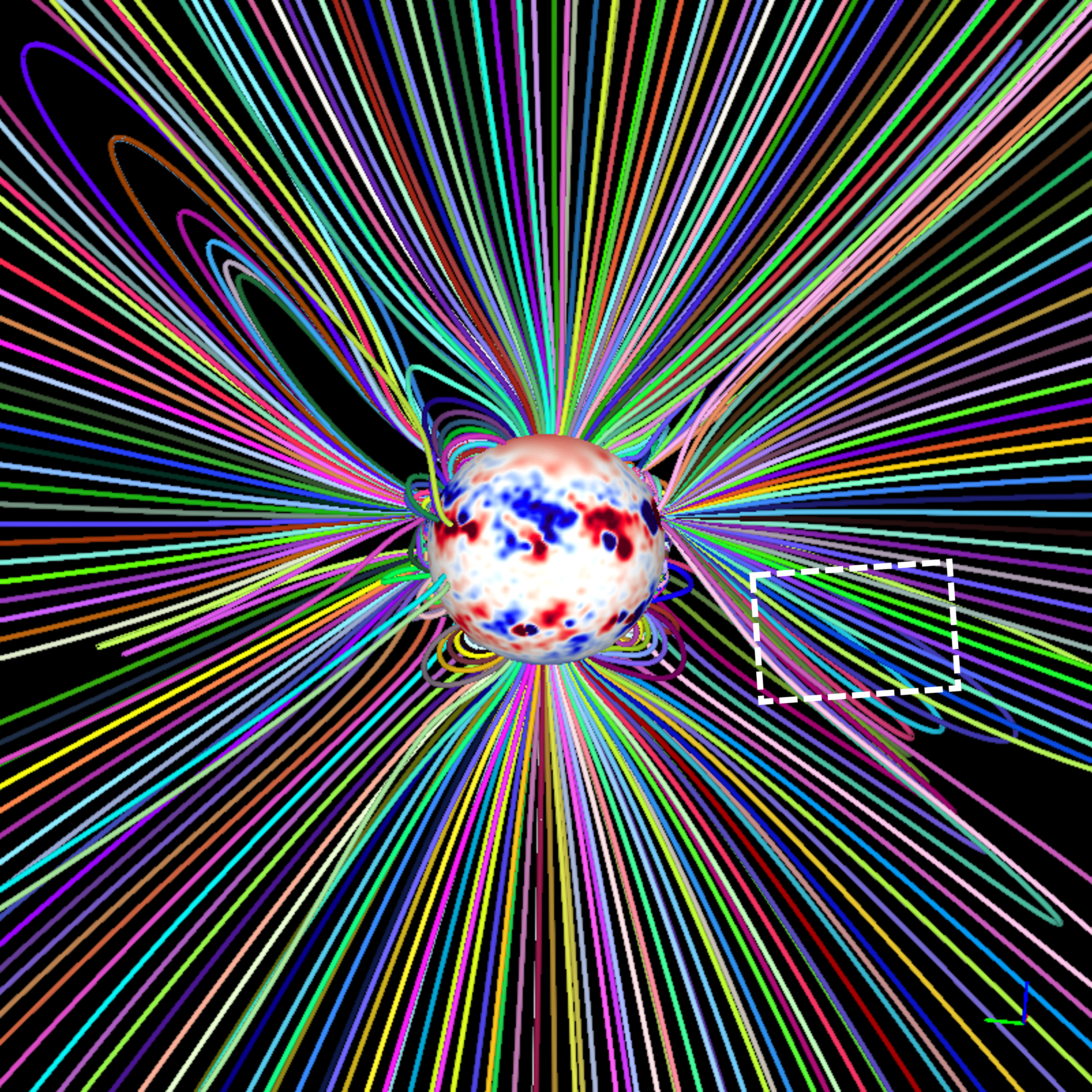}
    \includegraphics[width=0.2688\linewidth]{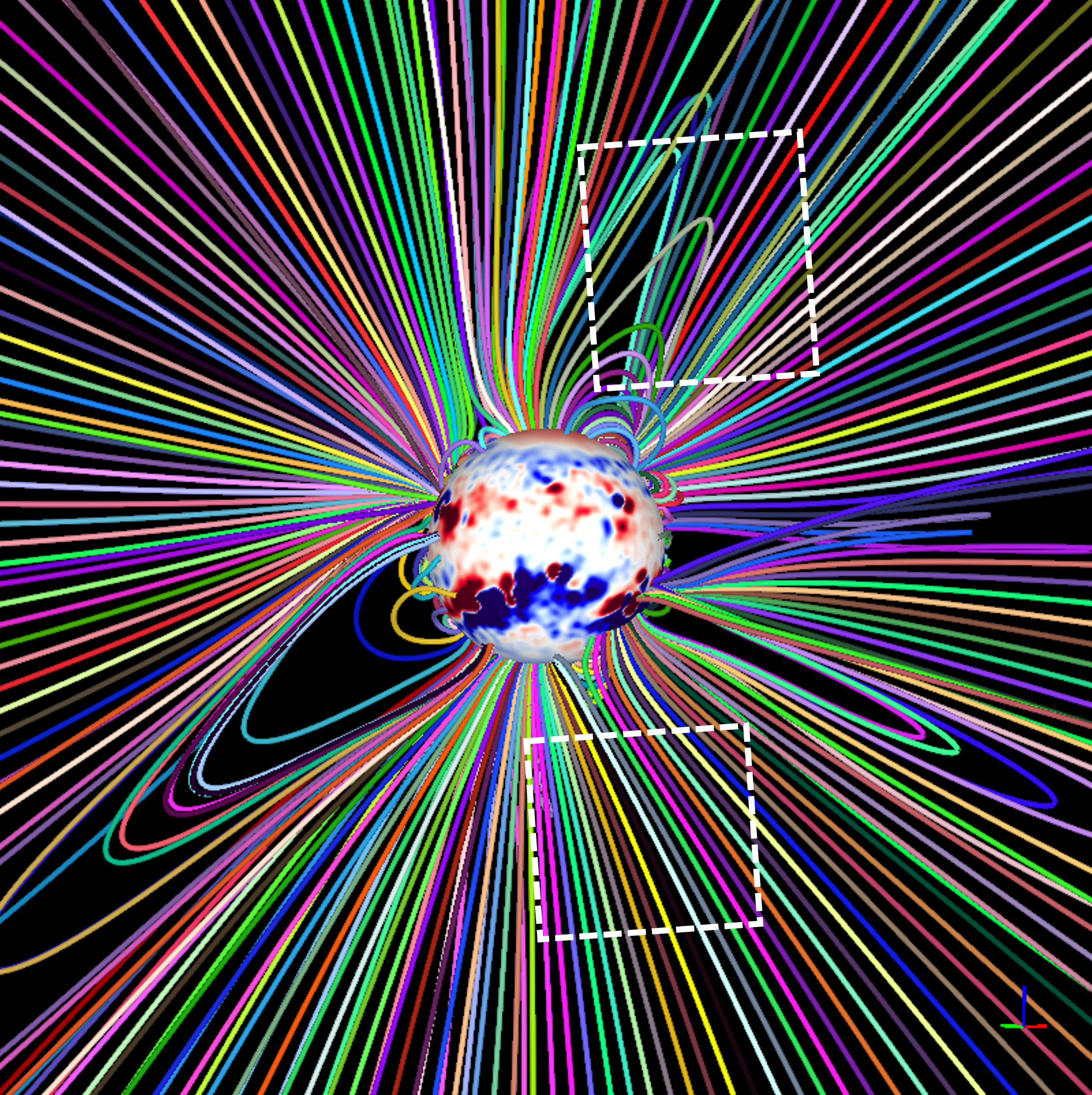}
    \includegraphics[width=0.2688\linewidth]{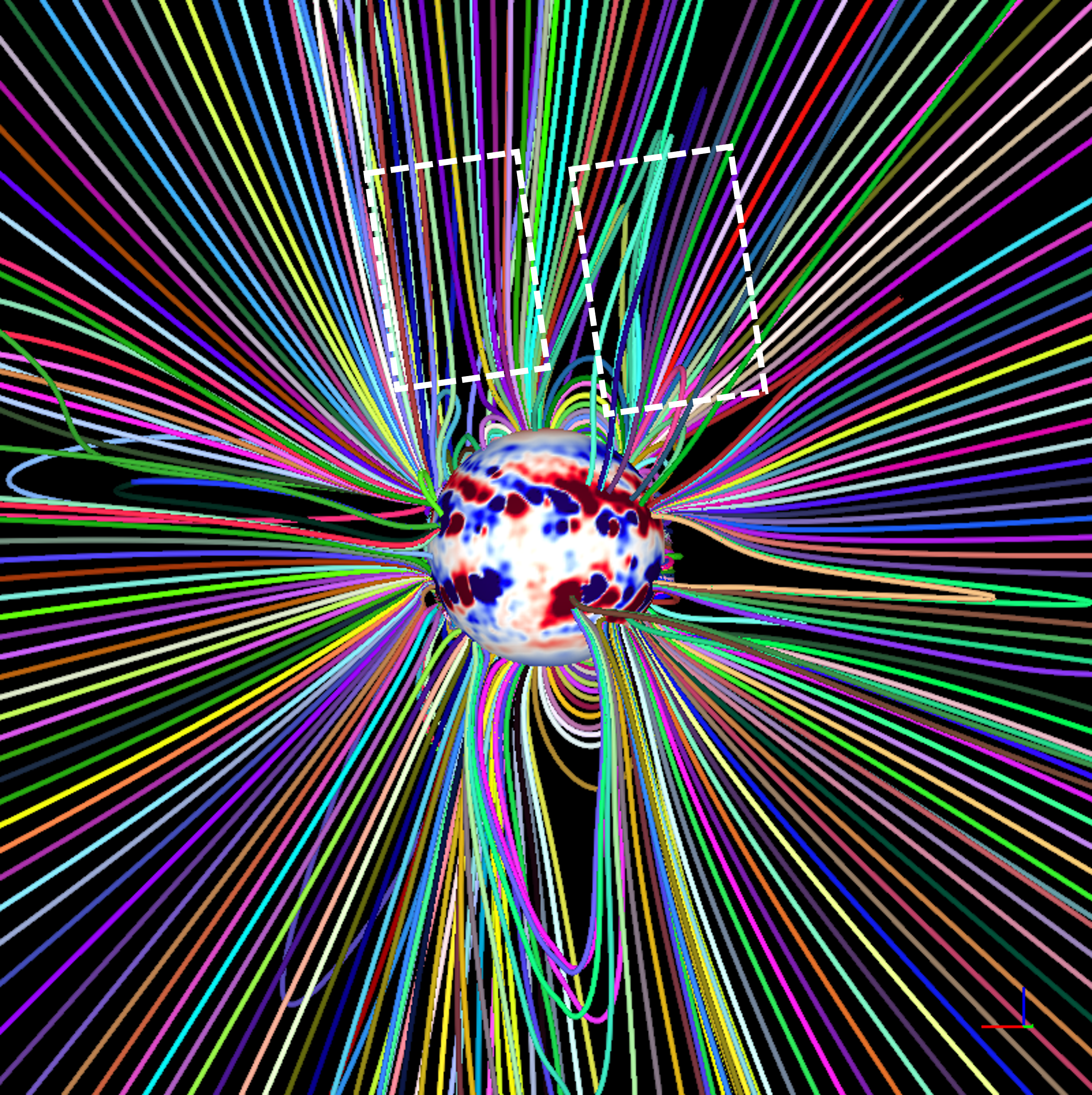}
    \caption{Magnetic extrapolations of the field lines with the 3D-MHD model developed by Predictive Science Incorporated (PSI) for the date on 25 March 2022 (left), on 8 October 2022 (centre) and on 13 April 2023 (right) from the Solar Orbiter point of view. The rectangular white boxes show the ROIs analysed in this work.}
    \label{Fig:Bfield}
  \end{figure*}

  Magnetic field extrapolations provide more information on the topology of the field in the corona observed by Metis.
  Figure~\ref{Fig:Bfield} shows the magnetic field extrapolations from the 3D MHD model developed by Predictive Science Incorporated~\citep[PSI; e.g.,][and references therein]{Mikic-etal:2018} for the analysed dates. The photospheric boundary conditions used for the extrapolation rely on Carrington maps of the measurements from HMI. The comparison between the observations and the extrapolated field lines allows us to identify more clearly the features, such as, for example, the bipolar streamers, the pseudo-streamers and the regions at the boundaries with the adjacent polar coronal holes.

  In particular, we can describe the magnetic configurations in the ROIs as follows:  the region \textit{A} observed at the West limb on 25 March 2022 corresponds to open field lines at the boundary of a streamer; the regions \textit{B} observed at North-West and \textit{C} observed at South-West on 8 October 2022, correspond to a quite complex bipolar streamer and a pseudo-streamer, respectively; the regions \textit{D} observed at North-West and \textit{E} observed at North-East on 13 April 2023, correspond to the transition from the streamer axis to the boundary and a pseudo-streamer, respectively.

    Inspection of the so-called squashing degree or factor $Q$ \citep{Titov-etal:2008,Titov-etal:2011} from the same MHD model provides further insights into the topology of the coronal magnetic field. Large values of the $Q$ factor, for instance, characterise separatrix layers or quasi-separatrix surfaces.  In particular, \cite{Titov-etal:2011} demonstrated that the signed $\log Q$, slog$Q$, is a convenient tool to visualise the topology of the coronal magnetic field.

\begin{figure*}[h!]
    \centering
    \includegraphics[height=0.2520\textwidth,trim=10 10 220 60,clip]{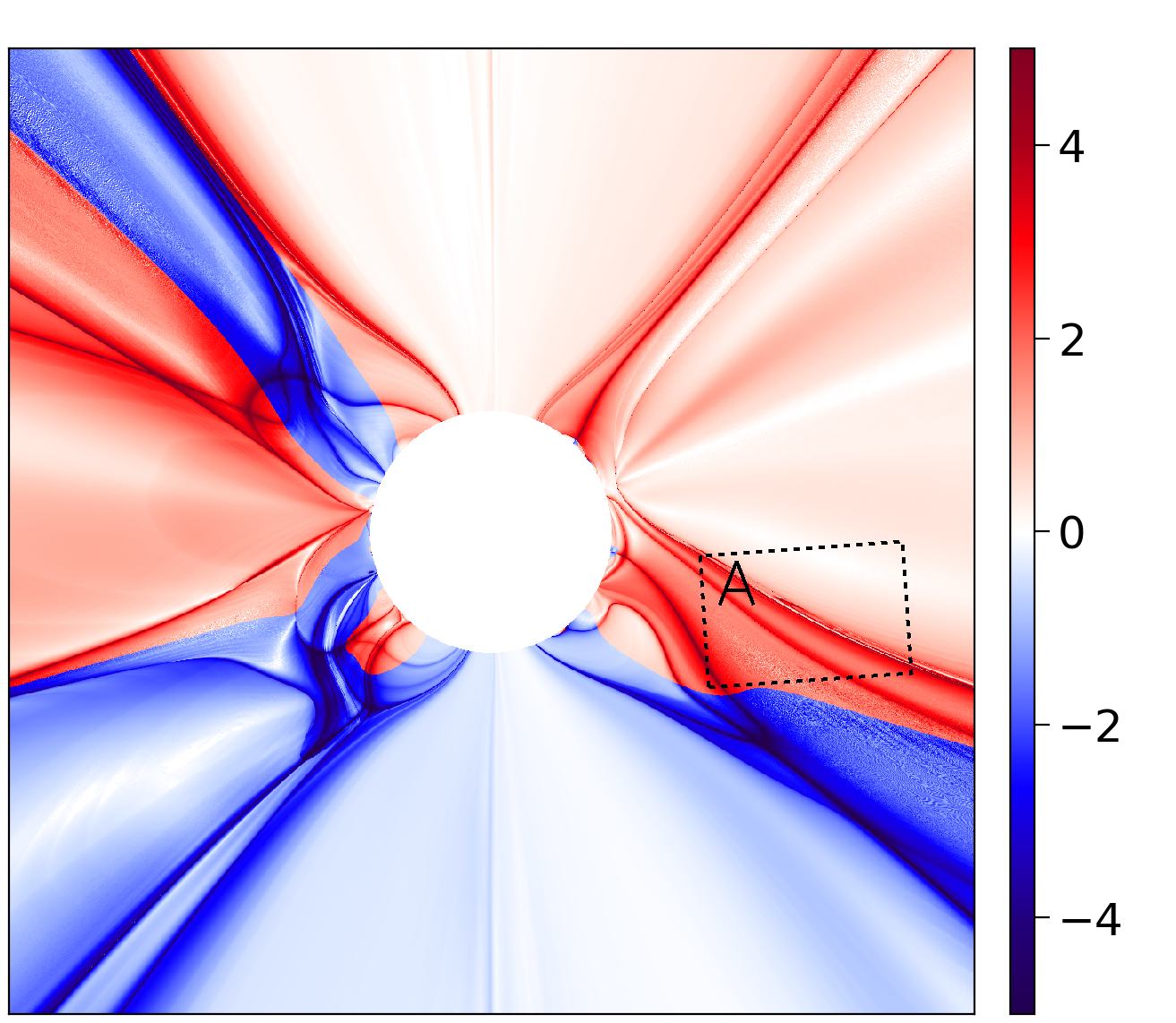}
    \includegraphics[height=0.2520\textwidth,trim=10 10 220 60,clip]{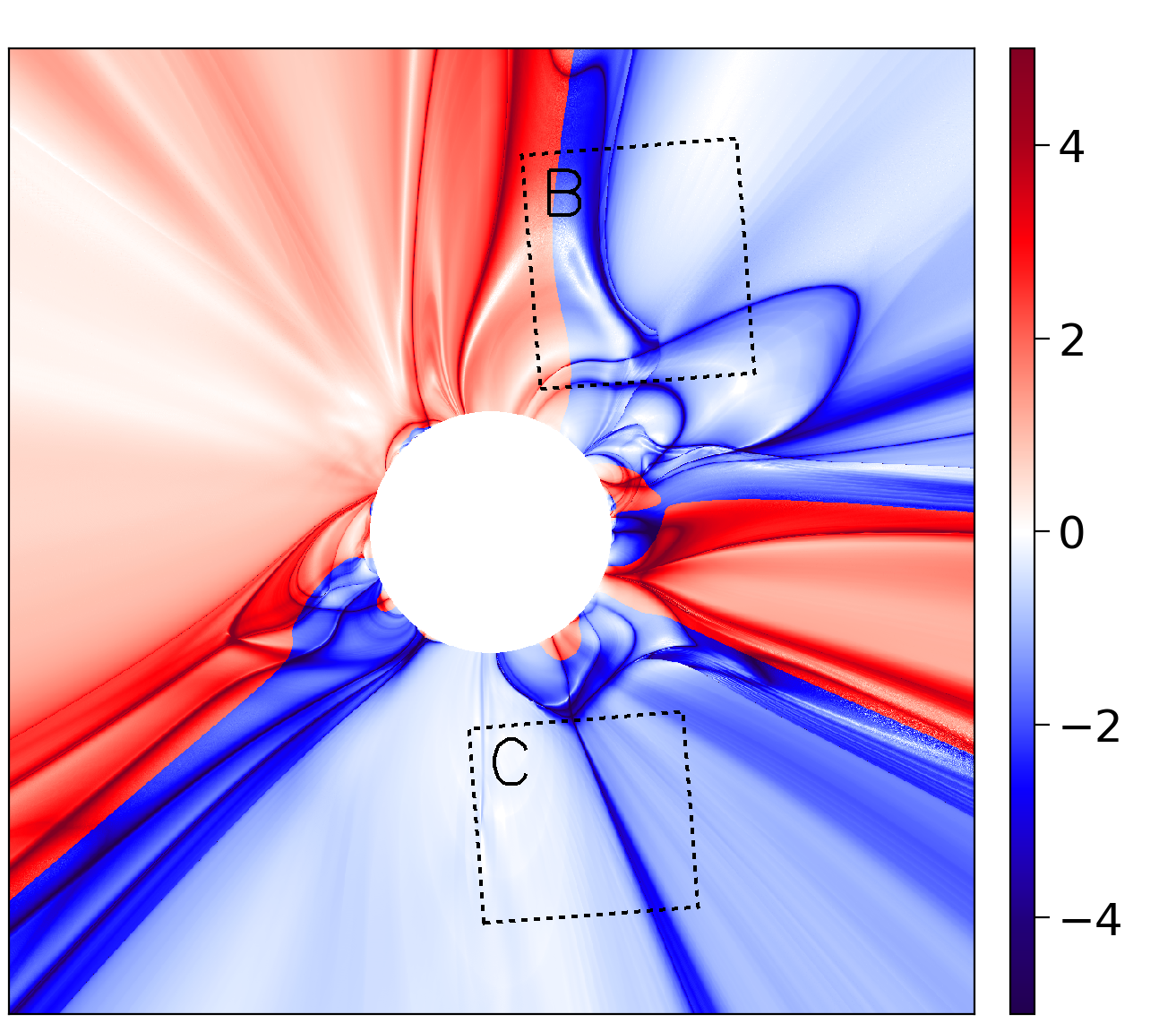}
    \includegraphics[height=0.2520\textwidth,trim=10 10  60 60,clip]{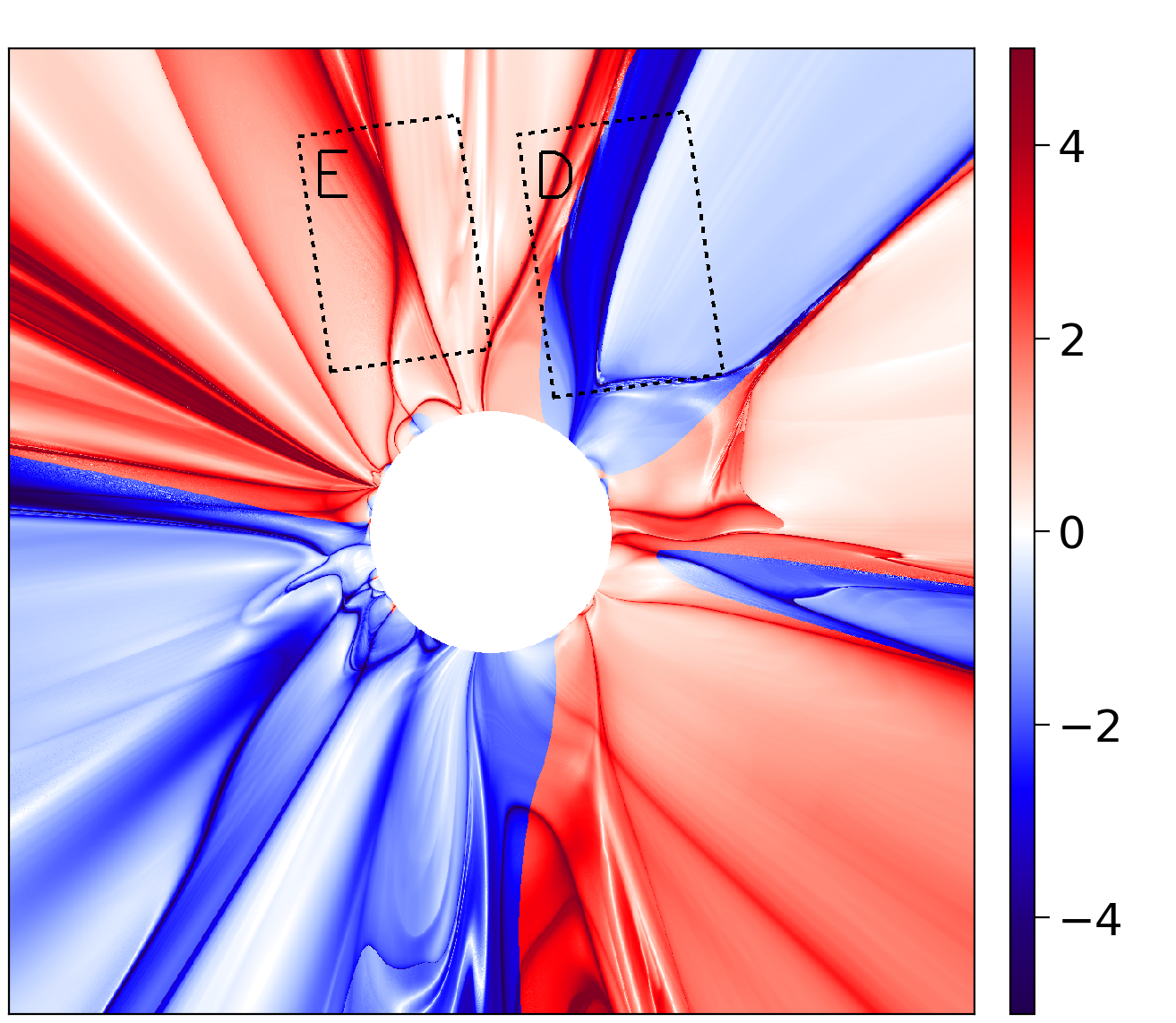}
    \caption{%
        Maps of slog$Q$ slices on the Metis plane of the sky for the dates considered, in the same order as in Fig.~\ref{Fig:Bfield}. The ROIs analysed in this work are also marked.
    }
    \label{Fig:slogQ}
  \end{figure*}

    Fig.~\ref{Fig:slogQ} shows slog$Q$ slices in the Metis plane of the sky for the analysed dates.  The left-hand map confirms that the magnetic field polarity region \textit{A} is positive, with some high-$Q$ structures suggesting intersection of unipolar 3D structures with the plane of the sky.  The centre map confirms that region $B$ includes a large bipolar (streamer) structure and that region $C$ is a strong negative unipolar region (pseudo-streamer).  Likewise, the right-hand map confirms again the interpretation of the bright structures in regions $D$ and $E$ as a streamer and a pseudo-streamer respectively.

  \subsection{Densities and speeds}
  \label{sec:roi_desc:characterisation}
    In addition to the analysis of the magnetic topology of the ROIs, we also estimated coronal electron densities, $N_\mathrm{e}$, by applying the \cite{vandeHulst:1950} inversion method to $pB$ images.  To do so, we considered the VL-pB (``polarised brightness'') acquisitions immediately following the VL-tB sessions analysed here (see Table~\ref{Tab:data}). In particular, we considered session no.~208408 starting at 20:54:01 UTC on 25 March 2022, session no.~228104 starting at 04:30:00 UTC on 8 October 2022, and session no.~310308 starting at 14:43:59 UTC on 13 April 2023.
  \begin{figure*}[h!]
    \centering
    \includegraphics[width=0.2730\textwidth,trim=160 40 50 40,clip]{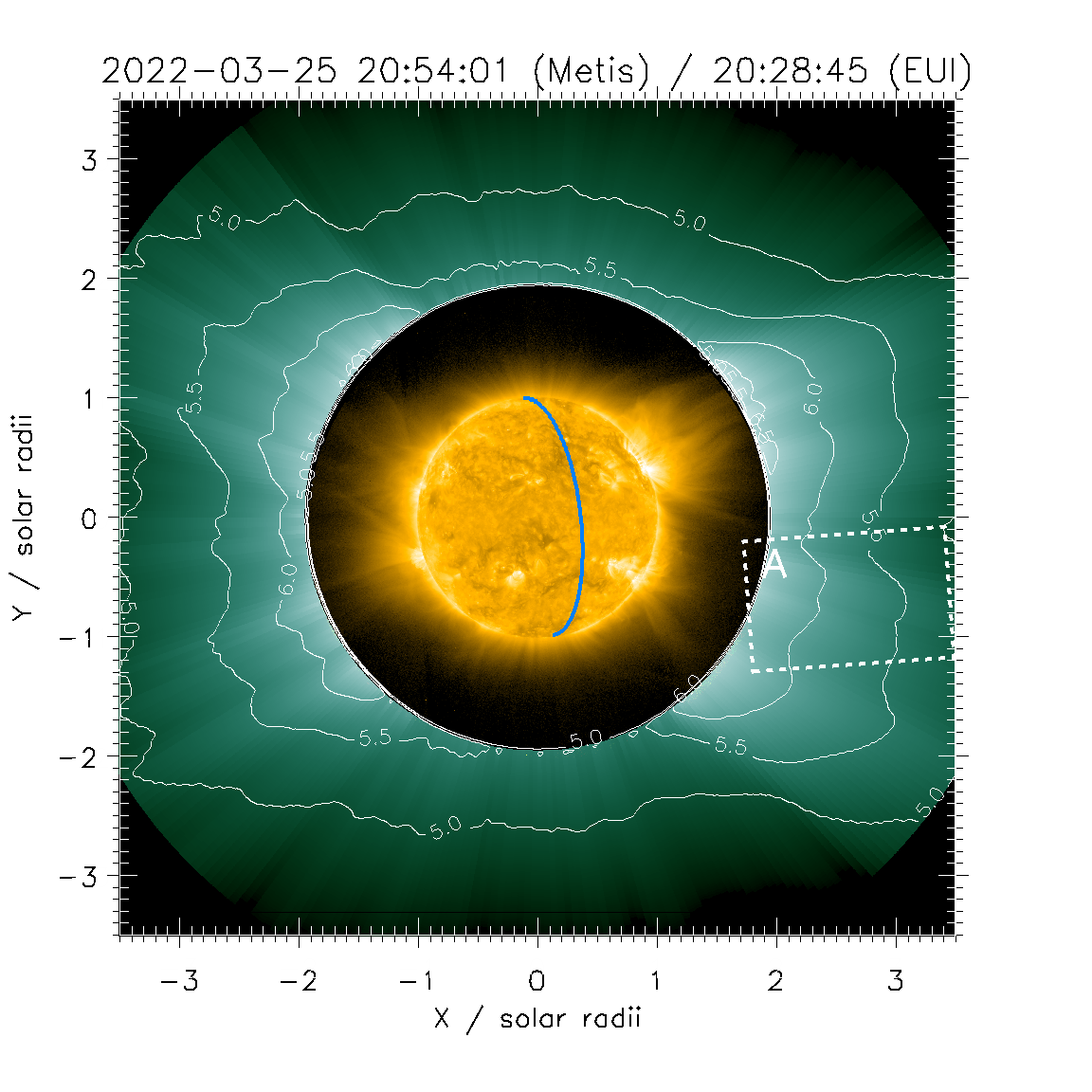}
    \includegraphics[width=0.2730\textwidth,trim=160 40 50 40,clip]{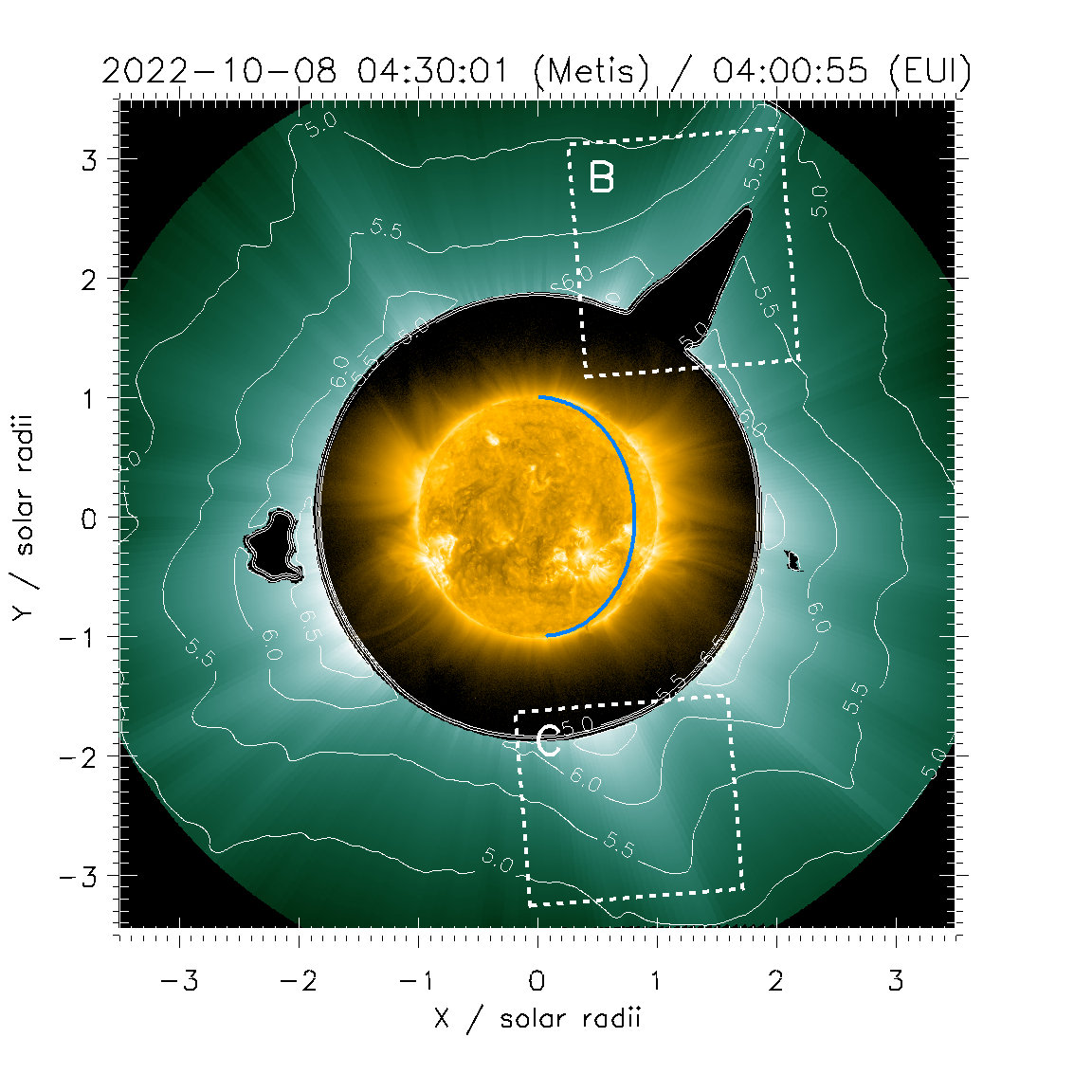}
    \includegraphics[width=0.2730\textwidth,trim=160 40 50 40,clip]{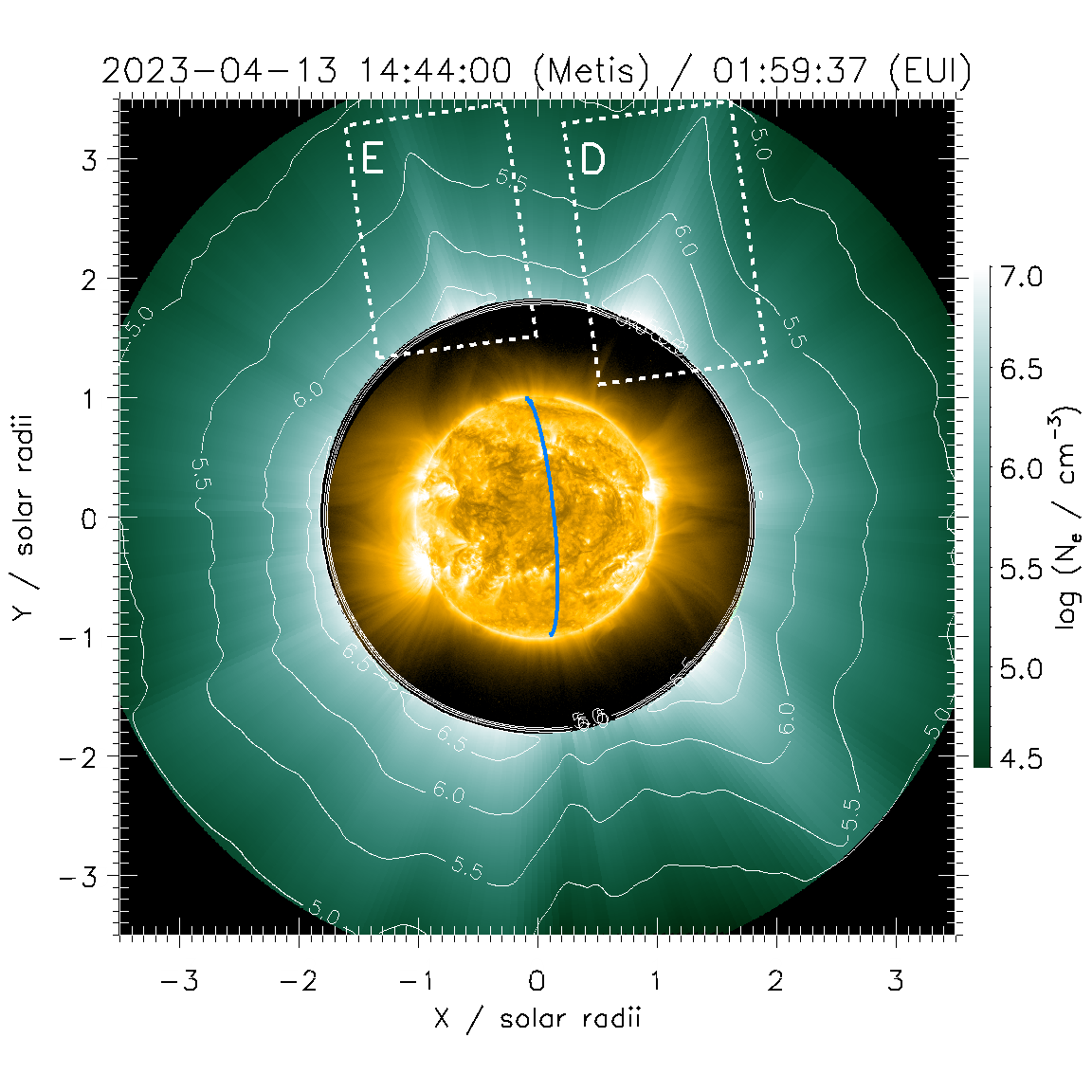}
    \caption{%
        Electron densities derived from Metis $pB$ measurements obtained right after the total brightness measurements described in this work.  In analogy with Fig.~\ref{Fig:overview}, this composite image shows also EUI/FSI174 images and the position of the solar limb (blue line) as seen from Earth.  Black areas in the 8 October 2022 Metis images mark regions where the $pB$ is saturated. Contours mark the values $\log_{10} N_\mathrm{e} = $ 5.0, 5.5, 6.0, and 6.5 (densities in cm$^{-3}$).}%
    \label{Fig:densities}
  \end{figure*}
    
    Figure~\ref{Fig:densities} shows the resulting electron densities for the three dates we have considered. For the 8 October 2022 date, the Metis $pB$ images are saturated in the bright streamer of region \textit{B}, hence no meaningful electron densities could be derived in that area.  In general, these values should be considered with all due caveats regarding the approximations used, especially the axisymmetry of the density distribution.  However, the values obtained in the streamer regions are roughly consistent with existing estimates, e.g. from \cite{Gibson-etal:1999} obtained at the minimum of the solar cycle, despite the relative complexity of the magnetic topology of the corona at the epochs of the Metis observations.  In particular, the densities along path \textit{A}-$\alpha$ are within 10\% of the \cite{Gibson-etal:1999} estimates, while the densities along the paths \textit{C}-$\alpha$ and \textit{E}-$\alpha$ (oriented along the axes of the respective pseudo-streamers) are about 35\% higher. The highest densities, roughly a factor 2 higher than the \cite{Gibson-etal:1999} estimates, are found along the streamer axis in region \textit{D} (path \textit{D}-$\alpha$).

    In summary, in the regions where the wave-like oscillations discussed in this work are observed, the electron densities are in the range 1 -- 3$\times 10^6$~cm$^{-3}$; in regions \textit{D} and \textit{E}, however, waves are visible at larger heights and therefore at lower densities, of the order of $3\times 10^5$~cm$^{-3}$ or smaller.

    In addition to the empirical estimates of quantities such as electron densities, the MHD modelling of the solar corona provide more detailed characterisation of the regions examined in this work.  It is useful, for instance, to place the propagation speeds reported in Sec.~\ref{sec:results:properties} and summarised in Table~\ref{Tab:speeds} in the context of the sound and Alfv\'en speeds in those regions.

\begin{figure*}[h!]
    \centering
    \includegraphics[height=0.2520\textwidth,trim= 0 10 240 60,clip]{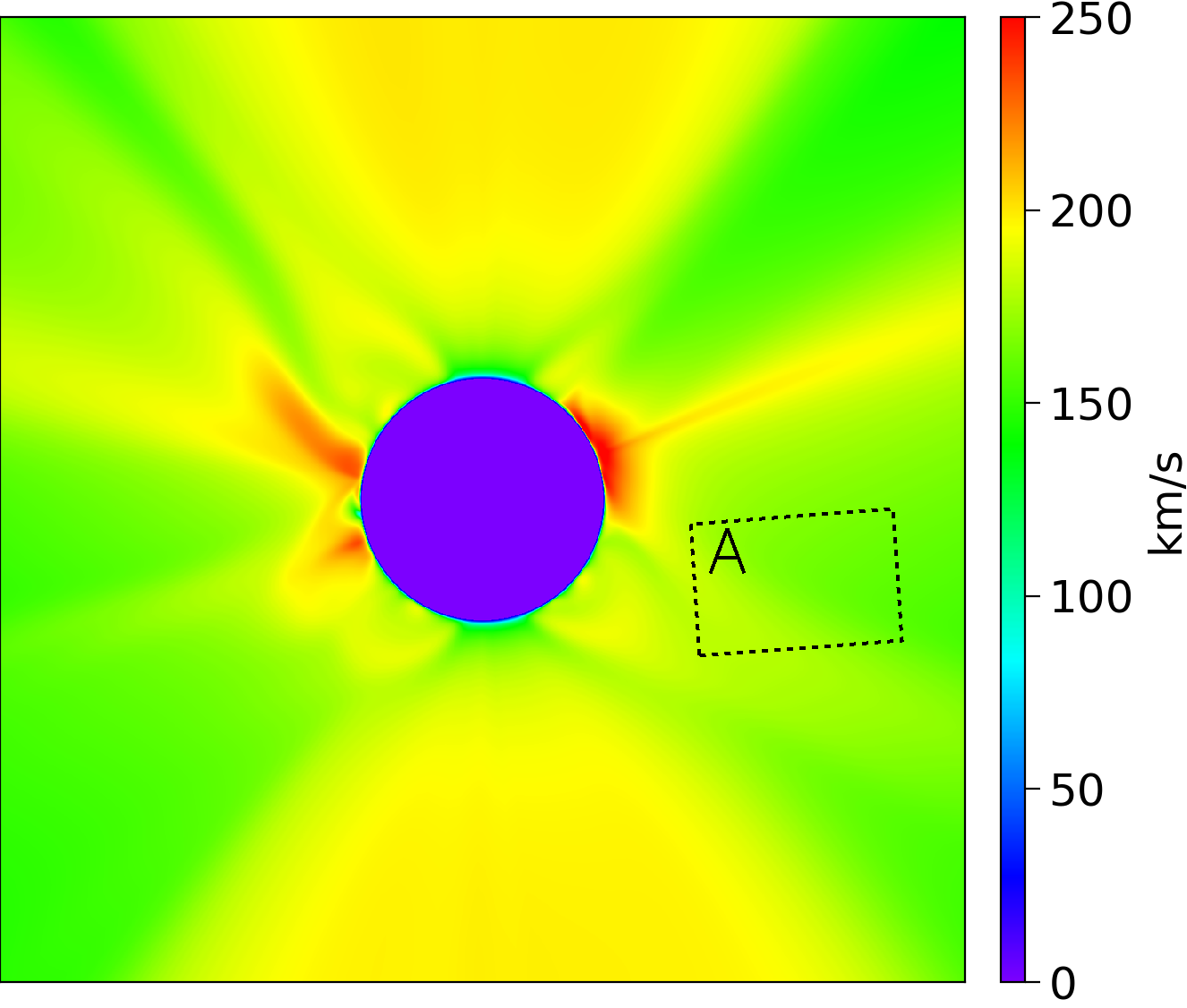}
    \includegraphics[height=0.2520\textwidth,trim= 0 10 240 60,clip]{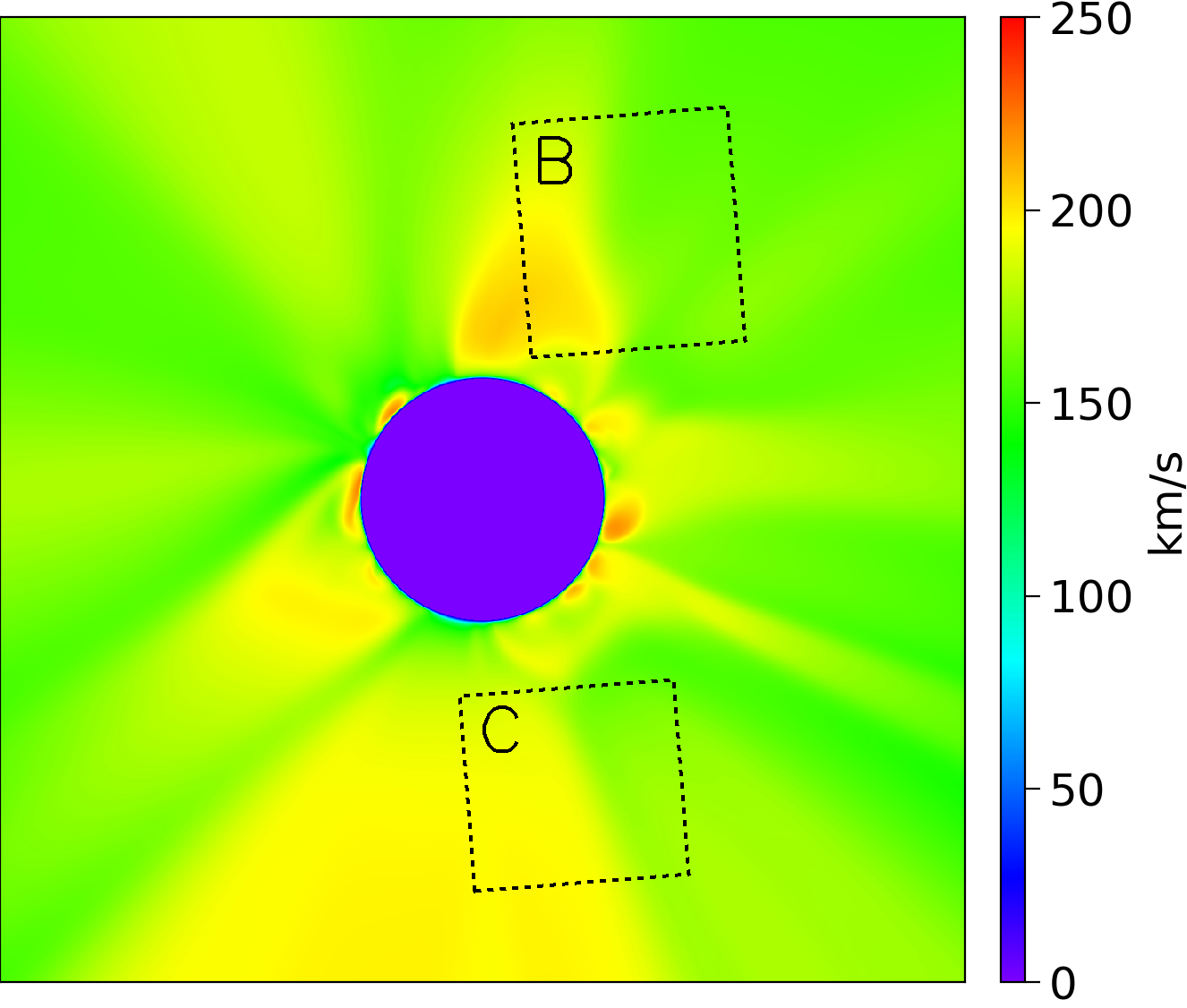}
    \includegraphics[height=0.2520\textwidth,trim= 0 10 -25 60,clip]{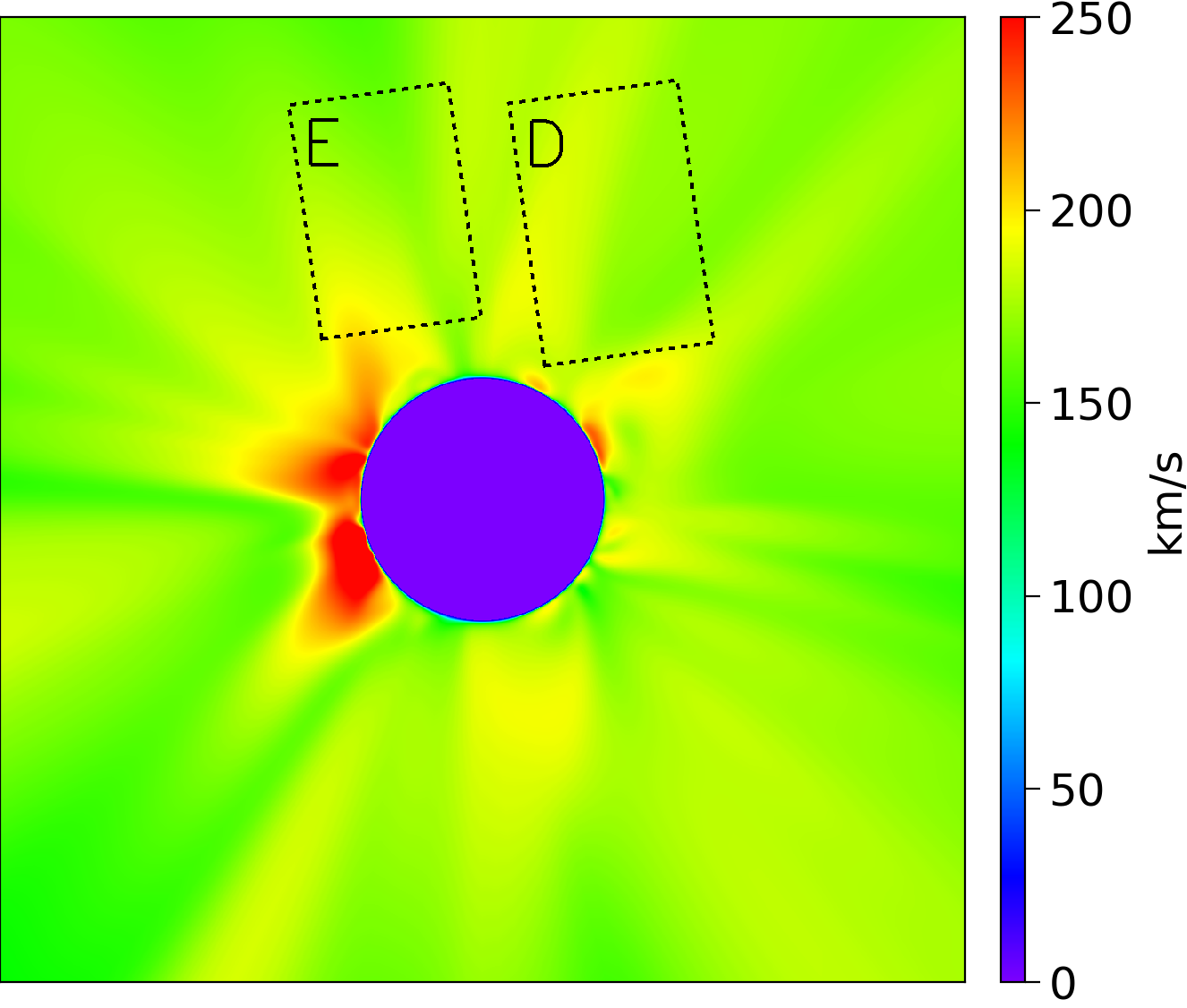}
    \includegraphics[height=0.2575\textwidth,trim=10 10 270 40,clip]{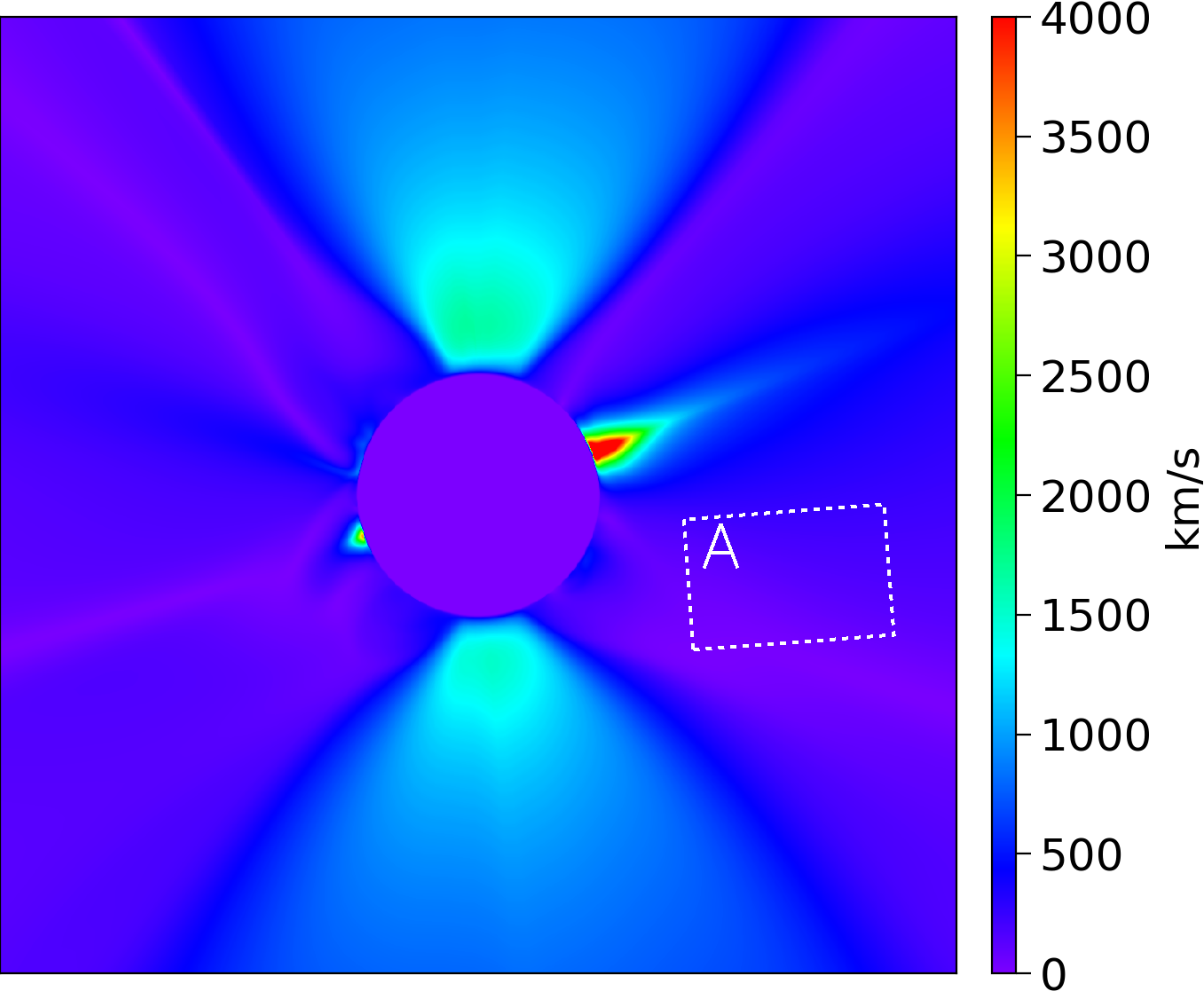}
    \includegraphics[height=0.2575\textwidth,trim=10 10 270 40,clip]{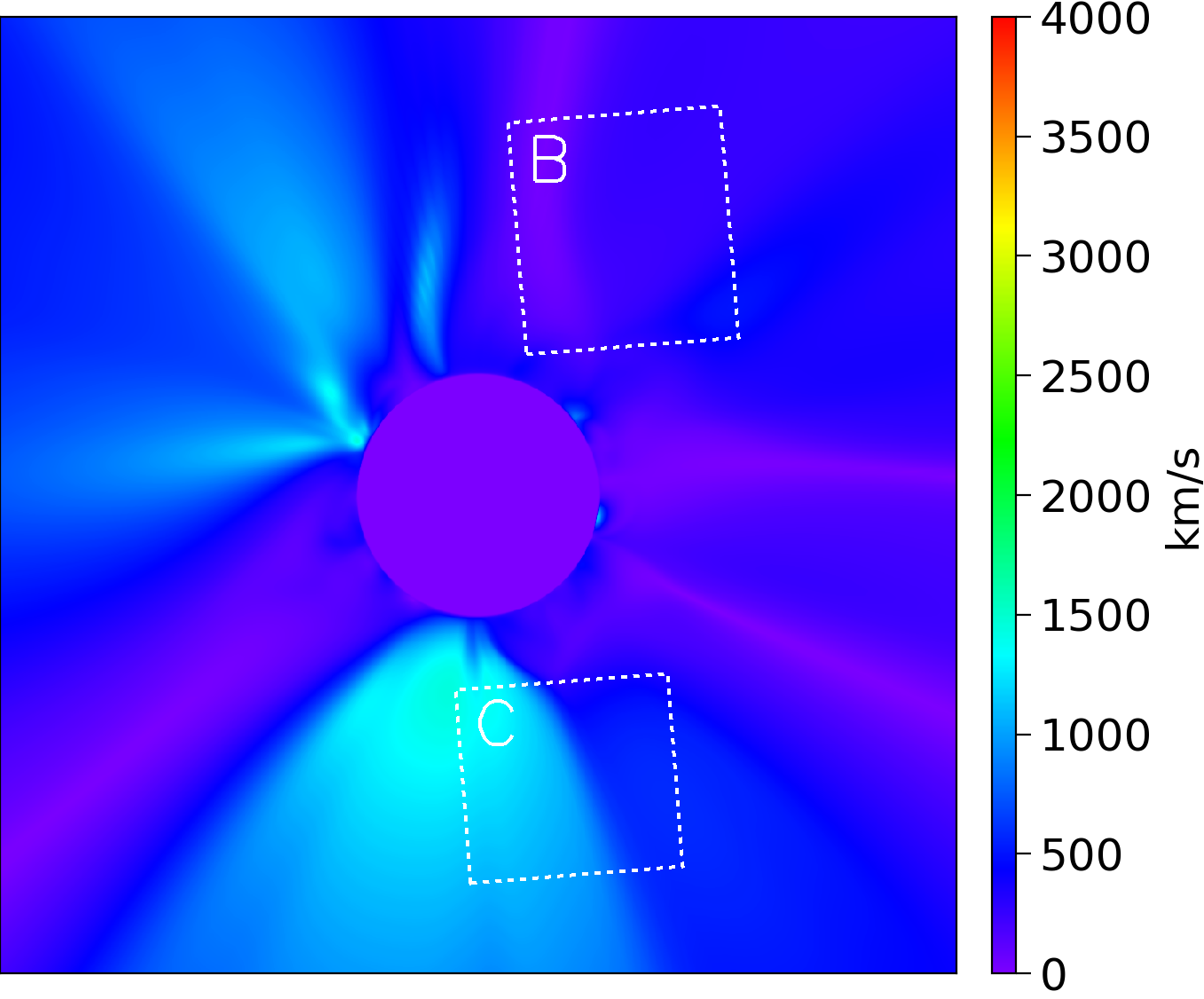}
    \includegraphics[height=0.2575\textwidth,trim=10 10   0 40,clip]{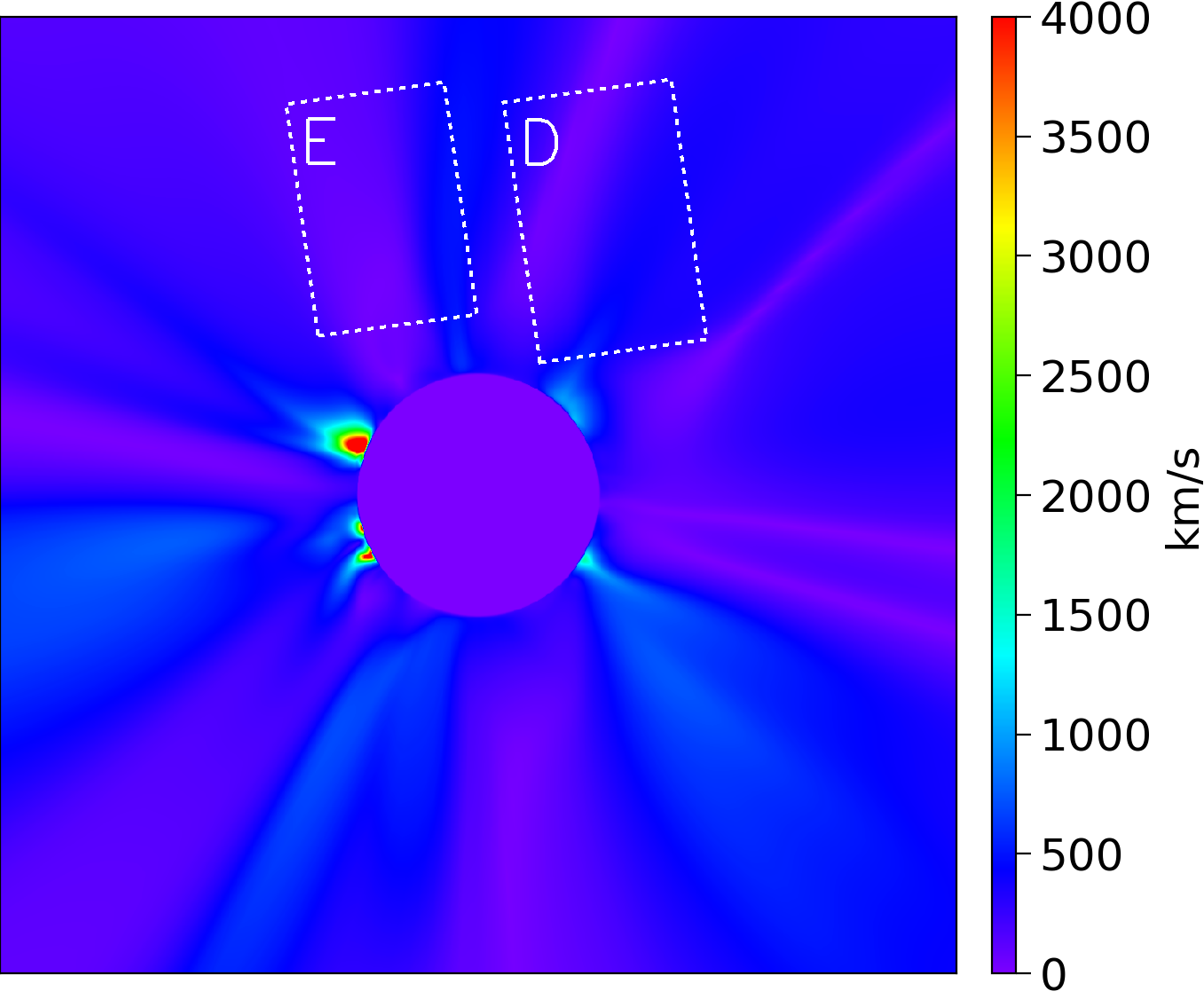}
    \caption{%
        Maps of sound and Alfv\'en speeds slices (top and bottom rows respectively) on the Metis plane of the sky for the dates considered with indication of the ROIs, in the same order as in the previous figures.
      }%
    \label{Fig:speeds_psi}
  \end{figure*}

    Figure~\ref{Fig:speeds_psi} shows in the top panels the computed sound speed in the Metis plane of the sky for the three dates considered.  The values in the regions of interest fall in the range $\sim$150 -- 200~\kms.  The computed Alfv\'en speed shown in the bottom panels are of the order of $\sim$1000~\kms for region $C$ and of the order of $\sim$500~\kms\ in all other regions.  These values are commented in Sec.~\ref{sec:discussion}.
  \begin{figure*}[h!]
    \centering
    \includegraphics[height=0.2436\textwidth,trim=10 10 240 60,clip]{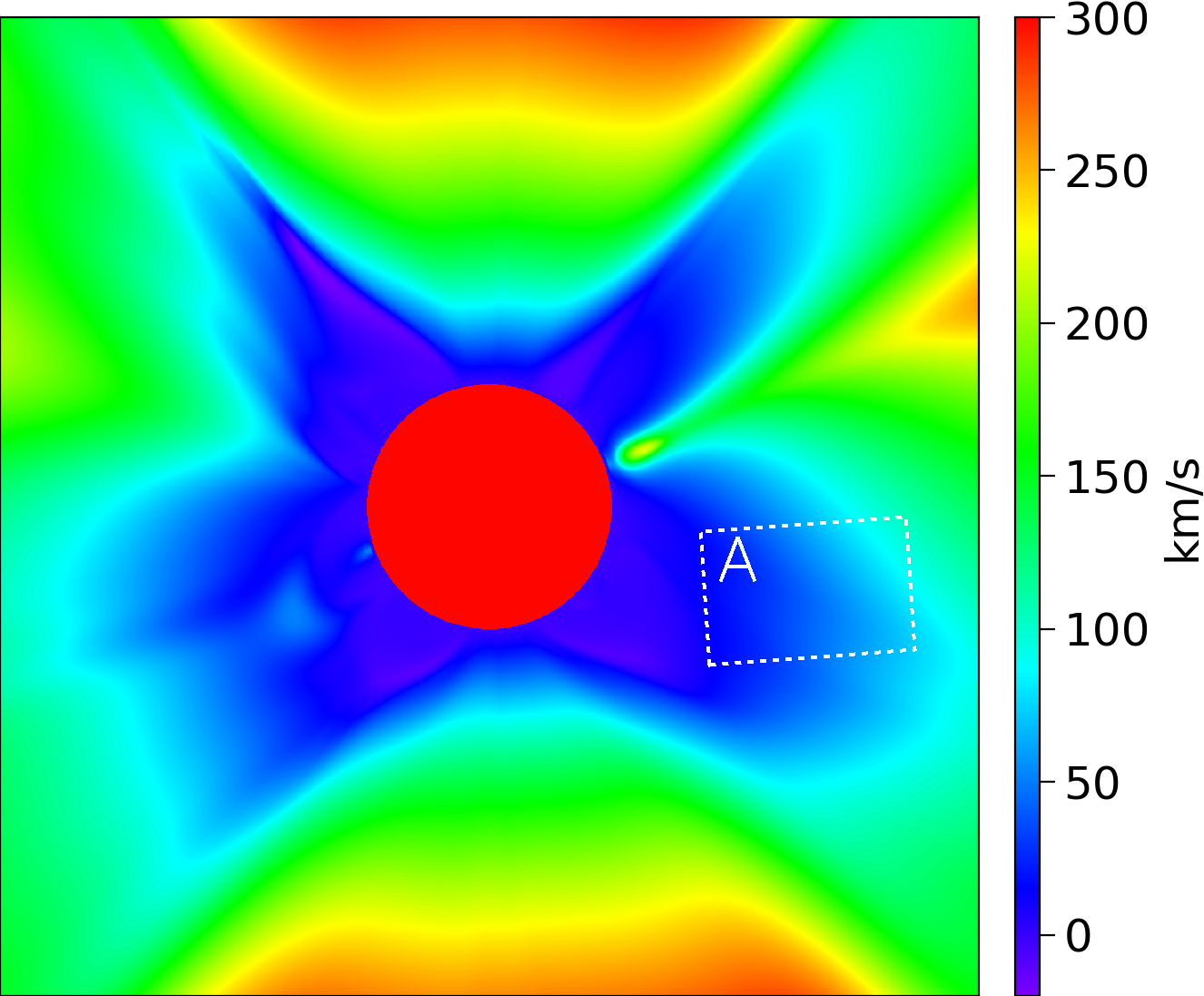}
    \includegraphics[height=0.2436\textwidth,trim=10 10 240 60,clip]{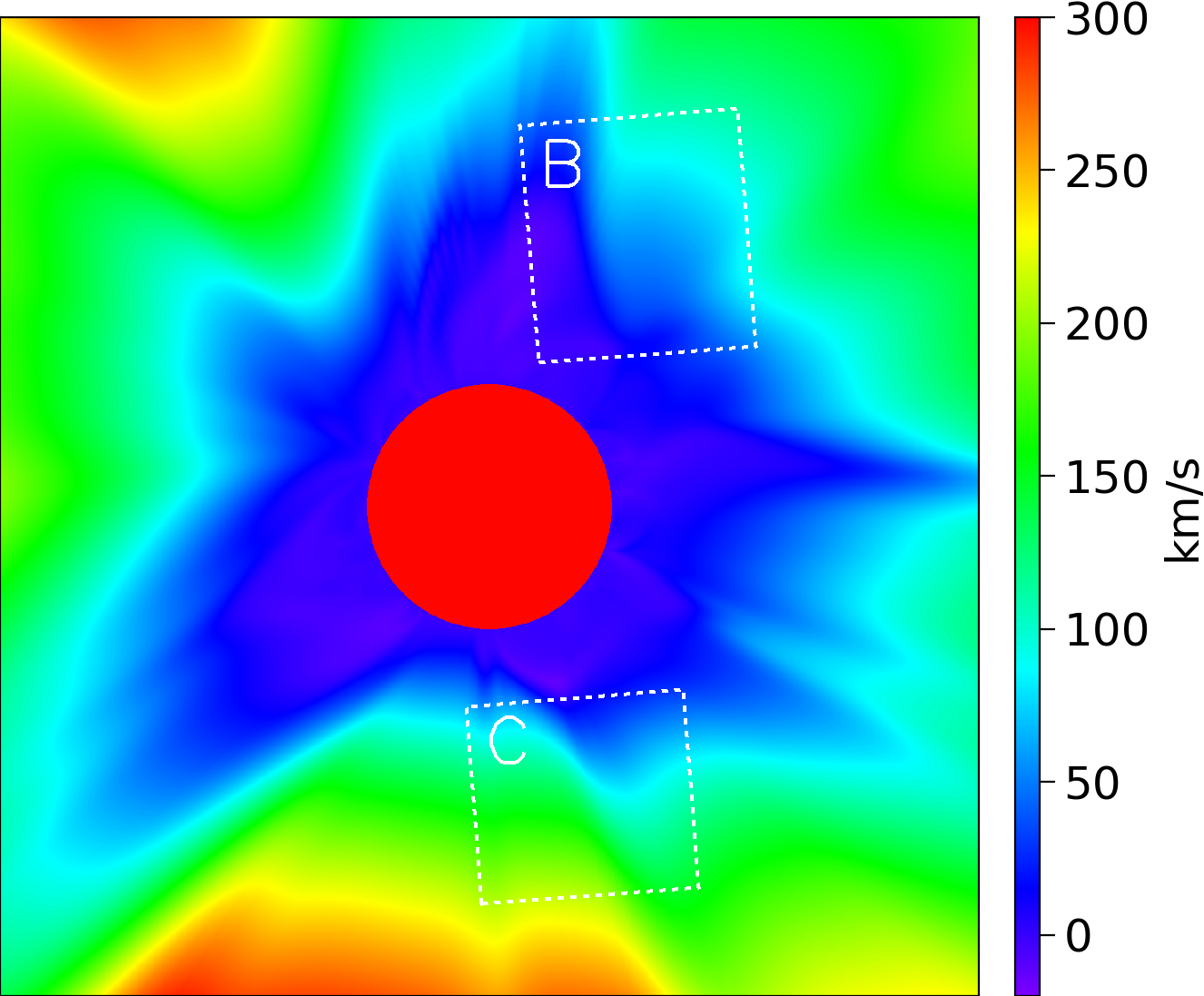}
    \includegraphics[height=0.2436\textwidth,trim=10 10   0 60,clip]{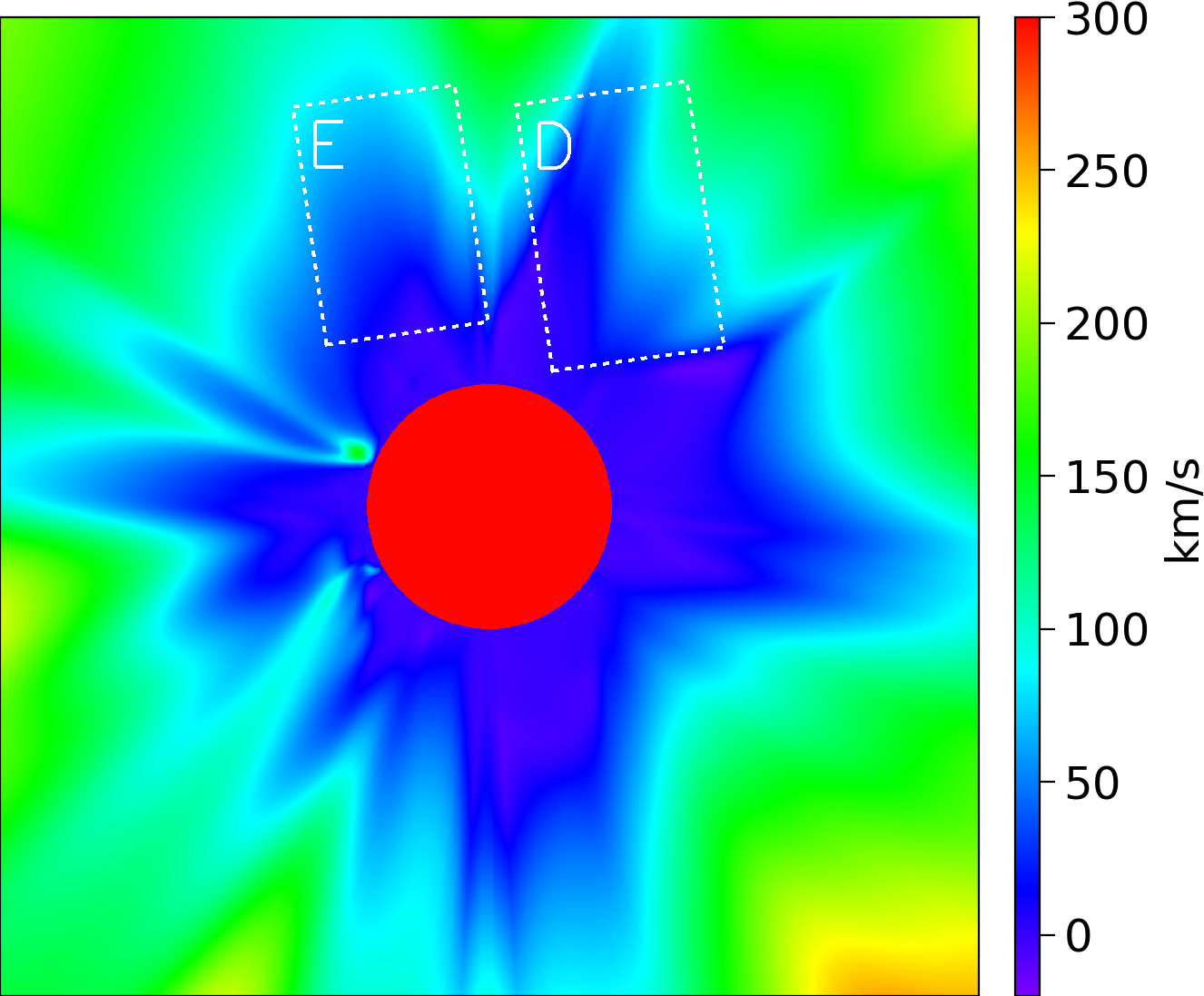}
    \caption{%
        Maps of radial speed slices on the Metis plane of the sky for the dates considered with indication of the ROIs, in the same order as in the previous figures.
      }%
    \label{Fig:wind_psi}
  \end{figure*}

    Finally, in order to support the discussion of the propagation speed of the wave-like oscillations with respect to the wind outflow speed (Sec.~\ref{sec:discussion}), we show in Fig.~\ref{Fig:wind_psi} the computed radial velocities in the Metis plane of the sky from the same MHD model.  Computed outflow speed in the pseudo-streamer of region $C$ are below $\sim$150~\kms; all other regions are characterised by even lower ($<$50~\kms) or vanishing outflow speeds.

    We also examined Metis UV data, when available, to investigate the possibility of deriving an empirical estimate of outflow velocities through the Doppler Dimming technique, i.e. through measurements of the dimming of \Lya\ radiance due to the Doppler effect \cite[e.g.][]{Withbroe-etal:1982,Noci-etal:1987}.

    The applicability of this technique to simultaneous \Lya\ and $pB$ coronagraphic images, like those obtained by Metis, has been discussed by \cite{Dolei-etal:2018,Dolei-etal:2019}.  Its successful implementation with Metis data was later demonstrated by \cite{Romoli-etal:2021} and \cite{Antonucci-etal:2023}.  A comprehensive analysis of the application of this technique to Metis data is the subject of a paper by Giordano et al., under review.

    As the Metis UV channel was not operating in early 2022, only the data sets acquired in October 2022 and April 2023 are suitable for this analysis.  As already mentioned above, $pB$ data in region $B$ are saturated, thus preventing the application of the technique to that area. We therefore only examined the data in region $C$ observed in October 2022, and in regions $D$ and $E$ observed in April 2023.  The optimal sets of parameters for this analysis is discussed in the work by Giordano et al.  Here, for simplicity, we adopted the distribution of coronal temperatures given by \cite{Gibson-etal:1999} and assumed an isotropic distribution of hydrogen kinetic temperatures, in addition to employing the electron densities derived from the co-temporal $pB$ maps.

    With these assumptions, we obtained wind outflow profiles along the axes of the bright structures of the considered regions.  In all three cases, no significant acceleration was measured.  The speed measured in region $C$ (the pseudo-streamer observed on 8 October 2022) falls in the range between 250 -- 300~\kms.  For the two regions observed in 13 April 2023 we derived lower outflow speeds: $<$150~\kms\ in the streamer of region $D$ and below 100~\kms\ in pseudo-streamer of region $E$.

    These results should be considered as indicative only: the large estimated uncertainties in particular of the results in regions $D$ and $E$ suggest that a wider set of assumptions should be considered.  A more detailed analysis is beyond the scope of this work. In any case, both the empirical estimates from the Doppler Dimming technique and the MHD models point to slow (subsonic) or negligible wind outflow speeds in the regions of interest.

\end{appendix}

\end{document}